\documentclass[useAMS,usenatbib]{mn2e}
\usepackage{txfonts}
\usepackage{graphicx}
\usepackage{pdflscape}
\usepackage{threeparttable}
\def\mum{\,$\mu$m}
\def\deg{^{\circ}}
\def\sun{$_{\odot}$}
\def\Hyp{\textit{Hyper}}
\def\Cut{\textit{Cutex}}
\def\Her{\textit{Herschel}}

\def\Msun{\,M$_{\odot}$}

\def\12co{$^{12}$CO}
\def\co13{$^{13}$CO}
\def\NH2{N$_{\mathrm{H}_{2}}$}
\def\n2h{N$_{2}$H$^{+}$}
\def\d2n{N$_{2}$D$^{+}$}
\def\nh3{NH$_{3}$}
\def\hco{HCO$^{+}$}
\def\h20{H$_{2}$O}

\def\avir{$\alpha_{vir}$}
\def\0avir{$\alpha_{0}$}
\def\aeq{$\alpha_{eq}$}

\title{Testing Larson's relationships in massive clumps}

\author[A. Traficante,  et al.]{A. Traficante$^{1}$\thanks{e-mail:alessio.traficante@iaps.inaf.it}, A. Duarte-Cabral$^{2}$, D. Elia$^{1}$, G. A. Fuller$^{3}$, M. Merello$^{1}$, S. Molinari$^{1}$, 
\newauthor
N. Peretto$^{2}$, E. Schisano$^{1}$, and A. Di Giorgio$^{1}$ \\
$^{1}$IAPS - INAF, via Fosso del Cavaliere, 100, I-00133 Roma, Italy \\
$^{2}$School of Physics and Astronomy, Cardiff University, Queens Buildings, The Parade, Cardiff CF24 3AA, UK \\
$^{3}$Jodrell Bank Centre for Astrophysics, School of Physics and Astronomy, University of Manchester, Oxford Road, Manchester M13 9PL, UK}

\begin{document}
\maketitle

\label{firstpage}

\begin{abstract}
We tested the validity of the three Larson relations in a sample of 213 massive clumps selected from the Herschel Hi-GAL survey and combined with data from the MALT90 survey
of 3mm emission lines. The clumps have been divided in 5 evolutionary stages to discuss the
Larson relations also as function of evolution. We show that this ensemble does not follow the three Larson relations, regardless of clump evolutionary phase. A consequence of this breakdown is that the virial parameter \avir\ dependence with mass (and radius) is only a function of the gravitational energy, independent of the kinetic energy of the system, and \avir\ is not a good descriptor of clump dynamics. Our results suggest that clumps with clear signatures of infall motions are statistically indistinguishable from clumps with no such signatures. The observed non-thermal motions are not necessarily ascribed to turbulence acting to sustain the gravity, but they may be due to the gravitational collapse at the clump scales. This seems particularly true for the most massive (M$\geq1000$ M\sun) clumps in the sample, where also exceptionally high magnetic fields may not be enough to stabilize the collapse.
\end{abstract}


\section{Introduction}
Massive star-forming regions are dominated by highly supersonic non-thermal motions. Velocity dispersions in giant molecular clouds \citep[GMCs, size $\simeq5-100$ pc,][]{Solomon87}, massive clumps \citep[regions with size $\simeq0.5-2$ pc, e.g.][]{Urquhart14,Traficante15b,Elia17} and massive cores \citep[size $\simeq0.1$ pc,][]{Zinnecker07} are of the order of 1-10 km s$^{-1}$, significantly higher than thermal motions ($\simeq0.25$ km s$^{-1}$ for hydrogen molecules at typical temperature T=15 K). 


The pioneering work of \citet{Larson81} investigated these motions in GMCs using the available \12co data and found that the non-thermal motions may be ascribed to internal turbulence acting to sustain the clouds against the gravitational collapse. In this work \citet{Larson81} showed that molecular clouds follow three fundamental relations: 

\begin{itemize}
\item[I]: a size-linewidth power-law relation which states that in molecular clouds the velocity dispersion $\sigma$ scales proportionally to the radius R. The first relation found by Larson was $\sigma\propto\mathrm{R}^{0.38}$. Later the analysis was refined and the relation modified to $\sigma\propto\mathrm{R}^{0.5}$ \citep[e.g.][]{Heyer04}.
\item[II]: clouds are in approximately virial equilibrium, with a virial parameter \avir=E$_{k}$/E$_{\mathrm{G}}$=5$\sigma^{2}\mathrm{R/GM}\simeq 1$, where M is the mass of the region and G the gravitational constant. This relation implies that the kinetic energy of the system E$_{k}\propto\sigma^{2}$M is of similar intensity as the gravitational energy of the system, E$_{\mathrm{G}}\propto$M$^{2}$/R. 
\item[III]: a volume density $n$-size relation, $n\propto\mathrm{R}^{-1.1}$. This relation implies that GMCs are universal structures, with a mostly uniform column density. From this relation follows indeed that the surface density $\Sigma$ is almost constant:  $\Sigma\propto\mathrm{R}^{-0.1}$. 
\end{itemize}

Early observations of GMCs confirm the validity of the three relations \citep[e.g.][]{Solomon87,Heyer04}, which were also observed in simulations of turbulent interstellar medium \citep[][and references therein]{MacLow04,McKee07}. 

These relations were questioned however over the years. For example, the validity of the third relation was attributed to selection effects \citep[e.g.][]{Kegel89}. The first and third Larson's relations in GMCs were questioned by e.g. \citet{Heyer09}. This work re-analised the GMCs using \co13\ data taken with the Boston University- FCRAO Galactic Ring Survey \citep[GRS,][]{Jackson06}. The higher critical density of \co13\ compared to \12co\ allowed to trace higher column density regions and these data demonstrate that the quantity $\sigma/\mathrm{R}^{0.5}$ and the surface density of GMCs are not constant. Nevertheless, the average value of the virial parameter in GMCs, \avir=1.9, is still consistent with virial equilibrium \citep{Heyer09}.

Challenging one of the Larson's relation has direct consequences on the other two as well: the three relations are algebraically linked. If two of them are true, the third is automatically implied \citep[e.g.][]{Kritsuk13}. At the same time, if one of the three is violated, necessarily (at least) one of the other two relations must not be true, with important implications on different star formation theories. The Larson's relationships are in fact assumed in models that predict the formation of massive stars through turbulent-regulated collapse \citep[e.g.][]{McKee03}, as opposed to gravity-dominated, almost free-fall collapse in which the theories predict for example values of the virial parameter \avir$<1$ \citep[e.g.][]{Bonnell04}.

While the validity of the Larson's relations has been widely investigated in GMCs, few and relatively small surveys have been dedicated to the study of non-thermal motions in massive star forming clumps and cores. For example, \citet{Ballesteros-Paredes11} showed that the first and third Larson's relations are violated in GMCs and massive clumps. Simulations of star forming regions showed that ensemble of clouds, clumps and cores do not follow the three Larson's relations, and this is particulalry true for the higher density regions \citep{Camacho16}. In a recent work we combined a survey of 16 massive 70\mum\ quiet clumps with several surveys of massive dense cores at different evolutionary phases and showed that the three Larson's relations seem to be violated in massive star-forming regions at the scales of clumps and cores \citep{Traficante17_PII}. However, a consistent analysis on a large sample of hundreds of massive star forming clumps at various evolutionary stages has not yet been performed. 

In this work we examine the three Larson's relations and their implications in a large sample of massive clumps obtained from the combination of the \citet{Elia17} catalogue of clumps extracted from the \Her\ Hi-GAL survey \citep{Molinari10_PASP}, with the sample of molecular lines observed at 3mm with the MALT90 survey \citep{Jackson13}. In Section \ref{sec:datasets} we describe the datasets used in this work and the selection of the final sample of 213 clumps with well defined dust and gas emission properties; in Section \ref{sec:clump_classification} we describe the classification scheme adopted for these clumps; in Section \ref{sec:gravo_turbulence} we explore in detail the three Larson's relations and we discuss the validity of these relations in massive clumps at different evolutionary stages; in Section \ref{sec:virial_dep} we analyze the implications of the previous results, in particular in the interpretation of the virial parameter; in Section \ref{sec:infall} we study the properties of the clumps that show signs of infall motions and we compare these results with the rest of the sample; in Section \ref{sec:non_thermal_motions_origin} we explore possible explanations for the observed non-thermal motions; finally, in Section \ref{sec:conclusions} we draw our conclusions.  

\section{Datasets and clumps selection}\label{sec:datasets}
In the following we will describe the main datasets we have considered in this work, the selection criteria used to obtain the final sample and the estimation of the uncertainties on the main parameters used in the rest of the paper.

\subsection{Hi-GAL data}
The \Her\ Hi-GAL survey observed the whole Galactic Plane in 5 wavelengths (70, 160, 250, 350 and 500\mum) using the two instruments, PACS \citep[][70 and 160\mum]{Poglitsch10} and SPIRE \citep[][250, 350 and 500\mum]{Griffin09}. This survey identified tens of thousands of filaments (Schisano et al. 2014) and point sources \citep{Molinari16_cat} across the Galaxy. The band-merged catalogue contains $\simeq100000$ sources in the longitude range $-71\deg\leq\ l\leq 67\deg$ with defined spectral energy distributions and clump properties \citep{Elia17}, from which we extracted the clumps used in this work.

\subsubsection{Complementary dust continuum datasets}
The Hi-GAL fluxes have been complemented at longer wavelengths with data at 870\mum\ taken from the ATLASGAL survey \citep{Schuller09}. This survey covers the Galactic longitudes $-80\deg\leq l\leq60\deg$ with a spatial resolution of 19.2\arcsec\ and a sensitivity of $\simeq70$ mJy/beam \citep{Csengeri14}. The ATLASGAL clumps catalogue \citep{Csengeri14} contains $\simeq10000$ sources, including all the sources presented in this work. 

The FIR-submm fluxes have been also complemented at shorter wavelengths with mid-infrared (MIR) data at 21\mum\ \citep[MSX,][]{Egan03}, 22\mum\ \citep[WISE,][]{Wright10} and 24\mum\ \citep[MIPSGAL,][]{Gutermuth15}. The MIR counterparts are described in the \citet{Elia17} catalogue.  We also used the results of the RMS survey \citep{Lumsden13} to classify the clumps. The RMS survey is a mid-infared (MIR) selection of massive, evolved young stellar object (YSO) candidates across the whole Galaxy identified in MSX. Details of the survey can be found in \citet{Lumsden13}. 
The source counterparts at 21, 22 and 24\mum\ have been used to determine the clumps evolutionary sequence according to the scheme proposed in Merello et al. (2018, in prep.) and summarized in Section \ref{sec:clump_classification}.

\subsection{MALT90 data}
The MALT90 survey \citep{Jackson13} is a large survey of 90 GHz ($\simeq$3 mm) emission lines associated with star forming regions. The survey observed 2012 clumps chosen from the ATLASGAL survey \citep[][]{Schuller09}. The clumps are distributed in the Galactic longitude ranges $3\deg\leq l\leq20\deg$ in the first quadrant and $300\deg\leq l\leq357\deg$ in the fourth quadrant. The survey has been carried out with the 22m Mopra telescope in on-the-fly mapping mode covering a region of 3.4\arcmin$\times$3.4\arcmin\ across each clump, centred in the ATLASGAL clump centroid position. The FWHM is 38\arcsec\ at 90 GHz and the velocity resolution is 0.11 km s$^{-1}$. Typical system temperatures were in the range $180\leq\mathrm{T}_{sys}\leq300$ K, for a typical \textit{r.m.s.} noise of $\simeq250$ mK per channel \citep{Jackson13}. The MALT90 survey observed 16 different species spanning from dense gas tracers of relatively quiescent gas as the \n2h\ ($1-0$), up to shock tracers as SiO $(1-0)$ and ionised gas tracers as H41$\alpha$ \citep{Jackson13}.

\subsection{Clumps selection}\label{sec:obs_data_reduction}
We combined the datasets provided by the Hi-GAL and MALT90 surveys to identify a statistically significant sample of clumps with known distances and well defined dust and line emission properties. 

From the 2012 Hi-GAL clumps also observed in the MALT90 survey we first excluded all clumps with longitudes l$\leq\vert10\deg\vert$, for which the distance estimation may be highly inaccurate. We also excluded all the clumps with a mass estimation M$\leq 5\times\sigma_{err}$, with $\sigma_{err}$ the error associated with the mass estimation as discussed in \citet{Elia17}. We obtain a first selection of 617 clumps.

\begin{figure}
\centering
\includegraphics[width=8cm]{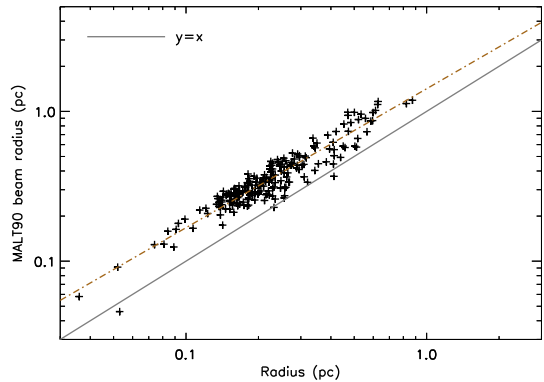} 
\caption{Comparison between radius of clumps as estimated at 250\mum\ and radius of a region equal to the size of a MALT90 beam. There is a strong correlation between the two values, and a systematic offset which shows that the MALT90 beam region is always bigger than the size of the Hi-GAL clump. The grey line is the y=x line. The red-dashed line is the fit of the distribution. The offset between the two lines has been used to estimate the filling factor, which is 0.64.} 
\label{fig:radius_malt90_radius}
\end{figure}

We further restricted our sample to well defined \n2h\ $(1-0)$ spectra that we used to estimate the gas velocity dispersion. The \n2h\ $(1-0)$ emission of each clump was evaluated from the MALT90 datacubes by averaging the spectrum across all the pixels within 1 MALT90 beam, $\simeq38\arcsec$. We assumed that all the \n2h\ emission comes from the clumps, and we estimate the filling factor from the comparison of the radius of each Hi-GAL clumps with the radius of a region equal to the MALT90 beam (Figure \ref{fig:radius_malt90_radius}). There is a strong correlation between these two quantities, and the size of the Hi-GAL clumps is systematically smaller than the radius estimated from the MALT90 beam for a factor of 0.64 on average. We assumed an average filling factor of 0.64 for the entire sample. The MALT90 datacubes are given in antenna temperature T$_{A}^{*}$ and they have been converted in main beam temperature T$_{MB}=$T$_{A}^{*}/\eta_{MB}$, assuming a mean beam efficiency $\eta_{MB}=0.49$ \citep{Miettinen14}. The properties of each \n2h\ $(1-0)$ averaged spectrum has been extracted in IDL using a hyperfine fitting routine and the \texttt{mpfitfun} algorithm \citep{Markwardt09}, after smoothing the data to a spectral resolution of 0.3 km s$^{-1}$ to enhance the S/N ratio. We excluded all the clumps with a S/N below 5, where the \textit{r.m.s.} in each smoothed datacube has been measured in a 100 km s$^{-1}$ wide spectral window near the \n2h\ emission. We further excluded clumps for which the fit converged but the spectrum was affected by spikes and/or by multiple components along the line of sight. Using this criteria, we obtained 308 clumps. We completed our selection excluding all clumps without a clear distance assignation, in particular without a well-defined resolution of the near-far distance ambiguity. First, we have refined the kinematic distances in the \citet{Elia17} catalogue (and the quantities that depend on them) with the newest set of distances developed for the Hi-GAL survey under the VIALACTEA project (Mege et al. 2018, in prep.). The method used by \citet{Elia17} was the same as in \citet{Russeil11}: the brightest emission line in the \12co or \co13 spectra along the line of sight of each source are used to estimate the velocities of the local standard of rest and converted in heliocentric distances using the \citet{Brand93} rotation curve. The Mege et  al. (2018, in prep.) distances have been determined with a similar approach, but including all the recent surveys of the Galactic Plane to trace structures along the line of sight, and using the more recent \citet{Reid09} rotation curve. Then, in order to identify only clumps with a well-defined distance estimation, we have compared the distances assigned to our 308 clumps with the distances of the MALT90 sample estimated in \citet{Whitaker17} and of the ATLASGAL sources published in \citet{Urquhart18}. We excluded from the sample all the sources with a difference in the distance estimation larger than 20\% among the three surveys.

We obtain a final selection of 213 clumps with well defined distances, dust properties and \n2h\ spectra. The properties of these clumps are summarized in Appendix \ref{app:clump_properties}.


\subsection{Estimation of uncertainties}\label{sec:uncertainties}
In this Section, we analyze the main source of uncertainties in both dust and gas properties of our 213 clumps. The results discussed in the next Sessions are significantly affected by the statistical uncertainties associated with each parameter, while they are not affected by uncertainties that produce systematic offsets.


The dust properties are mainly affected by the following source of uncertainties:

\begin{itemize}
\item Dust models for cold dust. The dust properties of the clumps in the \citet{Elia17} catalogue have been evaluated assuming a single-temperature greybody model with a spectral index $\beta=2.0$, an opacity $\kappa_{0}=0.1$ at $\lambda_{0}=300$\mum\ \citep{Beckwith90}, and a gas-to-dust ratio of 100. The commonly used model of \citet{Ossenkopf94}, assuming a thin ice mantle and a gas density of $10^{6}$ g cm$^{-3}$ leads to $\kappa_{0}=0.17$ at $\lambda_{0}=300$\mum, a difference of almost a factor of 2 from the \citet{Beckwith90} opacity. Since the mass estimate scales linearly with the opacity, using a different model would lead to a systematic offset in the mass, surface density (positive offset for higher values of $\kappa_{0}$) and virial parameter estimates (negative offset). Our results are however not affected by systematic offsets. The spectral index $\beta$ can vary across different sources as function of the dust temperature \citep{Paradis10}, as well as function of the dust column density \citep{Juvela15}. With a variation of the spectral index $\beta=2.0\pm0.3$ \citep[in line with the findings of][]{Paradis10}, the mass change for a factor of $\simeq30\%$. We consider this value as the uncertainties on the mass estimation due to the assumed dust model.


\item Errors in distance estimation.These uncertainties are due to: 1) the method used to estimate the radial velocities. The distances presented in the Mege et al. (2018, in prep.) catalogue are evaluated with a method similar to the one adopted in \citet{Urquhart18}. In this work the authors calculated an uncertainties on the distance estimation of $\simeq0.30$ kpc. Our final sample of 213 sources is at a mean distance of $\simeq4.2$ kpc, which gives an error of $\simeq7\%$ on the distance estimation due to the adopted method; 2) the rotation curve adopted to convert the radial velocities into kinematic distances. The work of \citet{Russeil11} compared the distance obtained using the \citet{Brand93} and the \citet{Reid09} rotation curves and showed that, within the uncertainties, the results are compatible. Therefore, the results are not greatly affected by different rotation curves; 3) the near-far distance ambiguity. As discussed in Section \ref{sec:obs_data_reduction}, we have selected only the conservative, but most reliable sub-sample of sources with the same solution for the near-far distance ambiguity in the Hi-GAL (Mege et al. 2018, in prep.), ATLASGAL \citep{Urquhart18} and MALT90 \citep{Whitaker17} catalogues. We therefore assumed that the distance ambiguity has been solved for our sub-sample of sources. 

These sources have a distance determination that differs up to 20\% with respect to the values in the ATLASGAL and MALT90 catalogues, but on average the difference is of only 4\%. Combining these results, we conservatively assume that the distance uncertainties are of the order of $15\%$. The same uncertainties are associated with the radius R. The mass depends on distance as M$\propto d^{2}$, so the mass uncertainties due to the distance uncertainties are $\simeq30\%$.

\item Uncertainties on radius estimation. The angular radius of each clump is defined as the geometrical mean R$_{eq}$ of the 2 FWHMs of the Gaussian fit done at 250\mum\ \citep{Elia17}. The majority of these sources are however elongated along one direction, and this asymmetry produces uncertainties in the definition of the source radius. We estimated these uncertainties by taking the differences between R$_{eq}$ and the minor and major axis of each source. The average differences are of the order of 10\% of the geometrical mean, with peaks up to 50\% and a standard deviation of $\simeq10\%$. We consider a conservative value of 20\% on the uncertainties associated with the radius estimation due to the geometrical mean approximation.

\item Results of the SED fitting routine. Mass and temperature are estimated only for clumps that have at least three consecutive fluxes in the Hi-GAL wavelengths $160\leq\lambda\leq500$\mum, and irregular SEDs are not considered \citep{Elia17}. The clumps in our sample have well defined properties and the uncertainties associated with the fitting routine are very small. They are of the order of 1.5\%, with a peak of 18\%. We assume an average error of 5\% associated with the SED fitting.

\item Choice of the photometry algorithm. In Appendix \ref{app:Hyper_photometry} we discuss how the estimates of properties such as the mass differ from the values of the \citet{Elia17} catalogue using a different algorithm to evaluate the source photometry \citep[\Hyp,][]{Traficante15a}. The differences in the estimation of the fluxes produce a systematic offset of $10\%$ in the mass values, which do not bias the results of this work. The statistical uncertainties are of the order of $\simeq25\%$, that we assume as the uncertainties on the mass estimation due to the photometry method. In Appendix \ref{app:Hyper_photometry} we also show that our results are robust against these differences, and they are not biased by the specific algorithm used to extract the properties of the clumps. 
\end{itemize}

The uncertainties associated with the estimation of the non-thermal velocity dispersion are dominated by the spectral resolution of our observations. The hyperfine fitting has been done on spectra smoothed three times, and the uncertainties on the fit are of the order of the smoothed spectral resolution, 0.3 km s$^{-1}$. The average non-thermal component of the velocity dispersion is $\simeq1.21$ km s$^{-1}$, so the error derived from the hyperfine fit is of the order of the 25\% of the measured non-thermal velocity dispersion.

The intensity of the thermal component to be subtracted from the observed velocity dispersion could be another source of uncertainties. We estimated this component assuming for the gas the same temperature of the clump, which span a range $8.5\leq\mathrm{T}\leq40$ K. The \n2h\ thermal component within this range of temperatures is much smaller than the non-thermal component, and varies in the range $0.05\leq\sigma_{th}\leq0.11$ km s$^{-1}$. Even accounting for a gas temperature which differs substantially from the estimated dust temperature, the error is $\sigma_{th,unc}\lesssim0.05$ km s$^{-1}$. This is smaller than 5\% of the measured velocity dispersion. We assume a conservative error of 5\% from the subtraction of the thermal component to the estimation of the non-thermal motions.

The uncertainties on the main parameters used in this work are summarized in Table \ref{tab:uncertainties}. The uncertainties on \avir\ has been evaluated using the standard formula for the propagation of uncertainties.

\begin{center}
\begin{table*}
\centering
\begin{tabular}{c|c|c}
\hline
\hline
Parameter & Relative uncertainties & Source of  uncertainties \\
	&	 (\%) &  \\
\hline
R & 25 & distance, geometrical mean \\
M & 50 & $\beta$ index, distance, SED fitting, photometry method \\
$\Sigma$ & 35$^{1}$ & $\beta$ index, SED fitting, photometry method \\
$\sigma$ & 30 & hyperfine fitting, thermal motions \\
\avir	 & 65 & M, R, $\sigma$ \\
\hline
\end{tabular}
\begin{tablenotes}
\scriptsize
\item $^{1}$ The uncertainties on $\Sigma$ depend only on the SED fitting and are independent of the source distance.
\end{tablenotes}
\caption{Relative uncertainties associated with the main parameters used in this work. Col. 1: Parameters; Col. 2: relative uncertainties, as estimated from the discussion in Section \ref{sec:uncertainties}; Col. 3: sources of uncertainties used to estimate the relative uncertainties on the parameters.}
\label{tab:uncertainties}
\end{table*}
\end{center}

\section{Clumps classification scheme}\label{sec:clump_classification}
The association between our 213 Hi-GAL clumps and counterparts in the MIPSGAL, WISE and RMS surveys are used to determine an evolutionary sequence for our sources. The association with MIPSGAL and WISE has been taken from the \citet{Elia17} catalogue. The association with RMS has been done looking for RMS counterparts of the Hi-GAL sample within a radius equal to the geometrical mean of the FWHMs of each source.

We adopted the evolutionary scenario based on the Hi-GAL survey and presented in Merello et al. (2018, in prep.). These authors analyzed $\simeq1000$ Hi-GAL clumps and followed a similar approach to the one used by \citet{Konig17}, but divided the clumps in 5 different evolutionary phases: 1) a starless phase, identified as bright regions at wavelengths $\lambda\geq160$\mum\ but still dark at wavelengths $\lambda<100$\mum; 2) protostar MIR dark, when a clump becomes visible at 70\mum\ but it is still dark in the MIR, or the emission too faint to be identified. These clumps are bright at all Hi-GAL wavelengths, with no counterparts in the MIPSGAL, WISE and RMS surveys; 3) protostar MIR bright, when the clumps become visible also in the MIR and their bolometric luminosity increases significantly. These clumps have at least a counterpart in one among MIPSGAL, WISE and MSX surveys but they do not pass the RMS criteria to be classified as YSOs \citep{Lumsden13}; 4) YSOs, when the protostars have reached the zero-age main sequence and become bright also in the NIR regime. They are classified as YSOs in the RMS survey; 5) HII regions, where the thermal bremsstrahlung emission of the gas ionized in the envelope of the more massive stars can be observed at radio wavelengths \citep{Wood89}. Radio observations are used to identify HII regions among YSOs \citep{Hoare07}. These sources have been classified as either UCHII or extended HII regions in the RMS catalogue. The classification scheme is summarized in Table \ref{tab:classification}.

\begin{figure}
\centering
\includegraphics[width=8cm]{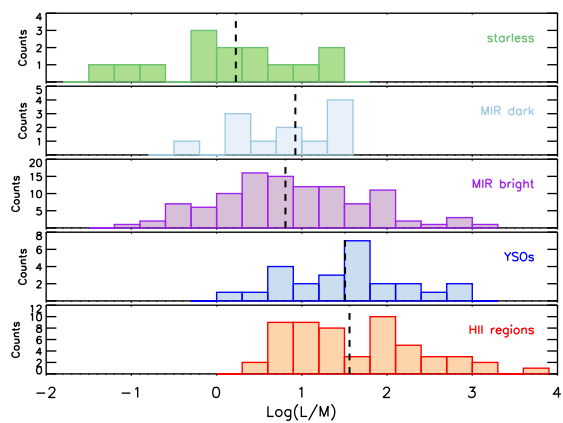} 
\caption{L/M ratio distribution of our 213 clumps divided for the various evolutionary phases. The entire sample spans more than 4 order of magnitudes and there is an evident increase of the L/M ratio going from starless to HII regions.} 
\label{fig:L_M_evolution}
\end{figure}

Based on this classification scheme, we have identified: 14 starless, 12 protostars MIR dark, 106 protostars MIR bright, 25 YSOs and 56 HII regions, among which 14 are extended HII regions. This classification is consistent with the results we obtain from a well-known indicator of clumps evolution, the luminosity over mass (L/M) ratio  \citep{Molinari08,Molinari16_l_m}. As showed in Figure \ref{fig:L_M_evolution}, the L/M ratio of these clumps spans more than four order of magnitudes in total, and increases from starless to HII regions. In agreement with the findings of Merello et al. (2018, in prep.), there is no significant differences in L/M between MIR dark and MIR bright sources, suggesting that the presence of a MIR source in a clump do not alter significantly the total luminosity of the cold dust envelope.

\begin{figure}
\centering
\includegraphics[width=8cm]{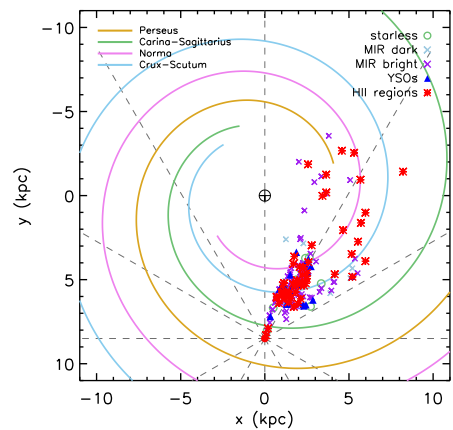} 
\caption{Galactic distribution of the 213 clumps investigated in this work. Colours and symbols represent clumps at different evolutionary stages. Overlaid is the Galactic model of \citet{Hou09}, the same used to discuss the clumps distribution in \citet{Elia17}. All clumps are in the IV Quadrant, mostly distributed across the Crux-Scutum  and Norma arms and in the inter-arms region. The red-filled dot at [0,8.5] kpc represents the position of the Sun. The black circle at [0,0] kpc is the Galactic Center.} 
\label{fig:Galactic_distribution}
\end{figure}

In Figure \ref{fig:Galactic_distribution} we present the Galactic distribution of our sources, overlaid with the four spiral-arms Galactic model of \citet{Hou09}. All sources are located in the IV Quadrant. They are mostly concentrated in a region between the Crux-Scutum and the Norma arms, and the inter-arms region.

In the next Sections we will investigate the gravo-turbulent properties of these clumps, in light of this classification scheme.

\begin{center}
\begin{table*}
\centering
\begin{tabular}{c|c|c|c|c}
\hline
\hline
Evolutionary & Bright  & Survey & L/M & Count \\
phase	&	wavelengths (\mum) &  & (L\sun/M\sun) & \\
\hline
Starless & $>$70 & Hi-GAL & 1.7 & 14 \\
Protostellar MIR dark & $\geq$70 & Hi-GAL & 8.4 & 12 \\
Protostellar MIR bright & $\geq$21 & Hi-GAL, MIPSGAL, WISE & 6.4 & 106 \\
YSOs	&	$\geq$8	& Hi-GAL, MIPSGAL, WISE, RMS & 32.1 & 25 \\
HII regions (UCHII + ext. HII) 	&	$\geq$8 \& radio emiss.	& Hi-GAL, MIPSGAL, WISE, RMS & 36.3 & 56 (42+14) \\
\hline
\end{tabular}
\caption{Classification scheme of our 213 clumps following the evolutionary scenario described in Merello et al. (2018, in prep.). Col 1: Evolutionary phase; Col. 2: wavelengths at which each evolutionary stage becomes bright; Col. 3: Survey with a visible counterpart in the data; Col.4: median value of L/M; Col. 5: Number of identified objects.}
\label{tab:classification}
\end{table*}
\end{center}

\section{Larson's relations in massive clumps}\label{sec:gravo_turbulence}
In this Section we analyse the three Larson's relations in our clumps and we discuss the implications of the results.

\subsection{Larson's first relation: linewidth-size}\label{sec:Larson_first}
The Larson's first relation shows the proportionality between the size of GMCs and the non-thermal motions of the gas in the region \citep{Larson81}. This relation has been often considered to be due to the interstellar turbulence. The interstellar medium modeled as a turbulent fluid dominated by shocks follows a power-spectrum relation of the form R$\propto\sigma^{0.5}$ \citep[i.e. a Burgers-like power-spectrum $E_{k}\propto k^{-2}$, e.g.][]{McKee07}, a scenario that reproduces the large-scale observations \citep{Padoan02,McKee07}. However, the relation seems to break in massive clumps embedded in molecular clouds. For example, \citet{Caselli95} observed the Orion A and B high-mass star forming regions and found a correlation between size and velocity dispersion of the form R$\propto\sigma^{0.21}$, significantly lower than that found in GMCs. Similar results have been obtained in the survey of high-mass star forming regions of \citet{Shirley03}. At the same time, other surveys of massive star-forming objects found no correlation \citep{Plume97,Ballesteros-Paredes11,Traficante17_PII} or even an inverse correlation \citep{Wu10} between size and linewidth.

\begin{figure}
\centering
\includegraphics[width=8cm]{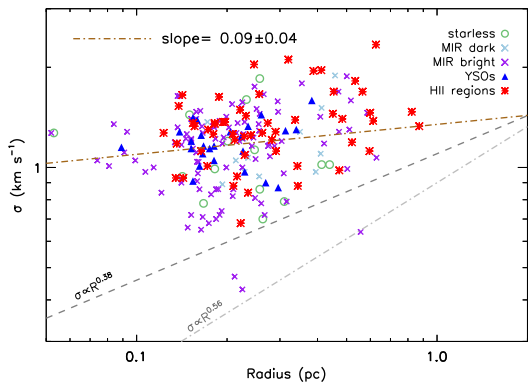} 
\caption{First Larson relation: velocity dispersion $\sigma$ as function of the radius R. The dark grey dashed line is the original Larson's relation, $\sigma\propto\mathrm{R}^{0.38}$, the light grey dash-dotted line is the revised \citet{Heyer04} relation, $\sigma\propto\mathrm{R}^{0.56}$. The correlation is weak, with a Pearson's coefficient of $\rho=0.26$.} 
\label{fig:velocity_size}
\end{figure}

In Figure \ref{fig:velocity_size} we report the velocity dispersion-radius relationship for our sample of 213 sources. The slope of the linear fit in the log-log space is $0.09\pm0.04$, suggesting that a correlation between velocity dispersion and radius, if present, is very low.  The fit in this plot (and for the rest of this work) has been obtained from a linear regression done with the \texttt{fitexy} IDL routine, which performs a chi-square approximation when uncertainties are known in both the x and y variables. The Pearson's correlation coefficient $\rho$, which measures the linear correlation between two variables and varies in the range $-1\leq\rho\leq1$, with $\rho=-1$ indicating total anti-correlation, $\rho=1$ total correlation and $\rho=0$ no correlation, has a value $\rho\simeq0.26$, also suggesting that a correlation between these two variables is rather weak.

\begin{figure}
\centering
\includegraphics[width=8cm]{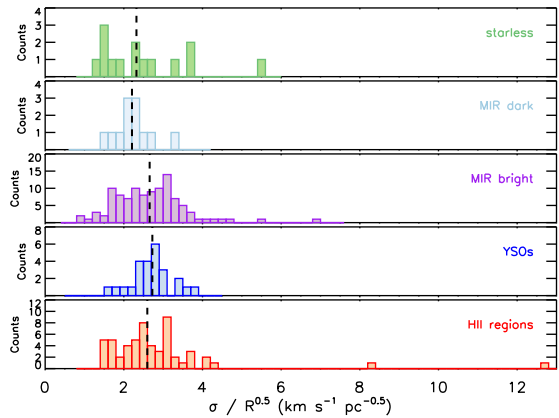} 
\caption{First Larson's relation distribution for our clumps, divided for different evolutionary stages. The relation states that the quantity $\sigma/\mathrm{R}^{0.5}$ should be a constant. Instead, the distributions span a range of values for each evolutionary phase.}
\label{fig:Larson_histo}
\end{figure}

In Figure \ref{fig:Larson_histo} we report the quantity $\sigma/\mathrm{R}^{0.5}$ divided for the different evolutionary stages. Following the first Larson's relation, this quantity should be a constant of the system. Instead, we find a distribution of this quantity across a range $0.9\leq\sigma/\mathrm{R}^{0.5}\leq12.8$ km s$^{-1}$ pc$^{-1/2}$, larger than the estimate uncertainties of size and velocity dispersion combined (see Section \ref{sec:uncertainties}). Also, within uncertainties there is no distinction between different evolutionary phases, with median values of [2.32,2.21,2.66,2.73,2.60] km s$^{-1}$ pc$^{-0.5}$ in the starless, protostar MIR dark, protostar MIR bright, YSOs and HII regions phases respectively. Altogether or divided for different evolutionary phases, these results suggest that the first Larson's relation typically breaks down at clump scales, and this break is not due to the different internal conditions of these objects.

The observed first Larson's relation implies that one, or both of the other two relations must not be followed by this ensemble of clumps.

\subsection{Larson's second relation: the virial equilibrium}\label{sec:Larson_second}
The second Larson's relation states that GMCs are approximately in virial equilibrium. The virial parameter \avir\ has been often interpreted as representative of the equilibrium between $\mathrm{E}_{k}$ and $\mathrm{E}_{\mathrm{G}}$ when all other forces such as magnetic fields are not involved \citep[and assuming spherical and homogeneous density distribution,][]{Bertoldi92}. The virial equilibrium implies \avir=\aeq=1 or, if a collapsing cloud is modeled as an isothermal (Bonnor-Ebert) sphere, the hydrostatic equilibrium is at \aeq$\simeq2$ \citep{Kauffmann13,Tan14}. GMCs are expected to be in virial equilibrium, with the kinetic energy due to local turbulence that provides support against the gravitational collapse \citep{McKee03,Heyer09}. The formation of massive clumps in a gravo-turbulent collapse is also predicted to happen in a state of global virial equilibrium \citep{Lee16}.

Alternatively, the observed non-thermal motions may partly be the result of the collapse itself, and not necessarily providing support against gravity. In this interpretation virial equilibrium looses its original meaning. The regions would be in approximately virial equipartition (which also implies \aeq=2), but misinterpreted as in virial equilibrium \citep{Ballesteros-Paredes06}.

Independently of the interpretation of the observed \avir, there is a general consensus that regions with \avir$<$\aeq\ are gravitationally bound and prone to collapse, if not sustained by strong magnetic fields which may stabilize them \citep[e.g.][]{Kauffmann13}. These regions do not follow the second Larson's relation.

\begin{figure}
\centering
\includegraphics[width=8cm]{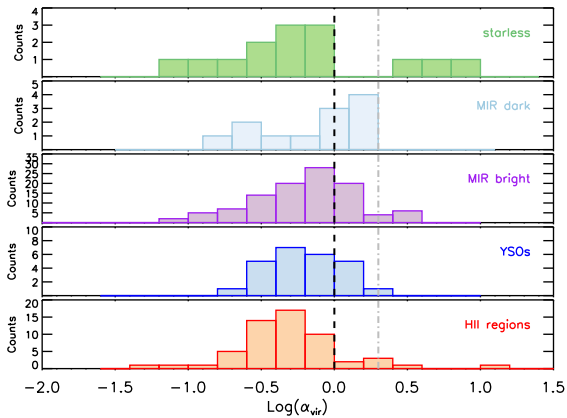} 
\caption{Histogram of the \avir\ distribution at different evolutionary stages. The majority of the clumps have \avir$\leq2$, independently from the evolutionary phase. The black dashed line is in correspondence of \avir=1, and the grey dash-dotted line is at \avir=2.} 
\label{fig:alpha_vir_histo}
\end{figure}

Figure \ref{fig:alpha_vir_histo} shows the distribution of the virial parameter of our 213 clumps divided in the five evolutionary stages, and in Table 3 we report the range of \avir\ for each phase. The virial parameter spans the range $0.05\leq\alpha_{vir}\leq12.8$, and each evolutionary phase spans at least one  order of magnitude, with no clear differences between the various stages of evolution. A total of 51 clumps have \avir$\geq1$, and only 14 have \avir$\geq2$. The majority of our clumps are gravitationally bound and these clumps, if not sustained by strong magnetic fields (see Section \ref{sec:magnetic_field}), are not in gravitational equilibrium. If the kinetic energy is due to turbulence acting to support gravity, its contribution is not sufficient to stop or slow-down the collapse at the clump scales.

\begin{center}
\begin{table}
\centering
\begin{tabular}{c|c}
\hline
\hline
Evolutionary phase & \avir \\
\hline
Starless & $0.07\leq\alpha_{vir}\leq7.98$  \\
MIR dark & $0.13\leq\alpha_{vir}\leq1.71$ \\
MIR bright &  $0.07\leq\alpha_{vir}\leq3.52$ \\
YSOs	&	 $0.20\leq\alpha_{vir}\leq2.03$ \\
HII regions 	& $0.05\leq\alpha_{vir}\leq12.81$ \\
\hline
\end{tabular}
\caption{Range of values of \avir\ in our clumps for each stage of evolution. Col. 1: Evolutionary phase; Col. 2: Range of values of \avir.}
\label{tab:alpha_vir}
\end{table}
\end{center}

\subsection{Larson's third relation: mass-radius diagram}\label{sec:Larson_third}
A practical form of the Larson's third relation states that molecular clouds have approximately the same surface density: $\Sigma\propto n\mathrm{R}\propto\mathrm{R}^{-0.1}$. This formulation is much easier to verify experimentally, since it does not require placing any constraints on the third dimension needed to evaluate the volume density of the observed regions.

The early observations of \citet{Larson81} suggest that GMCs have all similar column densities. However, the third Larson's relation may simply be an observational bias due to the molecular tracer used in early GMC observations \citep[e.g.][]{Kegel89,Ballesteros-Paredes06,Heyer09}. Using extinction as a tracer of molecular gas, \citet{Lombardi10} demonstrated that the third Larson's relation is observed in nearby molecular clouds only above a given surface density threshold. The relation does not hold in clumps and cores embedded in single clouds, and an apparent density-size relation may be observed as an artifact of clumps limited within column density thresholds \citep{Camacho16}. Indeed, several surveys of massive clumps have shown that they span almost two order of magnitude in surface densities \citep{Urquhart14,Traficante15b,Svoboda16,Elia17}.

\begin{figure}
\centering
\includegraphics[width=8cm]{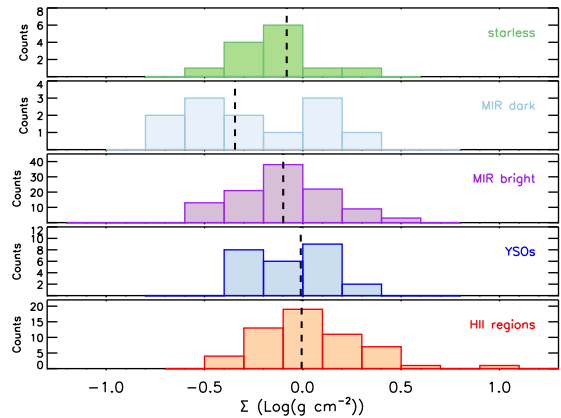} 
\caption{Surface density distribution of the clumps separated for each evolutionary phase. The black vertical lines are in correspondence of the median values of $\Sigma$ in each evolutionary stage. The surface density values span more than one order of magnitude, and there is no clear evidence of a trend with evolution.}
\label{fig:surface_density_histo}
\end{figure}

The surface densities of the 213 clumps analised in this work are in Figure \ref{fig:surface_density_histo}. The surface densities are in the range $0.13\leq\Sigma\leq8.57$ g cm$^{-2}$, spanning more than one order of magnitude in each evolutionary phase. We found median values of [0.83, 0.45, 0.80, 0.98, 0.99] g cm$^{-2}$ in starless, protostar 24\mum\ dark, protostar 24\mum\ bright, YSOs and HII regions respectively. Within the uncertainties in the estimation of the surface densities (35\% of their value, Table \ref{tab:uncertainties}), the median distribution are likely indistinguishable, in agreement with with the findings of \citet{Urquhart14} and \citet{Svoboda16}, and with the results presented in Merello et al. (2018, in prep.). The surface density of each clump may be more likely related with the density properties of the local environment, regardless of its evolution.

\subsubsection{Mass-radius relationship}
An alternative way to look at the third Larson's relation is through the mass-radius diagram. A sample of star-forming regions with roughly constant column density should have a mass distribution M$\propto$R$^{\delta}$ with $\delta\simeq2$. Previous surveys of massive clumps have found a large range of values for $\delta$, which is strongly dependent of the different strategies used to extract the dust properties of the clumps. Mass-radius diagrams have been observed with slopes in the range $\delta\simeq1.6-1.7$ \citep{Lombardi10,Kauffmann10b,Urquhart14}, as well as with $\delta\geq2$ \citep{Ellsworth-Bowers15} or even greater \citep[$\delta\geq2.7$,][]{Ragan09}.

\begin{figure}
\centering
\includegraphics[width=8cm]{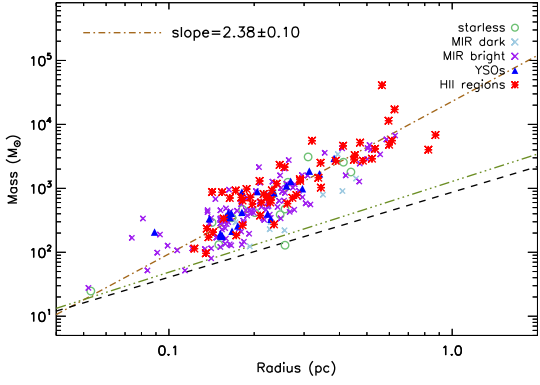} 
\caption{Mass versus radius distribution. The green dash-dotted line is the \citet{Kauffmann10} massive star formation threshold (M(r)$>870$ M\sun\ ($\mathrm{R/pc})^{1.33}$), and the black dashed line is the revised \citet{Baldeschi17} threshold (M(r)$>1282$ M\sun\ ($\mathrm{R/pc})^{1.42}$). The red dashed line is the best fit to our sample, M$\propto\mathrm{R}^{2.38\pm0.10}$.}
\label{fig:mass_radius}
\end{figure}

In Figure \ref{fig:mass_radius} we show the mass-radius diagram of our clumps. The fit has a slope $\delta=2.38\pm0.10$, not in agreement with constant $\Sigma$. We investigated how much this result is sensitive to the estimated uncertainties by modifying the errors associated with the parameters of $\pm10\%$. We obtained a difference of up to 5\% in the value of the slope, which varies in the range $2.19\pm0.09\leq\delta\leq2.53\pm0.13$. Also accounting for these variations, the slope is not consistent with constant surface density, as expected from the large spread showed in Figure \ref{fig:surface_density_histo}.

The mass-radius diagram is also a useful tool to investigate clumps which may likely form high-mass stars. The vast majority of these clumps may form high-mass objects, following the empirical mass-radius thresholds determined by \citet{Kauffmann10} and \citet{Baldeschi17}. All but two clumps (G333.449-00.183 and G338.917+00.382) are above the former, and all but three clumps (G333.449-00.183, G338.917+00.382 and G343.938+00.097) are above the latter, which is a more stringent threshold.

To conclude this Section, we have shown that the Larson's relations do not describe the dynamical properties of an ensemble of (massive) clumps. In the next Section we will explore some implications of this evidence.

\section{Virial parameter dependences}\label{sec:virial_dep}
A consequence of the breakdown of the Larson's relations in massive clumps is the dependence of the virial parameter with the dust properties of these regions (mass and radius). There is an observed trend of decreasing virial parameter at increasing mass, which is interpreted as the most massive regions are also the more gravitationally bound. For example, the massive star-forming regions analyzed by \citet{Urquhart14} showed a power-law form \avir$\propto\mathrm{M}^{\alpha}$ with $\alpha=-0.53\pm0.16$. Similarly, \citet{Kauffmann13} found a slope varying in the range $-1\leq\alpha\leq-0.4$ among various surveys of massive clumps and cores.

As noted by \citet{Kauffmann13}, the slope $\alpha$ depends also on both the first and third Larson's relationships:

\begin{equation}\label{eq:avir_mass_slope}
\frac{\mathrm{d\ log(\alpha)}}{\mathrm{d\ log(M)}}=\frac{2\frac{\mathrm{d\ log(\sigma)}}{\mathrm{d\ log(R)}}+1-\frac{\mathrm{d\ log(M)}}{\mathrm{d\ log(R)}}}{\frac{\mathrm{d\ log(M)}}{\mathrm{d\ log(R)}}}
\end{equation}

If the size and velocity dispersion in star forming regions are not correlated, $\mathrm{d\ log(\sigma)}/\mathrm{d\ log(R)}=0$. Equation \ref{eq:avir_mass_slope} becomes:

\begin{equation}\label{eq:avir_mass_slope_no_larson}
\frac{\mathrm{d\ log(\alpha)}}{\mathrm{d\ log(M)}}=\frac{1-\frac{\mathrm{d\ log(M)}}{\mathrm{d\ log(R)}}}{\frac{\mathrm{d\ log(M)}}{\mathrm{d\ log(R)}}}
\end{equation}

which implies that the slope of the \avir-mass diagram depends \textit{only} on the slope of the mass-radius diagram. 

\begin{figure}
\centering
\includegraphics[width=8cm]{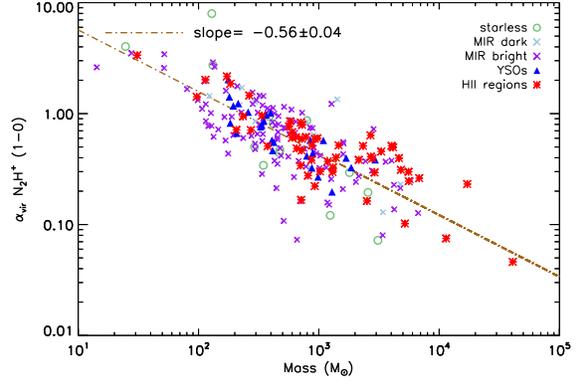} 
\caption{Virial parameter distribution as function of mass. The red dashed line is the best-fit to our data, which gives a slope $\alpha=-0.56\pm0.04$.}
\label{fig:mass_virial}
\end{figure}

Applying Equation \ref{eq:avir_mass_slope_no_larson} to our clumps, since the mass-radius slope is $\delta=2.38\pm0.10$ (see Section \ref{sec:Larson_third}), the predicted \avir-mass slope is $\alpha-0.58\pm0.02$. In Figure \ref{fig:mass_virial} we show the \avir\ vs. mass diagram. We found a slope $\alpha=-0.56\pm0.04$, in agreement with the prediction of Equation \ref{eq:avir_mass_slope_no_larson}. The fit is robust against the estimation of the uncertainties, with a variation of less than $\simeq1\%$ assuming a variation of $\pm10\%$ on the errors associated with M and \avir.

Similarly, the slope of the virial parameter-radius diagram is

\begin{equation}\label{eq:alpha_radius_no_larson}
\frac{\mathrm{d\ log(\alpha)}}{\mathrm{d\ log(R)}}=2\frac{\mathrm{d\ log(\sigma)}}{\mathrm{d\ log(R)}}+1-\frac{\mathrm{d\ log(M)}}{\mathrm{d\ log(R)}}
\end{equation}

In the case where the first Larson's relation is not valid, the slope depends, again, only on the slope of the mass-radius diagram.

\begin{figure}
\centering
\includegraphics[width=8cm]{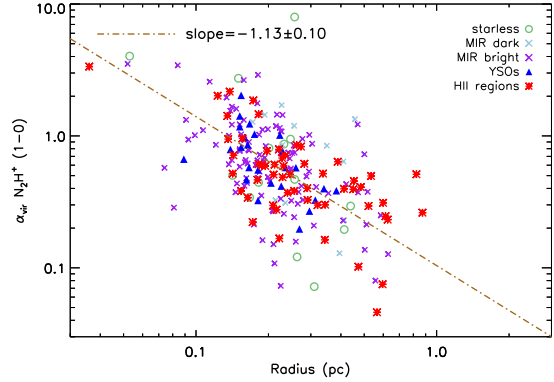} 
\caption{Distribution of the virial parameter as function of clumps radius R. The best-fit line has a slope of $\alpha_{r}=-1.13\pm0.10$.}
\label{fig:radius_virial}
\end{figure}

The slope of the \avir-radius diagram predicted from Equation \ref{eq:alpha_radius_no_larson} is  $\alpha_{r}=-1.38\pm0.06$. The \avir-radius diagram is in Figure \ref{fig:radius_virial} and the slope is $\alpha_{r}=-1.13\pm0.10$, slightly lower than the predicted value. This fit is also the most sensitive to the estimation of the uncertainties. A variation of $\pm10\%$ on the errors associated with R and \avir\ leads to a difference of  more than 50\% in $\alpha_{r}$, which varies in the range $-0.97\pm0.10\leq\alpha_{r}\leq-1.50\pm0.12$, within the prediction of Equation \ref{eq:alpha_radius_no_larson}.

These results show that in our sample the kinematic properties of the clumps do not affect the distribution of the virial parameter, which are driven by the dust properties (mass and size) of these sources. In other words, the concept of virial equilibrium may be misleading: in a sample that violates the three Larson's relations the virial parameter varies independently from the kinetic energy of the sources.

Note that these results are valid for an ensemble of clumps, where the non-thermal motions are estimated with a single gas tracer. This common choice however biases the observations toward regions with similar volume densities within each clump, regardless of the clumps physical properties (i.e. of mass and size). The correlation between \avir\ and mass seems indeed to disappear when different surveys of clumps and cores observed with different tracers are combined together \citep{Kauffmann13}. These results may differ from the analysis of the energy balance within single star-forming regions. They may all be near virial equilibrium, as predicted by e.g. \citet{Lee16}, but the kinetic energy in the more massive clumps could not be properly measured.

The virial parameter determined for an ensemble of massive clumps may not be a good descriptor of the clumps dynamics, an hypothesis that we will explore in the next Section.

\section{Infalling properties}\label{sec:infall}
If the virial parameter is not a good descriptor of the energy balance in an ensemble of massive clumps, these regions may be gravitationally bound and collapsing independently from the value of \avir. One way to explore this hypothesis is to compare the properties of clumps with signs of infall motions against clumps with no such signatures.

The MALT90 survey observed also the \hco\ $(1-0)$ transition, an optically thick line and a good tracer of infall motions \citep{Fuller05,Rygl13,He15,Traficante17_PI}. These motions can be identified looking at blue asymmetries in the spectra in correspondence of single-peaked \n2h\ $(1-0)$ spectra, which avoid the risk that the asymmetries in the \hco\ $(1-0)$ spectra may be due to the obscuration by surrounding filaments \citep{Chira14}.

We identified by eye all clumps with well defined infall signatures, i.e. all clumps with double-peaked blue-asymmetric \hco\ $(1-0)$ spectra, and we found 21 clumps with such properties. This number likely represents a sub-sample of clumps that have parsec-scale infall motions. Many clumps may present red-asymmetric \hco\ $(1-0)$ spectra despite the presence of infall motions that can be identified with blue-asymmetric spectra in higher density tracers \citep{Wyrowski16}. Also, infalling clumps may not always show the expected blue-asymmetric line profiles \citep{Smith13}. At the same time, well-defined blue-asymmetries in the \hco\ $(1-0)$ spectra are clear signatures of infall motions and can be modeled to infer infall parameters (see next Section), therefore we restrict the analysis to this sub-sample of 21 clumps to analyze the main infall properties of our sample. These clumps are divided in 10 protostar MIR bright, 4 YSOs, and 7 HII regions. 

The clumps with infall signatures span 2 order of magnitudes in mass, and the virial parameter varies in the range $0.10\leq\alpha_{vir}\leq1.96$, similar to the \avir\ distribution of the rest of the sample, with four clumps that have $\alpha_{vir}\geq1$. The differences between these clumps and the sub-sample of clumps with no blue-asymmetric \hco\ spectra are not statistically significant. A Kosmogorov-Smirnov test, which gives the probability that two samples come from the same distribution, gives a 71\% probability that the distribution of the virial parameter of infalling clumps is similar to that of the other clumps. A t-Student test, which gives the probability that two samples have the same mean, gives a probability of 61\% that the means of the virial parameter distributions in the two sub-samples are the same. 

\begin{figure}
\centering
\includegraphics[width=8cm]{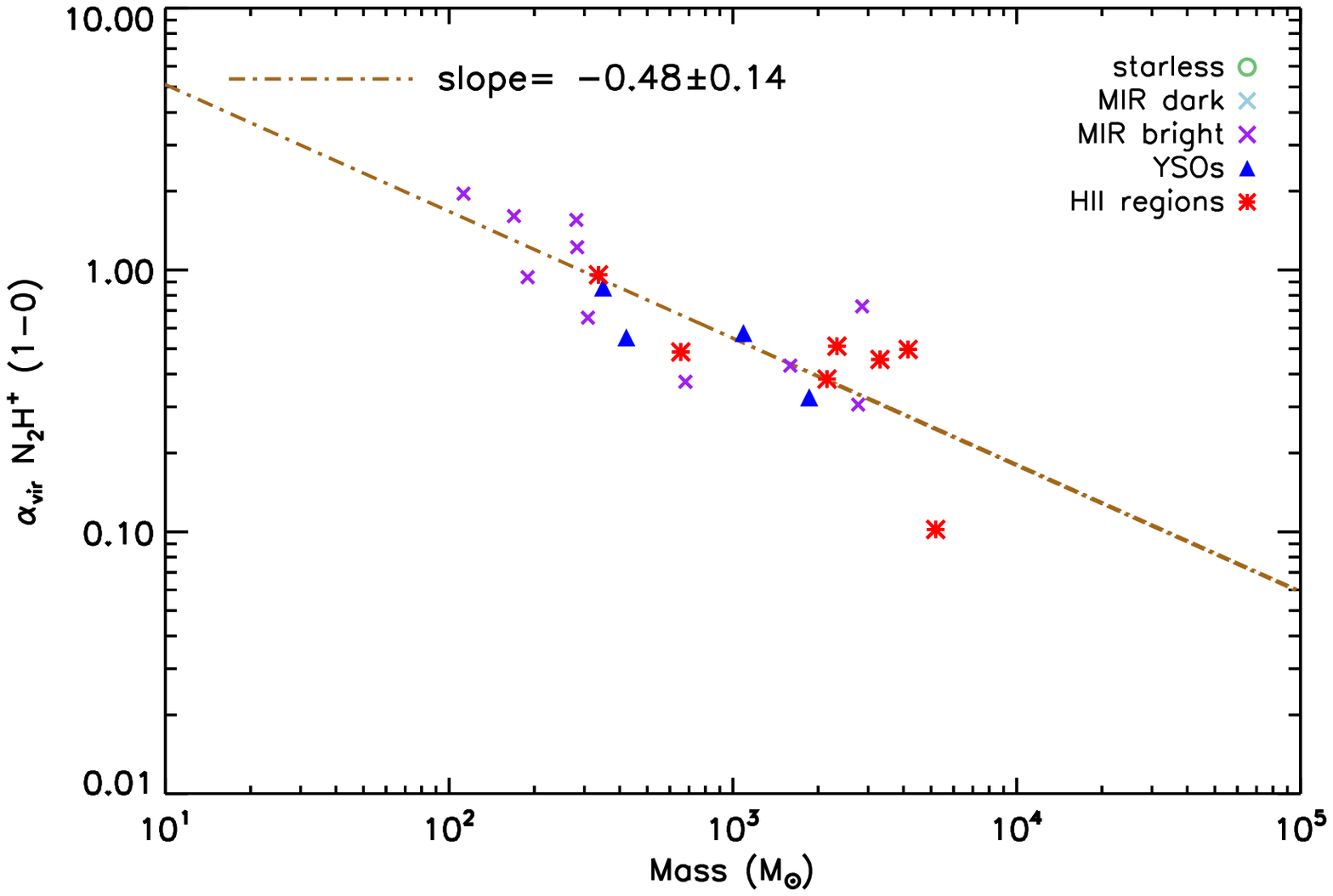}\\ 
\includegraphics[width=8cm]{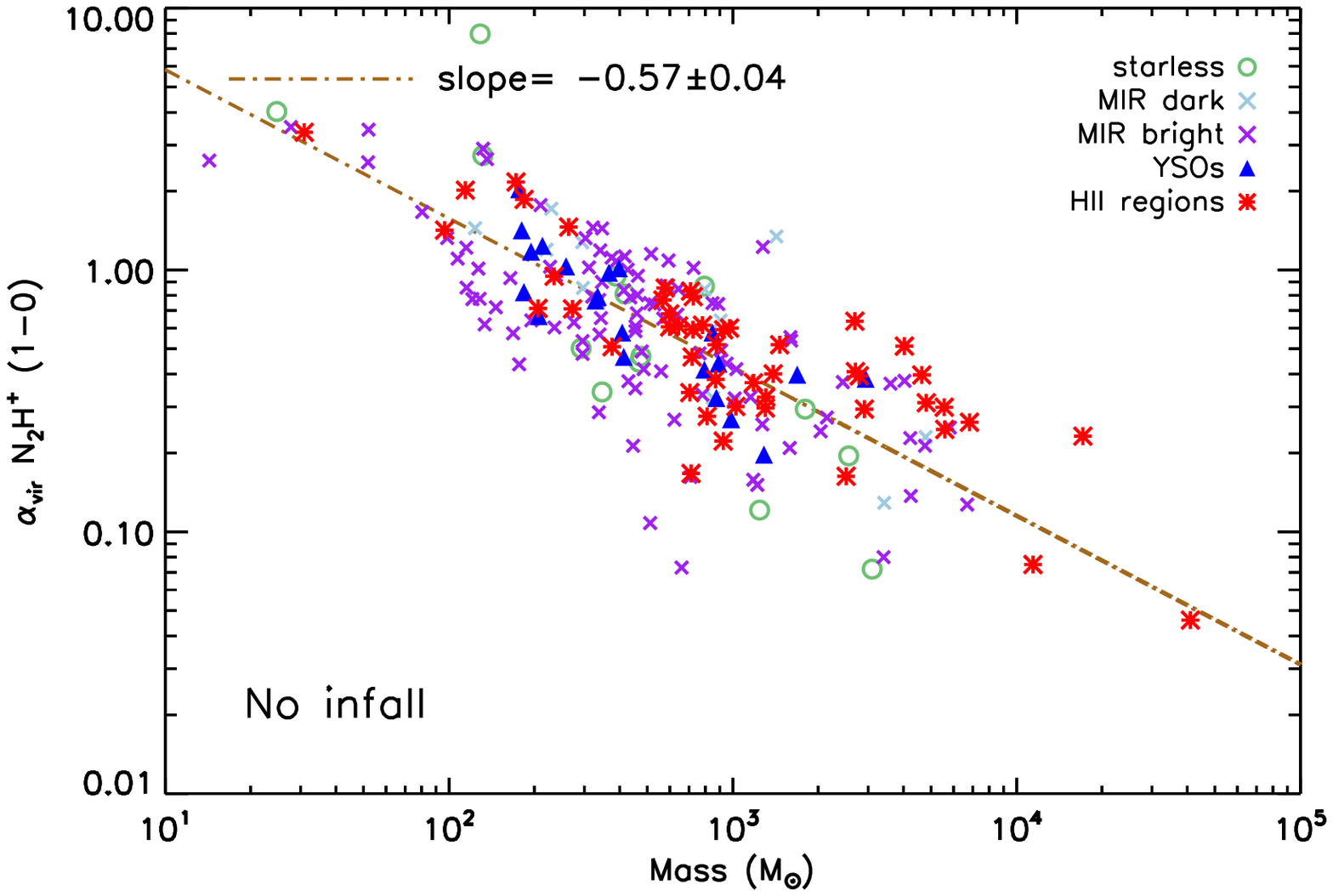} 
\caption{Upper panel: \avir\ vs. mass diagram for the 21 clumps with evidence of infalling motions in their \hco\ $(1-0)$ spectra. Lower panel: same diagram for the remaining 192 clumps of the sample. The slopes of the best linear fit in the two plots are the same within the uncertainties, suggesting that the clumps dynamics is independent of the values of \avir.}
\label{fig:mass_virial_infall}
\end{figure}

In Figure \ref{fig:mass_virial_infall} we show the \avir\ vs. mass diagrams for the 21 clumps with blue-asymmetric \hco\ spectra (upper panel) and for the remaining 192 clumps (lower panel), with the respective fits. The linear fits in the diagrams in Figure \ref{fig:mass_virial_infall} have very similar slopes within the uncertainties, suggesting that the observed values of the virial parameter are independent of the dynamics of the clumps.


\begin{center}
\begin{table}
\centering
\begin{tabular}{c|c|c|c}
\hline
\hline
 Clump  & $\mathrm{v}_{in}$ & $\dot{\mathrm{M}}$ & $\epsilon$ \\
  & (km s$^{-1}$) & (10$^{-3}$ M\sun\ yr$^{-1}$) & \\
\hline
G309.422-00.622  &    1.01(0.51)  &   7.22(4.33)  &    0.11(0.08)  \\
G316.085-00.674   &    0.26(0.13)  &    0.68(0.41)  &    8.80(6.60)  \\
G316.140-00.504  &    0.85(0.43)  &   28.65(17.19)  &    0.15(0.11)  \\
G320.285-00.309  &    0.53(0.27)  &   11.90(7.14)  &    3.08(2.31)  \\
G321.935-00.007  &    0.26(0.13)  &    1.63(1.98)  &   11.39(8.54)  \\
G322.520+00.637  &    0.22(0.11)  &    1.27(0.76)  &   10.81(8.11)  \\
G327.393+00.199  &    0.19(0.10)  &    2.45(1.47)  &   41.37(31.03)  \\
G327.403+00.444  &    0.66(0.33)  &   17.11(10.27)  &    1.50(1.13)  \\
G331.132-00.245  &    0.21(0.11)  &   6.11(3.67)  &   87.48(65.61)  \\
G331.708+00.583  &    0.55(0.28)  &    9.65(5.79)  &    3.80(2.85)  \\
G331.723-00.203   &    0.74(0.37)  &   4.23(2.54)  &    0.65(0.49)  \\
G332.604-00.168   &    0.21(0.11)  &    0.73(0.44)  &   21.47(16.10)  \\
G338.927+00.632  &    1.43(0.72)  &   45.83(27.50)  &    0.15(0.11)  \\
G339.476+00.185  &    1.51(0.76)  &   36.06(21.64)  &    0.17(0.13)  \\
G339.924-00.084  &    0.51(0.26)  &   9.21(5.53)  &    1.60(1.20)  \\
G341.215-00.236  &    0.20(0.10)  &    1.73(1.04)  &   16.45(12.34)  \\
G342.822+00.382  &    0.33(0.17)  &    3.87(2.32)  &    4.26(3.20)  \\
G343.520-00.519  &    0.33(0.17)  &    2.13(1.28)  &    5.39(4.04)  \\
G343.756-00.163  &    0.23(0.12)  &    3.30(1.98)  &   14.20(10.65)  \\
G344.101-00.661  &    1.13(0.57)  &    7.00(4.20)  &    0.30(0.23)  \\
G344.221-00.594  &    0.24(0.12)  &    1.61(0.97)  &   15.72(11.79)  \\
\hline
\end{tabular}
\caption{Parameters of the 21 clumps derived from double-peaked, blue-shifted \hco\ $(1-0)$ spectra. Col.1: Clump name; Col. 2: Infall velocity; Col. 3: Mass accretion rate; Col. 4: Efficiency. The uncertainties have been propagated from the uncertainties on M, R and $\sigma$ showed in Table \ref{tab:uncertainties}, assuming a further uncertainties of 25\% on the estimation of $\mathrm{v}_{red}$ and$\mathrm{v}_{blue}$ due to the resolution of the smoothed spectra used to estimate the velocities (see Section \ref{sec:uncertainties}). We obtained uncertainties of 50\%, 60\% and 75\% to the estimation of $\mathrm{v}_{in}$, $\dot{\mathrm{M}}$ and $\epsilon$ respectively.}

\label{tab:infall_params}
\end{table}
\end{center}

\subsection{Properties of infalling gas}\label{sec:infall_velo_accr_rate}
Clumps with double-peaked blue-asymmetric spectra can be modeled to obtain properties such as the infall velocity $\mathrm{v}_{in}$ and mass accretion rate $\dot{\mathrm{M}}$ using e.g. the two-layers model of \citet{Myers96}. In this model the line intensity of the red and blue peaks, as well as of the dip in correspondence of the central velocity of the clump (deduced from an optically thin line such as \n2h) are used to estimate the infall velocity:

\begin{equation}\label{eq:infall_velocity}
\mathrm{v}_{in}=\frac{\sigma^{2}}{\mathrm{v}_{red}-\mathrm{v}_{blue}}\ \mathrm{ln}\bigg(\frac{1+e^{\mathrm{(T_{blue}-T_{dip})/T_{dip}}}}{1+e^{\mathrm{(T_{red}-T_{dip})/T_{dip}}}}\bigg).
\end{equation}
where $\mathrm{v}_{red}$ and $\mathrm{v}_{blue}$ are the velocities of the blue and red peaks respectively, $T_{blue}$ and $T_{red}$ are the main beam temperatures of the blue and red peaks respectively and $T_{dip}$  is the main beam temperature of the valley between the two peaks. To obtain these parameters, the spectra of the 21 clumps have been fitted with a double-Gaussian model using the \texttt{mpfitfun} IDL routine \citep{Markwardt09}. The spectra with the double-Gaussian fits are in Appendix \ref{app:hco_spectra}. 

The infall velocity varies in the range $0.03\leq\mathrm{v}_{in}\leq2.75$ km s$^{-1}$, with an average value of $\mathrm{v}_{in}=0.55$ km s$^{-1}$, in line with similar estimates in massive star forming regions \citep{Fuller05,Rygl13,Traficante17_PI}. The value for each clump is in Table \ref{tab:infall_params}. 

The mass accretion rate $\dot{\mathrm{M}}$ can be evaluated assuming spherical geometry as $\dot{\mathrm{M}}=4\pi\mathrm{R}^{2}n_{\mathrm{H}_{2}}\mu m_{\mathrm{H}}\mathrm{v}_{in}$ \citep{Myers96}, where $m_{\mathrm{H}}$ is the hydrogen mass, $\mu=2.33$ is the molecular weight and $n_{\mathrm{H}_{2}}$ the volume density. It ranges in the limits $0.68\leq\dot{\mathrm{M}}\leq45.8\times10^{-3}$ M\sun\ yr$^{-1}$ (Table \ref{tab:infall_params}), with an average value of $\dot{\mathrm{M}}=9.6\times10^{-3}$ M\sun\ yr$^{-1}$, comparable with similar results for massive protostellar clumps \citep[e.g.][]{Rygl13,Peretto13}. 

\begin{figure}
\centering
\includegraphics[width=8cm]{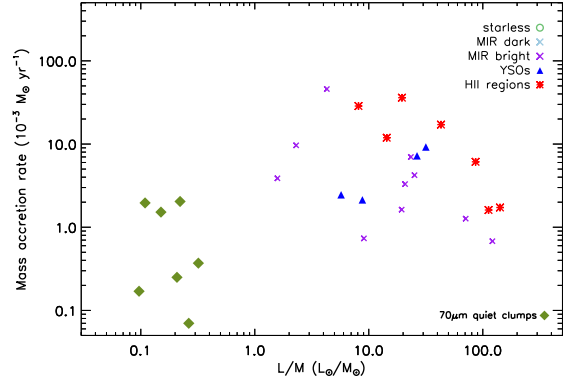} 
\caption{Mass accretion rate as function of the L/M ratio, an indicator of clumps evolution. Our sample of 21 clumps has been combined with the survey of 70\mum\ quiet clumps in \citet[][green diamonds]{Traficante17_PI}. Altogether, these data show a moderate correlation ($\rho=0.44)$ and suggest that the mass accretion rate may increase with evolution.}
\label{fig:L_M_infall_rate}
\end{figure}

In Table \ref{tab:average_infall_params} we show the average values of $\mathrm{v}_{in}$ and $\dot{\mathrm{M}}$ for the various evolutionary phases. There is an indication that $\dot{\mathrm{M}}$ is higher in HII regions than in the rest of the sample, suggesting an increasing of the accretion rate with evolution. Since we do not observe clumps at the earliest evolutionary stages with clear hint of infall motions in our data, in order to investigate this trend we have combined our data with the sample of seven 70\mum\ quiet massive clumps studied in \citet{Traficante17_PI}, for which they have measured the mass accretion rates. In Figure \ref{fig:L_M_infall_rate} we plot $\dot{\mathrm{M}}$ against the quantity L/M. There is a large scatter among the sources, but the correlation is not irrelevant ($\rho=0.44)$. With an average accretion rate for the 70\mum\ quiet clumps of $0.91\times10^{-3}$ M\sun\ yr$^{-1}$, this diagram support the indication that $\dot{\mathrm{M}}$ increases with evolution in these massive objects.

\begin{figure}
\centering
\includegraphics[width=8cm]{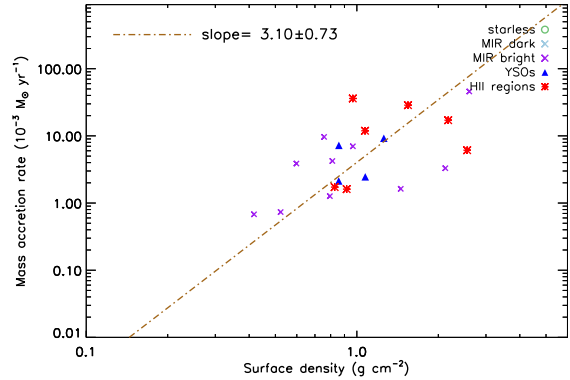} 
\caption{Mass accretion rate as function of surface density. The Pearson's correlation coefficient is 0.61, suggesting a good correlation between these two quantities. The best-fit gives a slope of 3.10$\pm0.73$, suggesting an increasing of $\dot{\mathrm{M}}$ as function of $\Sigma$.} 
\label{fig:surf_dens_infall_rate}
\end{figure}

 
The mass accretion rate is instead proportional to the surface density ($\rho=0.61)$, as showed in Figure \ref{fig:surf_dens_infall_rate}. This result implies that higher density regions sustain a higher accretion rate, a point that we will further discuss in Section \ref{sec:gravity_driver}.

\begin{center}
\begin{table}
\centering
\begin{tabular}{c|c|c|c}
\hline
\hline Clump phase & Count & $\bar{\mathrm{v}}_{in}$ & $\bar{\dot{\mathrm{M}}} $ \\
  
  & & (km s$^{-1}$) & (10$^{-3}$ M\sun\ yr$^{-1}$) \\
\hline

24\mum\ bright &	10 &	0.54	&	7.82 \\
YSOs &	4	&	0.51		&	5.25		\\
HII regions	&	7	& 0.60 & 14.74	\\

\hline
\end{tabular}

\caption{Mean infall parameters of the 21 clumps with infall signatures divided by different evolutionary phases. Col.1: Clump evolutionary phase; Col 2: number of clumps in each phase; 
Col. 3: Mean infall velocity; Col. 4: Mean mass accretion rate.}
\label{tab:average_infall_params}
\end{table}
\end{center}

\section{Origin of non-thermal motions}\label{sec:non_thermal_motions_origin}
In comparison with the non-thermal motions found in GMCs, our clumps have an excess of kinetic energy at small radii (Figure \ref{fig:velocity_size}), in agreement with the findings of \citet[][and references therein]{Ballesteros-Paredes11}. \citet{Larson81} himself noted that the relation breaks down at the size of the clumps/cores and the inner part of massive star forming regions tends to have higher velocity dispersion at a given radius. The kinetic energy excess in this ensemble of clumps should have a different origin from shock turbulence. In this Section we investigate possible origins of the observed non-thermal motions in these objects.

A possible explanation for the origin of non-thermal motions in massive clumps is given by the model of \citet{Murray15}. These authors break down the assumption that collapsing regions are in hydrostatic equilibrium. Instead, the turbulent velocity is adiabatically heated by the collapse itself, following the evolution of the system. Combined with the back-pressure generated by turbulence, \citet{Murray15} predicted a power-law form for the Larson's first relation of R$\propto\sigma^{0.2-0.3}$. This model successfully predicts a deviation of Larson's first relation in massive star forming regions as found by e.g. \citet{Caselli95} and \citet{Shirley03}. However our observations, in agreement with the findings of \citet{Ballesteros-Paredes11} and \citet{Traficante17_PII}, suggest that there is no correlation between velocity dispersion and size of an ensemble of clumps.

Non-thermal motions in massive star forming objects may also be driven by stellar feedbacks such as protostellar jets/outflows \citep[e.g.][]{Federrath16}. Figures \ref{fig:velocity_size} and \ref{fig:Larson_histo} show, however, that the velocity dispersion in starless clumps, the less affected by stellar feedbacks, is similar to the one observed in more evolved clumps, and the quantity $\sigma/\mathrm{R}^{0.5}$ is not constant. Even if stellar feedbacks play an important role in the observed non-thermal motions of protostars, it cannot alone explain the observed linewidth-size relation in all these clumps.

\subsection{Accretion-driven turbulence}\label{sec:accretion_driven}
An alternative explanation to the observed supersonic motions is that these non-thermal velocities are the result of accretion-driven turbulence \citep{Klessen10}. This model predicts that (at least part of) the energy injected by the accretion into the system is converted into turbulent motions, which set up a Kolmogorov-like turbulent cascade. The large-scale fed accretion generates enough turbulence to produce supersonic motions in the high-density clumps. 

If the energy injected by the infall motions is much lower than the turbulent dissipation rate, the conversion of these motions into turbulent energy cannot maintain the turbulent cascade which will rapidly dissipate. The observed non-thermal motions would therefore not be the result of a turbulent cascade, and they would not follow a Larson-like relation.

The key parameters to evaluate the energy injected by the accretion and the turbulent dissipation rate are the scale at which the turbulence is driven and the mass of the infalling gas. We consider that the clumps are globally collapsing as a whole \citep[as observed in e.g.][]{Traficante17_PI}, and the representative scale is the scale of the clumps.  
 
With these assumptions, the turbulent dissipation rate is
$\mathrm{\dot{E}}_{dis}=\frac{1}{2}\mathrm{M}\sigma^{2}/\tau_{d}=\frac{1}{2\sqrt{3}\mathrm{L}_{d}}\mathrm{M}\sigma^{3}$ \citep{Hennebelle12}, where M is the total mass of the clump. The turbulence decays in a turbulent crossing time $\tau_{d}=\mathrm{L}_{d}/\sigma_{3D}$, with $\sigma_{3D}=\sqrt{3}\sigma$ being the 3-dimensional velocity dispersion and L$_{d}$, the turbulence driving scale \citep[][and references therein]{Hennebelle12}, the size of the clump. The energy injected by the accretion is $\mathrm{\dot{E}}_{inj}=\frac{1}{2}\mathrm{\dot{M}}\mathrm{v}_{in}^{2}$ \citep{Klessen10}, where $\mathrm{\dot{M}}$ has been evaluated from the mean density of the clumps as in Section \ref{sec:infall_velo_accr_rate}. Defining the efficiency $\epsilon$=$\mathrm{\dot{E}}_{dis}/\mathrm{\dot{E}}_{inj}$, the conditions for accretion-driven turbulence are satisfied if $\epsilon\leq1$ \citep{Klessen10}.

\begin{figure}
\centering
\includegraphics[width=8cm]{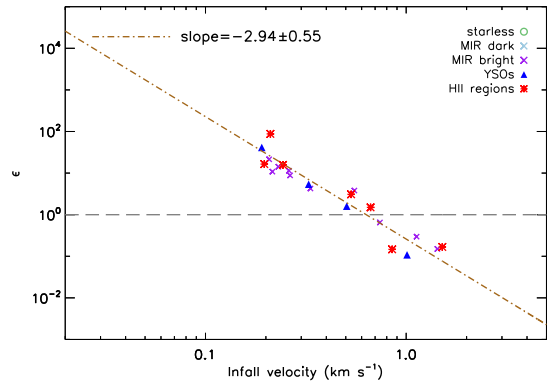} 
\caption{Efficiency $\epsilon$ as function of the infall velocity. The red-dashed line is the best-fit to the data and shows that the efficiency goes rapidly down with the increase of the infall velocity. The black dashed line is in correspondence of $\epsilon=1$. The non-thermal motions observed in clumps with efficiency below this value may be due to accretion driven turbulence.}
\label{fig:E_inj_eff}
\end{figure}

We evaluate $\mathrm{\dot{E}}_{dis}$, $\mathrm{\dot{E}}_{inj}$ and $\epsilon$ for the 21 clumps with defined infall velocities (Section \ref{sec:infall_velo_accr_rate}). The efficiency as function of the infall velocity is in Figure \ref{fig:E_inj_eff}. The efficiency goes rapidly down as the infall velocity increases, and becomes less than 1 for the six clumps with the highest accretion rates and with infall velocities $\mathrm{v}_{in}\geq0.75$ km s$^{-1}$ (Table \ref{tab:infall_params}). For the majority of the clumps the turbulent dissipation rate seems to be sufficiently high to dissipate the energy injected by the accretion.

The observed non-thermal motions can partly originate from turbulence driven by the accretion in clumps with high infall velocity and accretion rates, but, under the hypothesis that the driving scales are the clump scales that are globally collapsing as a whole, this mechanism alone cannot explain the supersonic non-thermal motions observed in clumps with infall velocity $\mathrm{v}_{in}<0.75$ km s$^{-1}$.

\subsection{Gravity-driven non-thermal motions}\label{sec:gravity_driver}
Non-thermal motions in star forming regions may originate from gravity itself, which seems to play a dominant role in the evolution of molecular clouds able to form high-mass stars, down to $\simeq0.1$ pc scales \citep{Li16,Li17}. In particular, non-thermal motions may originate from a hierarchical, global collapse of clouds and clumps \citep{Ballesteros-Paredes11}. In this picture the supersonic motions are not hydrodynamical turbulence, but organized motions driven by gravity in multiple centers of collapse. This hyphothesis implies that massive regions should develop a larger velocity dispersion for larger column densities \citep{Ballesteros-Paredes11}, which may also explain the higher accretion rates observed in higher surface density clumps (Figure \ref{fig:surf_dens_infall_rate}). 

\begin{figure}
\centering
\includegraphics[width=8cm]{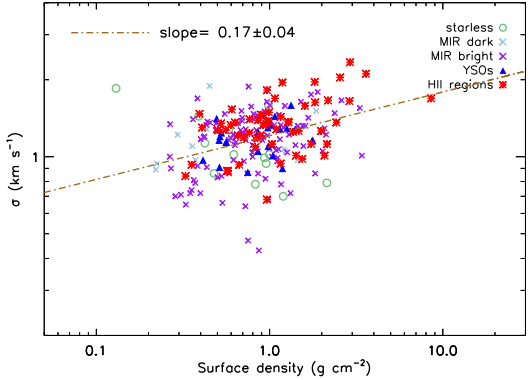} 
\caption{Velocity dispersion $\sigma$ as function of surface density $\Sigma$. The positive correlation is weak, with a Pearson's correlation coefficient of $\rho=0.30$, and the best-fit gives a slope of 0.17$\pm0.04$.}
\label{fig:Sigma_sigma}
\end{figure}

In Figure \ref{fig:Sigma_sigma} we show the $\sigma$ vs. $\Sigma$ diagram. The correlation is not strong, but there is a weak positive correlation ($\rho=0.30$) and regions with higher surface density have on average higher velocity dispersion, suggesting that non-thermal motions may partly result from the large gravitational force acting in the system.

This model also considers that the system develops a Heyer-like relation $\sigma/\mathrm{R}^{1/2}\propto\Sigma^{1/2}$ \citep{Heyer09}. This relation is equivalent to a generalized first Larson's relation for regions with different surface densities \citep{Ballesteros-Paredes11,Camacho16}. In this global collapse model, rather than the virial equilibrium what counts is the conservation of the total energy of the system. The virial parameter represents energy equipartition, which numerically is equivalent to set \aeq=2 \citep{Ballesteros-Paredes11}. 

\begin{figure}
\centering
\includegraphics[width=8cm]{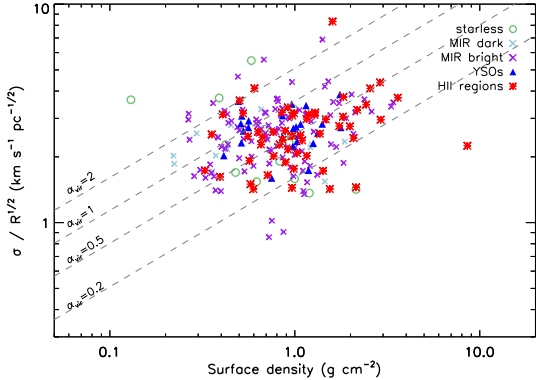} 
\caption{Heyer plot, i.e. the quantity $\sigma/\mathrm{R}^{1/2}$ versus the surface density $\Sigma$. The dashed lines are in correspondence of constant values of the virial parameter, from \avir=2 (highest line) to \avir=0.2 (lowest line). The correlation is weak, the Pearson's coefficient is $\rho=0.18$.}
\label{fig:Heyer}
\end{figure}

The Heyer plot for our clumps is in Figure \ref{fig:Heyer}. The correlation between $\Sigma$ and the quantity $\sigma/\mathrm{R}^{1/2}$ is relatively weak ($\rho=0.18$). The majority of the clumps lie below the equipartition value and there is a significant spread across the diagram, which reflects the different values of the virial parameter (Section  \ref{sec:Larson_second}).

\begin{figure}
\centering
\includegraphics[width=8cm]{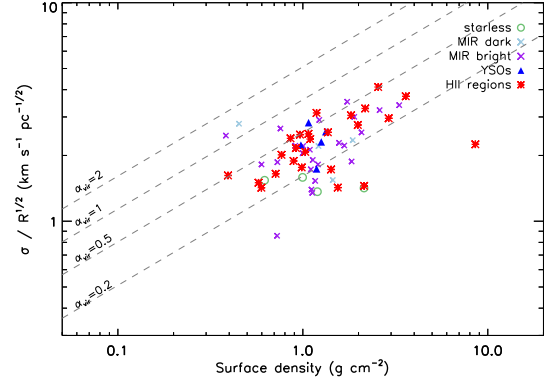} 
\caption{Same as in Figure \ref{fig:Heyer} but for clumps with M$\geq1000$\Msun. The correlation is higher, with a Pearson's coefficient of $\rho=0.40$.}
\label{fig:Heyer_high_mass}
\end{figure}

In Figure \ref{fig:Heyer_high_mass} we show the same Heyer plot, but limited to clumps with mass M$\geq1000$\Msun. The correlation in this case is higher ($\rho=0.40$). We interpret this result as an indication that gravity drives at least partially the observed non-thermal motions, in particular in the more massive clumps of the sample. If gravity contributes to the generation of the observed kinetic energy, this contribution is more dominant at higher masses, although all these massive clumps lie below the equipartition value. A possible explanation for the observed departure from the equipartition, at least for the sub-virial clumps at the earliest phases of evolution, is that a collapsing region with sufficiently low level of local turbulence can start in a sub-virial state, and it can reach the equipartition during its evolution \citep{Ballesteros-Paredes17}. As discussed in Section \ref{sec:virial_dep}, there may also be a fraction of kinetic energy not properly traced in these massive regions, which may explain the departure from the energy equipartition expected from the models. In the next Section we explore if the magnetic pressure in these clumps is another valid explanation for this observed departure.

\subsection{Role of magnetic fields}\label{sec:magnetic_field}
In this Section we estimate the possible contribution of the magnetic fields to the stability of these clumps. 

In accordance with the findings of the previous Section these clumps, in particular the more massive ones, may be undergoing gravitational collapse. However, these clumps may be sustained against the collapse by strong magnetic pressure. 

\citet{Crutcher12} showed that, observationally, the magnetic fields strength may not be sufficient to balance gravity in high density regions ($n\geq300$ cm$^{-3}$), although they may give a significant contribution in lower density ones. There is an expected upper limit to the intensity of the magnetic fields B$_{Cr}$ which increases at increasing density as B$_{Cr}\simeq\Sigma^{0.65}$ \citep{Crutcher12}. At the same time, the work of \citet{Kauffmann13} showed that the magnetic fields strength required to maintain a clump in a hydrostatic equilibrium (equivalent to \avir=2) is proportional to the observed non-thermal motions following the relation:

\begin{equation}
B_{\mathrm{M_{BE}}}=81\mu\mathrm{G}\frac{\mathrm{M}_{\phi}}{\mathrm{M_{BE}}}\bigg(\frac{\sigma}{\mathrm{km s^{-1}}}\bigg)^{2}\bigg(\frac{\mathrm{R}}{\mathrm{pc}}\bigg)^{-1}
\end{equation}

where $\mathrm{M}_{\phi}/\mathrm{M_{BE}}$ is the ratio between the magnetic flux mass and mass of sphere in hydrostatic equilibrium (a Bonnor-Ebert sphere), and it is proportional to $2/$\avir-1 \citep{Kauffmann13}.

\begin{figure}
\centering
\includegraphics[width=8cm]{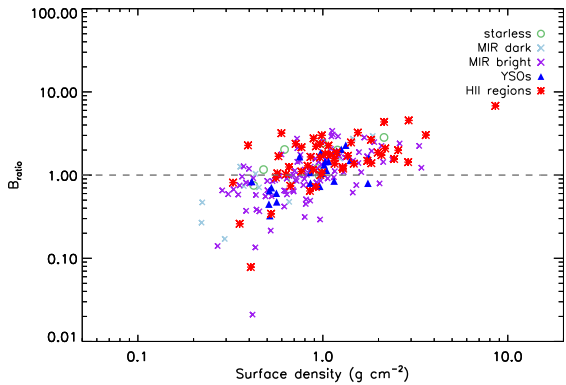} 
\caption{B$_{ratio}$=B$_{\mathrm{M_{BE}}}/$B$_{Cr}$ ratio as function of surface density. To be stabilized by magnetic fields, clumps with B$_{ratio}>1$ (black dotted line) requires magnetic fields stronger than the maximum values estimated by \citet{Crutcher12}.}
\label{fig:surf_density_vs_magnetic_field}
\end{figure}

In Figure \ref{fig:surf_density_vs_magnetic_field} we show the quantity B$_{ratio}$=B$_{\mathrm{M_{BE}}}/$B$_{Cr}$ as function of the surface density for the 199 clumps with \avir$\leq2$. Almost $40\%$ of these clumps (80) lies below the threshold B$_{ratio}=1$. The majority of them have B$_{ratio}>1$, for which the implication is that
only exceptionally high magnetic fields can stabilize their collapse.

\begin{figure}
\centering
\includegraphics[width=8cm]{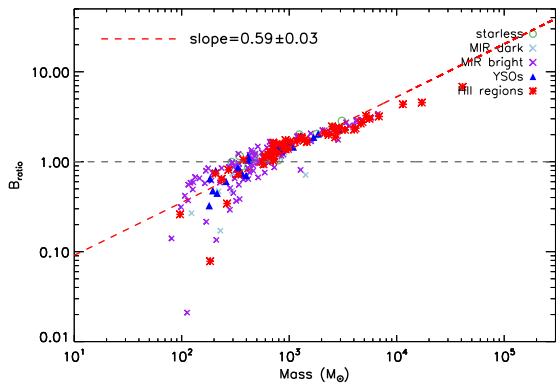} 
\caption{B$_{ratio}$ as function of the clump mass. The correlation is strong (Pearson's coefficient $\rho=0.80$) and shows that the magnetic fields required to stabilize the clumps increase significantly with the mass of the clumps. For M$\geq1000$ M\sun\ exceptionally high magnetic fields are required to slow down the collapse.}
\label{fig:mass_vs_magnetic_field}
\end{figure}

In Figure \ref{fig:mass_vs_magnetic_field} we show B$_{ratio}$ as function of the mass of the clumps. The correlation is strong ($\rho=0.80$). This diagram shows that the intensity of the magnetic fields required to stabilize a collapsing clump exceeds the threshold estimated by \citet{Crutcher12} if the clump has a mass M$\geq1000$ M\sun. 

The results suggest that magnetic fields may be relevant in some of these clumps but for the majority of them, in particular the more massive ones, this mechanism alone cannot sustain the collapse at clump scales. This analysis does not exclude that at the scales of the inner cores the magnetic fields may play a relevant role and act against the gravitational collapse \citep[e.g.][]{Fontani16}. Further observations at high resolution with instruments like ALMA or NOEMA are required to investigate the properties of these clumps at the core scales.

\section{Conclusions}\label{sec:conclusions}
We have discussed the validity of the three Larson's relations and the implications of the results for a large sample of 213 massive clumps at different evolutionary stages. These clumps have been obtained combining the Hi-GAL clumps catalogue \citep{Elia17} with the MALT90 survey of 3mm emission lines \citep{Jackson13} and selected to be a sample of sources with well-known distances, dust emission properties and \n2h\ $(1-0)$ emission, the latter used to extract the gas kinematics. The sample has been divided in five evolutionary stages, and we have obtained: 14 starless, 12 protostar MIR dark, 106 protostar MIR bright, 25 YSOs, and 56 HII regions. They are all located in the IV Quadrant and the vast majority of these clumps will likely form high-mass stars, based on the \citet{Kauffmann13} and \citet{Baldeschi17} massive star formation selection criteria. 

We have shown that the three Larson's relations do not describe the properties of an ensemble of massive clumps, independently from the evolutionary stage of these objects. At these scales $\sigma$ is not proportional to the radius R (first Larson's relation), these clumps are not in virial equilibrium (second Larson's relation), and these clumps have no constant surface density (third Larson's relation).

We demonstrated that the absence of a scaling relation between $\sigma$ and R implies that the the virial parameter \avir, defined as the ratio between kinetic energy $\mathrm{E}_{kin}$ and gravitational energy $\mathrm{E}_{\mathrm{G}}$, decreases with mass and radius only as function of the gravitational content of the clumps, independently of their kinetic energy. 

A consequence of these findings is that the measured virial parameter is not a good descriptor of the clumps dynamics. In fact, the virial values in clumps with evidence of infalling motions (measured from blue-asymmetric \hco\ $(1-0)$ spectra) are statistically indistinguishable from the values of the rest of the sample. This also suggests that all these clumps may be dynamically active, even the ones without clear evidence in their \hco\ $(1-0)$ spectra. 

We showed that the observed non-thermal motions in massive clumps are not likely due to turbulent cascade, collapse in adiabatically heated regions, nor accretion-driven turbulence. The velocity dispersion and mass accretion rate moderately correlate with surface density, which suggest that the gravitational collapse contribute at least partially at the variation of the observed non-thermal motions, in agreement with global collapse models \citep{Ballesteros-Paredes11}. The gravitational collapse seems to play a dominant role particularly in the more massive clumps of our sample (M$\geq1000$ M\sun), although they all are sub-virial and not in energy equipartition as predicted in many gravitationally-driven collapse models.

We also showed that, on average, magnetic fields may not contribute significantly to the stability of these clumps, and exceptionally strong magnetic fields would be required to stabilize the clumps with M$\geq1000$ M\sun.

\section*{acknowledgements}
AT wants to thanks J. Kauffmann for the useful and stimulating discussions that lead to some of the results in this work. This work has benefited from research funding from the European Community's Seventh Framework Programme. ADC acknowledges the support of the UK STFC consolidated grant ST/N000706/1.

\bibliographystyle{mn2e}
\bibliography{bibliography.bib}

\appendix

\section{Alternative source photometry}\label{app:Hyper_photometry}
In order to estimate the uncertainties associated with the chosen source extraction and photometry strategy described in \citet{Elia17} and obtained using the \Cut\ algorithm \citep{Molinari11}, we compared the results with the photometry obtained using an alternative method.

\begin{figure*}
\centering
\includegraphics[width=8cm]{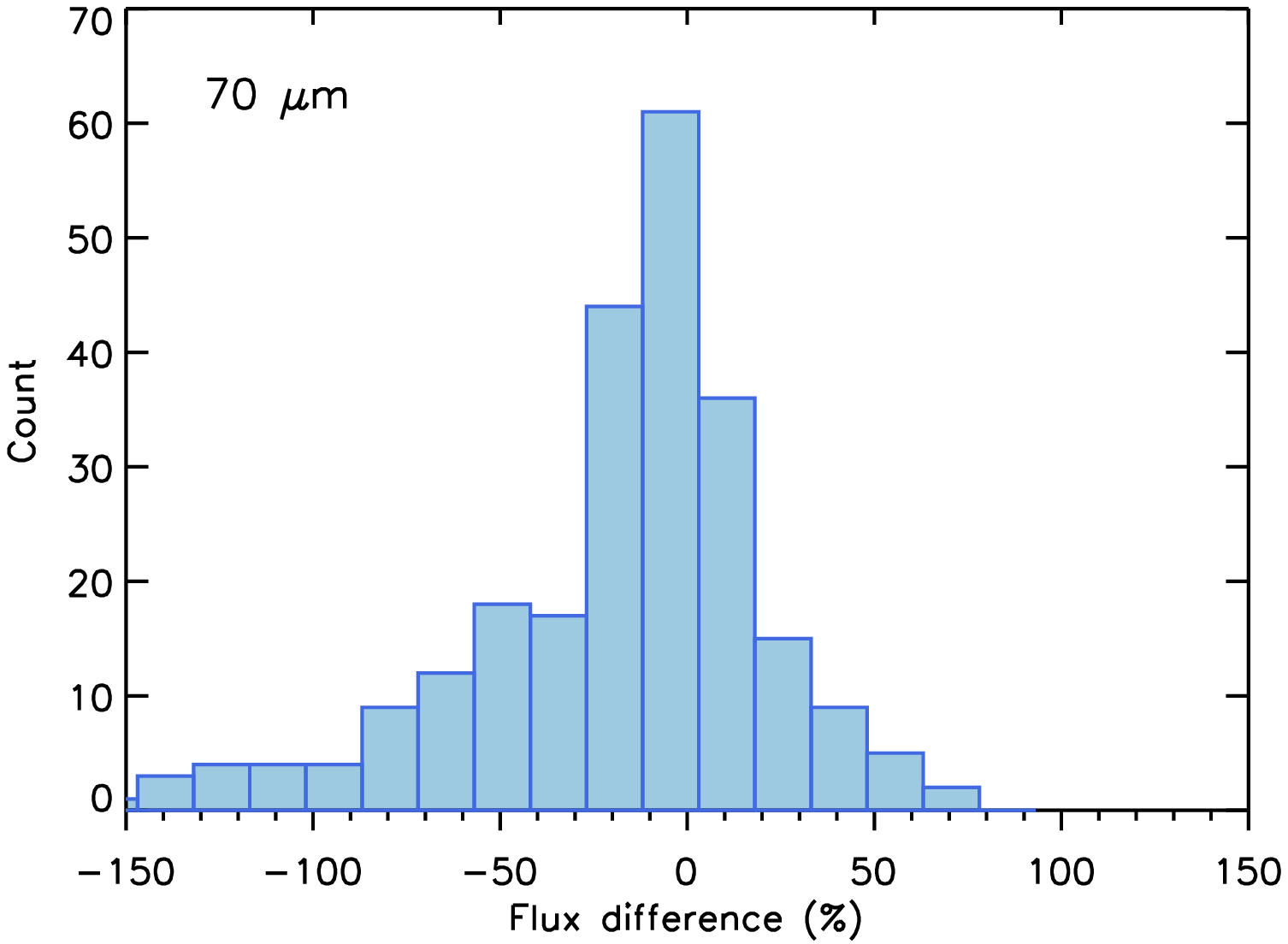} 
\includegraphics[width=8cm]{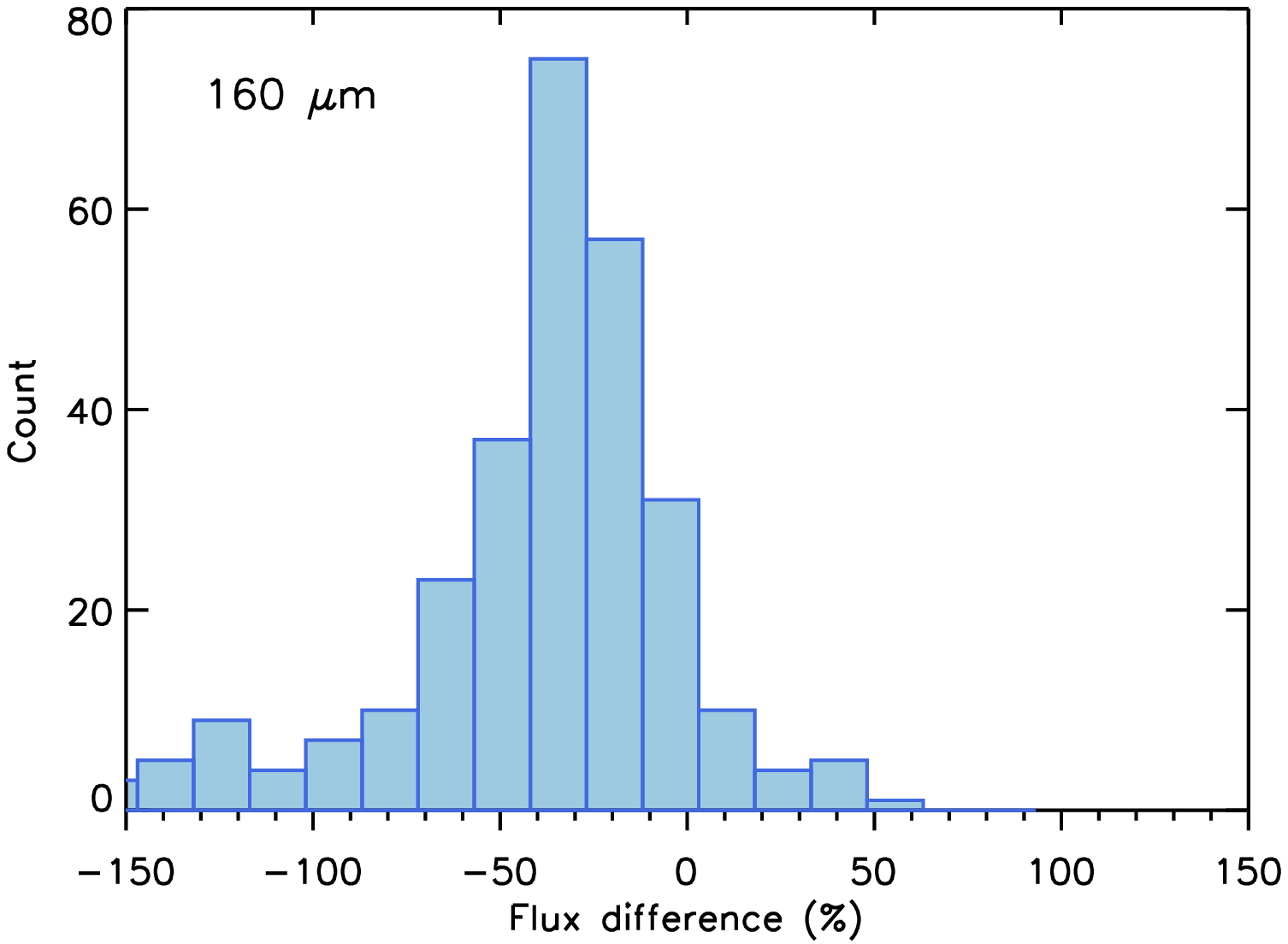} 
\includegraphics[width=8cm]{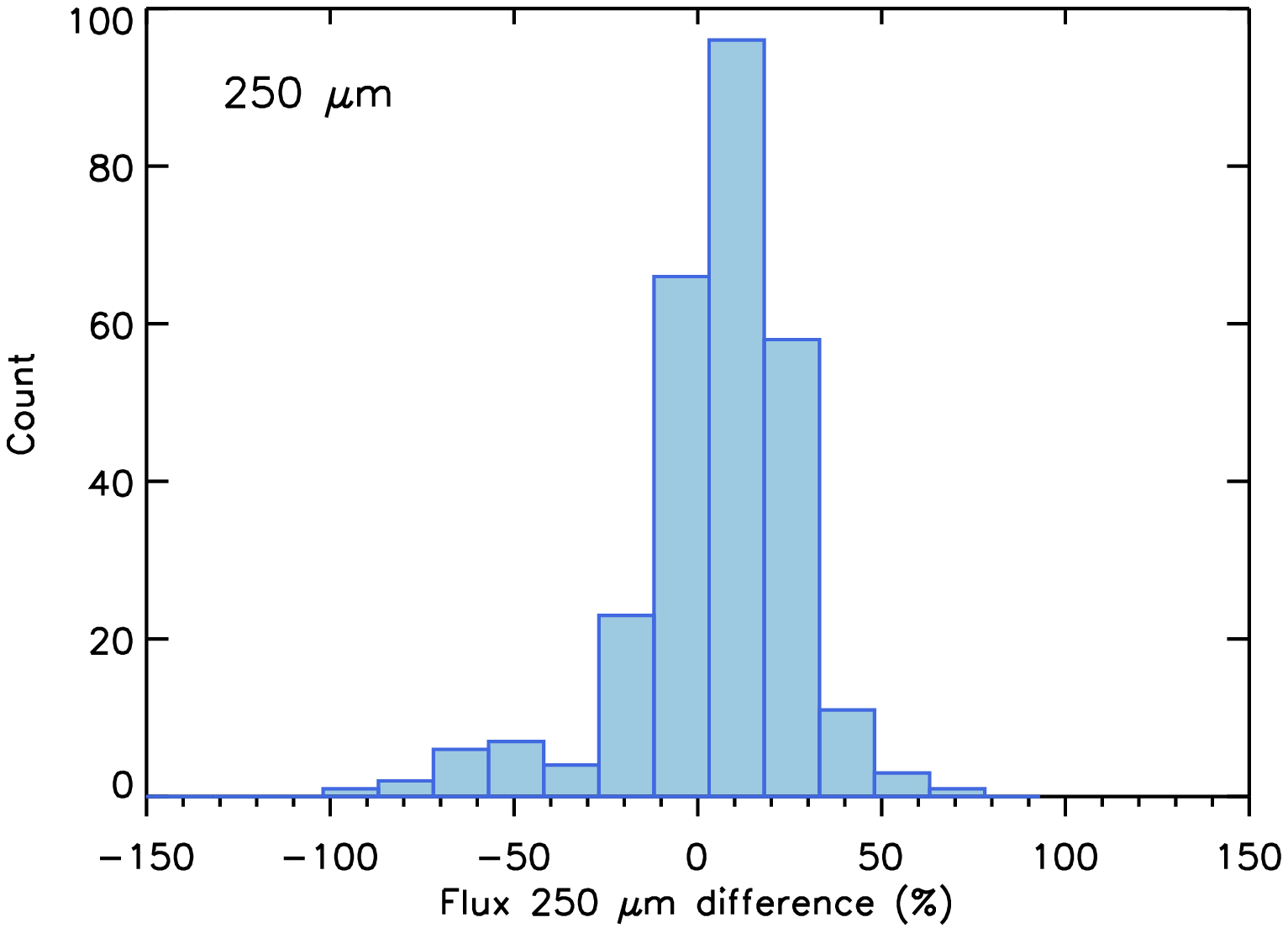} 
\includegraphics[width=8cm]{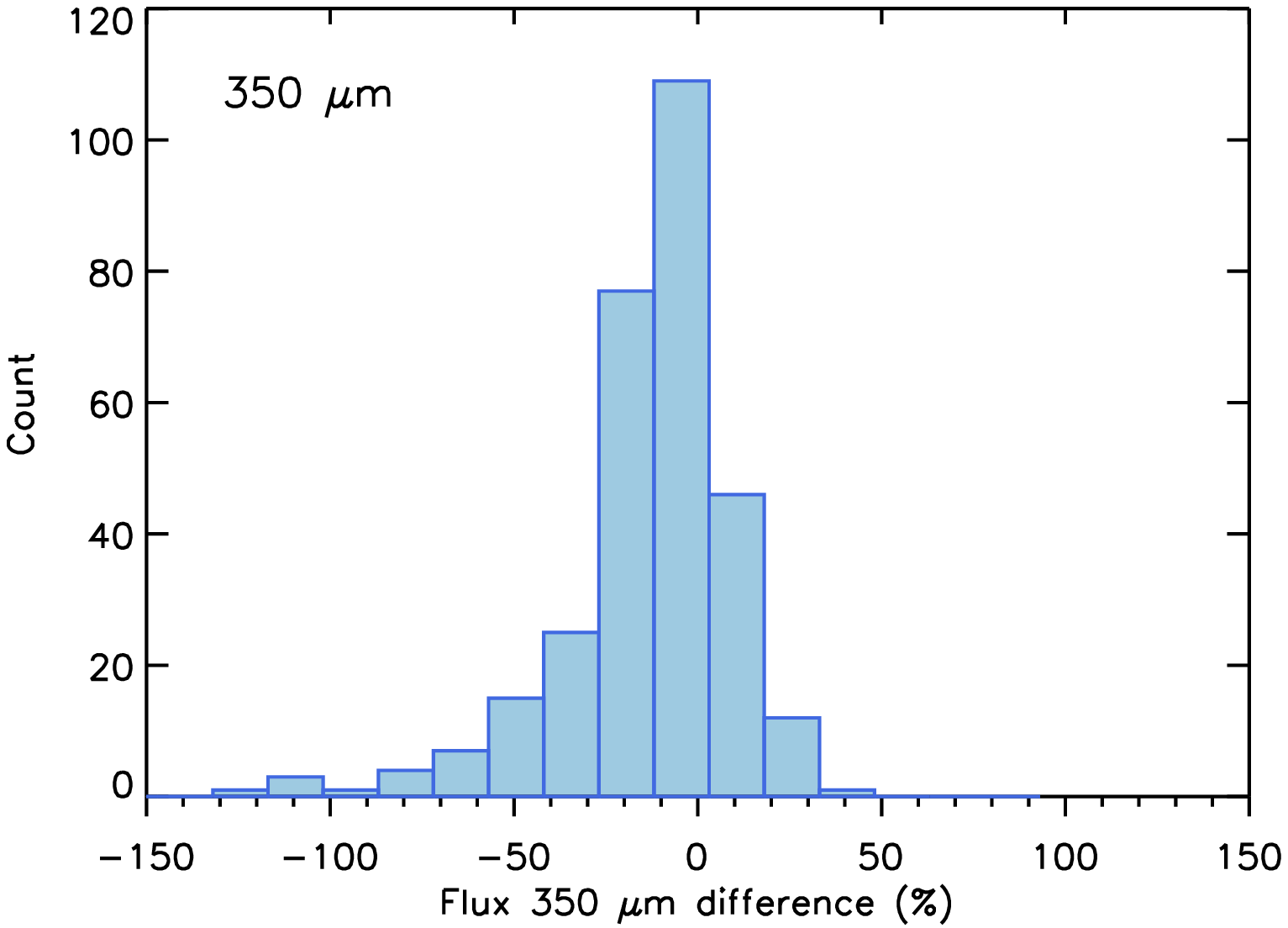} 
\includegraphics[width=8cm]{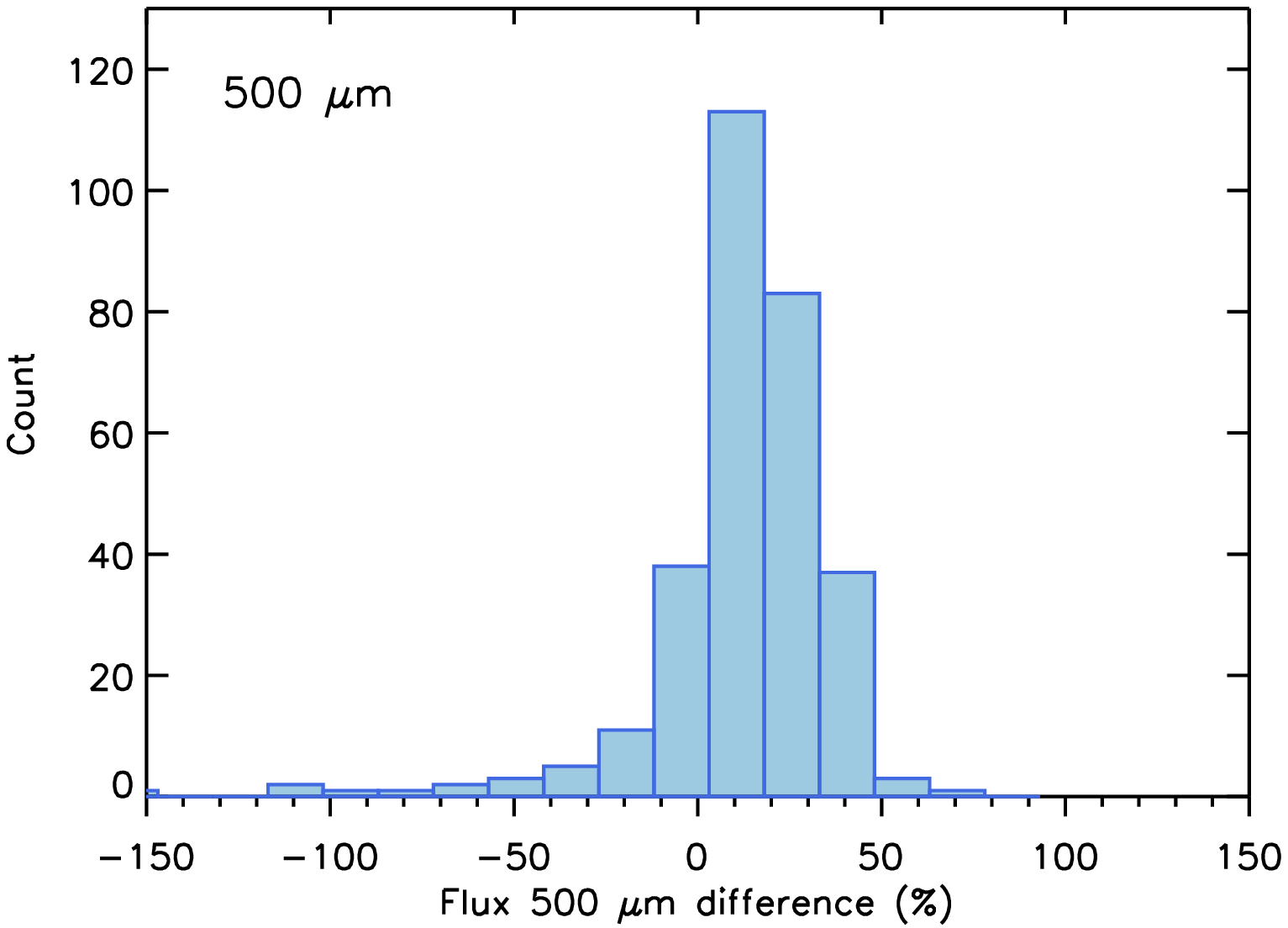} 
\caption{Difference in percentage between the fluxes of the clumps as estimated in \citet{Elia17} and estimated using the \Hyp\ algorithm. \textit{From the top:} Flux difference at 70, 160, 250, 350 and 500\mum\ respectively.}
\label{fig:flux_histo_Hyper}
\end{figure*}

The alternative photometry has been done using \Hyp, which performs elliptical aperture photometry in presence of highly variable backgrounds \citep{Traficante15a}. The 2 FWHMs and the position angle of the clumps in the \citet{Elia17} catalogue, estimated from the 250\mum\ fit, have been used to define the radius of the ellipses over which perform the \Hyp\ aperture photometry at all wavelengths. This approach is substantially different from the method used in the Hi-GAL catalogue, since \Cut\ estimates the flux as the integral of the 2d-Gaussian fitted at each wavelength. The flux differences in percentage at each wavelength are in Figure \ref{fig:flux_histo_Hyper}. 

As expected, at the reference wavelength of 250\mum\ the flux differences are minimal. At $\lambda<250$\mum\ the \Cut\ fluxes are lower than the \Hyp\ counterparts on average. This is particularly true at 160\mum, where the diffuse emission contributes substantially to the integrated flux. At 70\mum\ the measured emission is dominated by the emission from the central protostars, and the differences are less sensitive to the photometry method. At $\lambda>250$\mum\ the \Cut\ fluxes are re-scaled according to the 250\mum\ size in the Hi-GAL catalogue \citep{Elia17}, leading to a small differences between the \Hyp\ and \Cut\ photometry.

In order to add an additional point in the SED fitting, for each source we also extracted the fluxes at 870\mum, evaluated from the ATLASGAL calibrated maps using the same aperture adopted to extract the Hi-GAL fluxes.


\begin{figure}!th
\centering
\includegraphics[width=8cm]{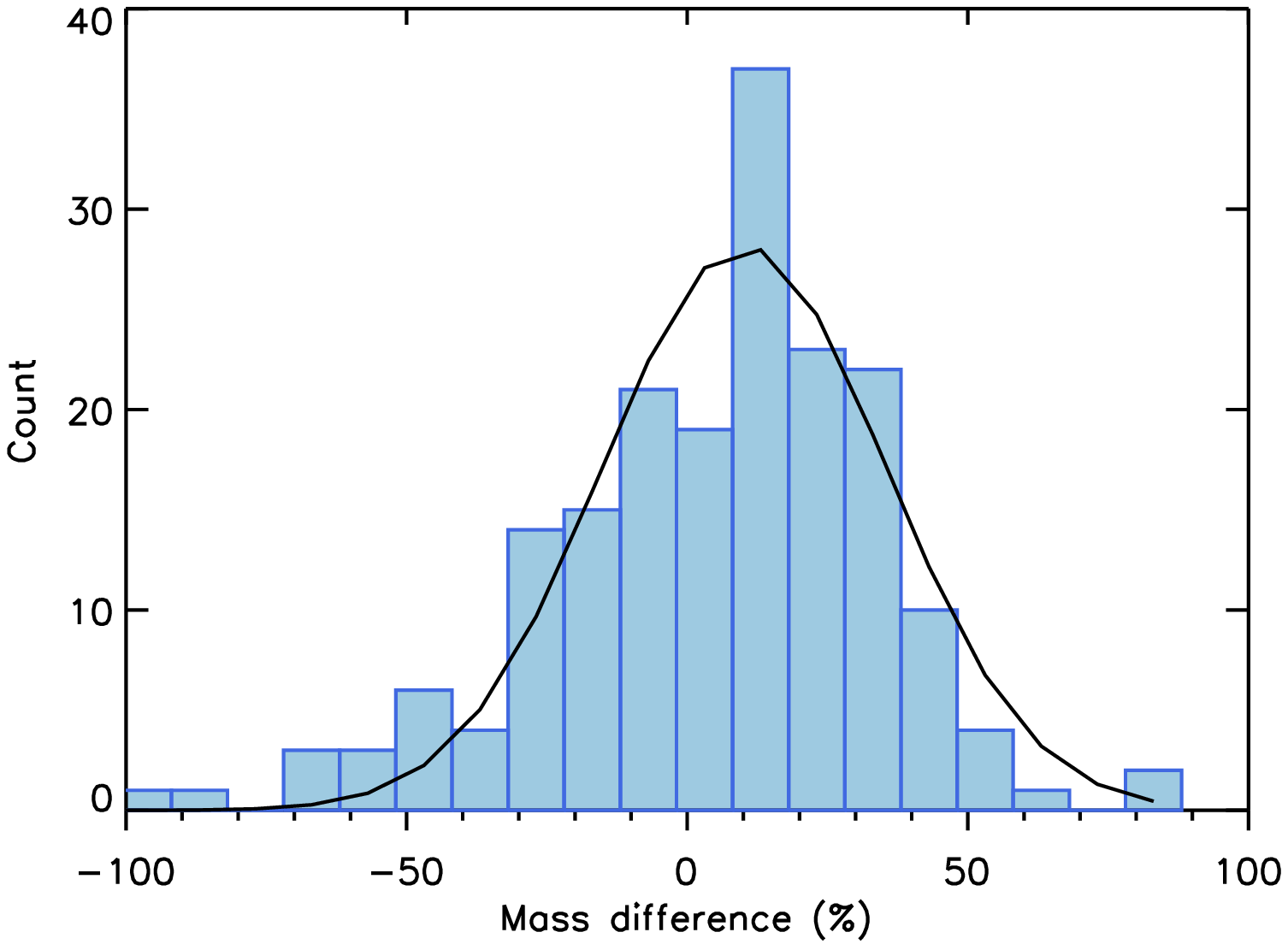} 
\caption{Mass differences (\Cut\ $-$ \Hyp\ values) for the 192 clumps with properties measured as in the \citet{Elia17} catalogue and using \Hyp. The black curve is the Gaussian fits to the distribution, used to estimate the uncertainties associated with the dust photometry.}
\label{fig:mass_radius_histo_Hyper}
\end{figure}

Few sources (21) are saturated at 250\mum\ in the Hi-GAL maps. This is due to a combination of the strong flux emission of some sources whose SED peaks at around the 160-250\mum\ wavelengths, and of the higher dynamical range of the PACS instrument at 160\mum\ which allows the 160\mum\ band to saturate at higher fluxes than the SPIRE bands. While these sources still have a flux estimation in the Hi-GAL catalogue, no aperture photometry can be reliably performed, and we excluded these clumps from the comparison. The source properties have been evaluated using the same greybody model described in Section \ref{sec:datasets}. We obtained a final sample of 192 clumps with well-defined dust properties that we used for the comparison. The mass differences in percentage are showed in Figure \ref{fig:mass_radius_histo_Hyper}.

To estimate the uncertainties on the mass due to the source photometry, we fit a Gaussian to the histogram of the mass differences. The standard deviation of the Gaussian is $\simeq25\%$, that we assume as mass uncertainties. Note that the masses estimated with \Hyp\ are $\simeq10\%$ systematically lower than the \Cut\ counterparts, likely as consequence of the different photometry at $\lambda=160$\mum. This systematic offset however does not influence the main results of this work, as discussed in Section \ref{sec:uncertainties}. 

\begin{figure}!th
\centering
\includegraphics[width=8cm]{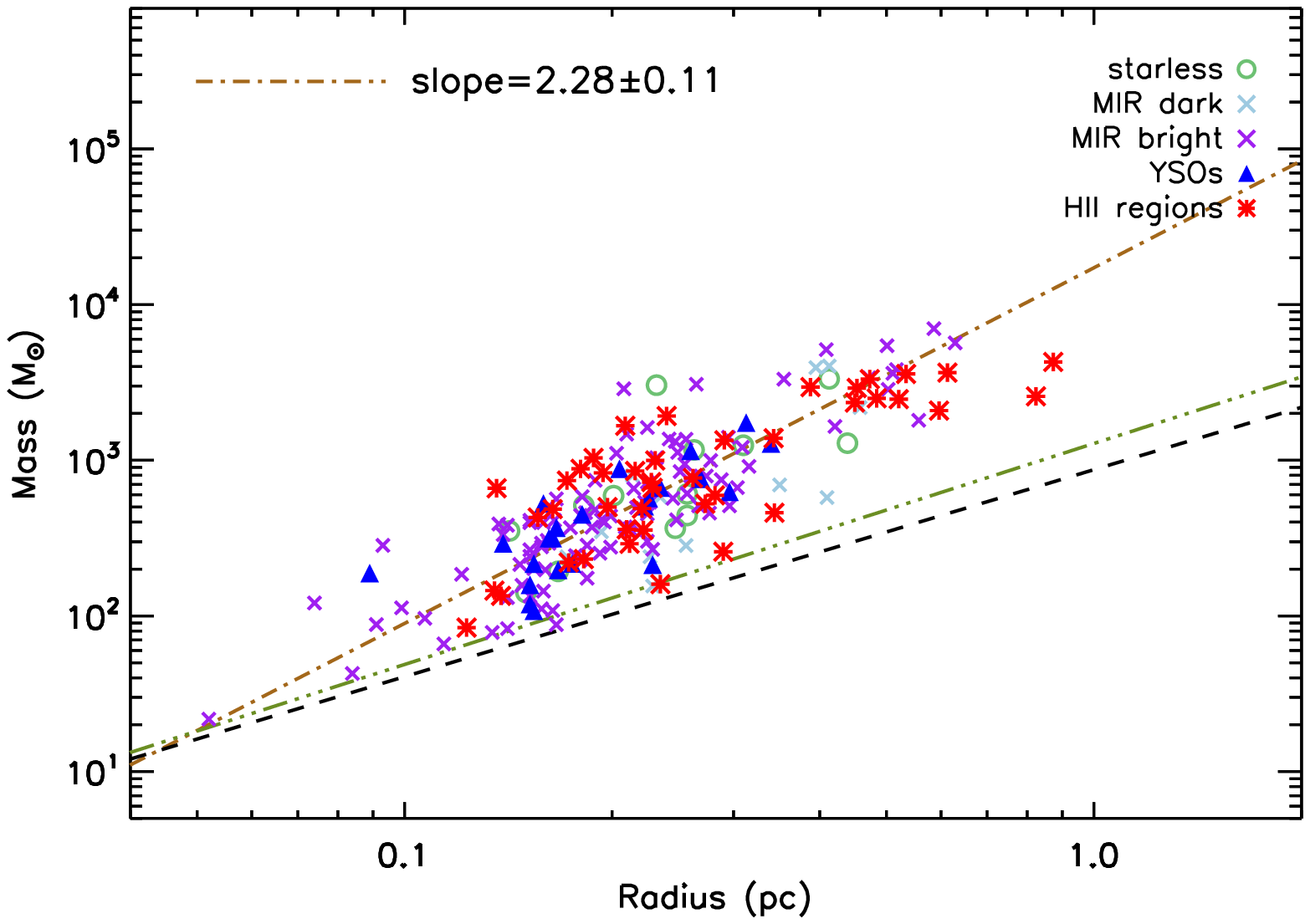} 
\includegraphics[width=8cm]{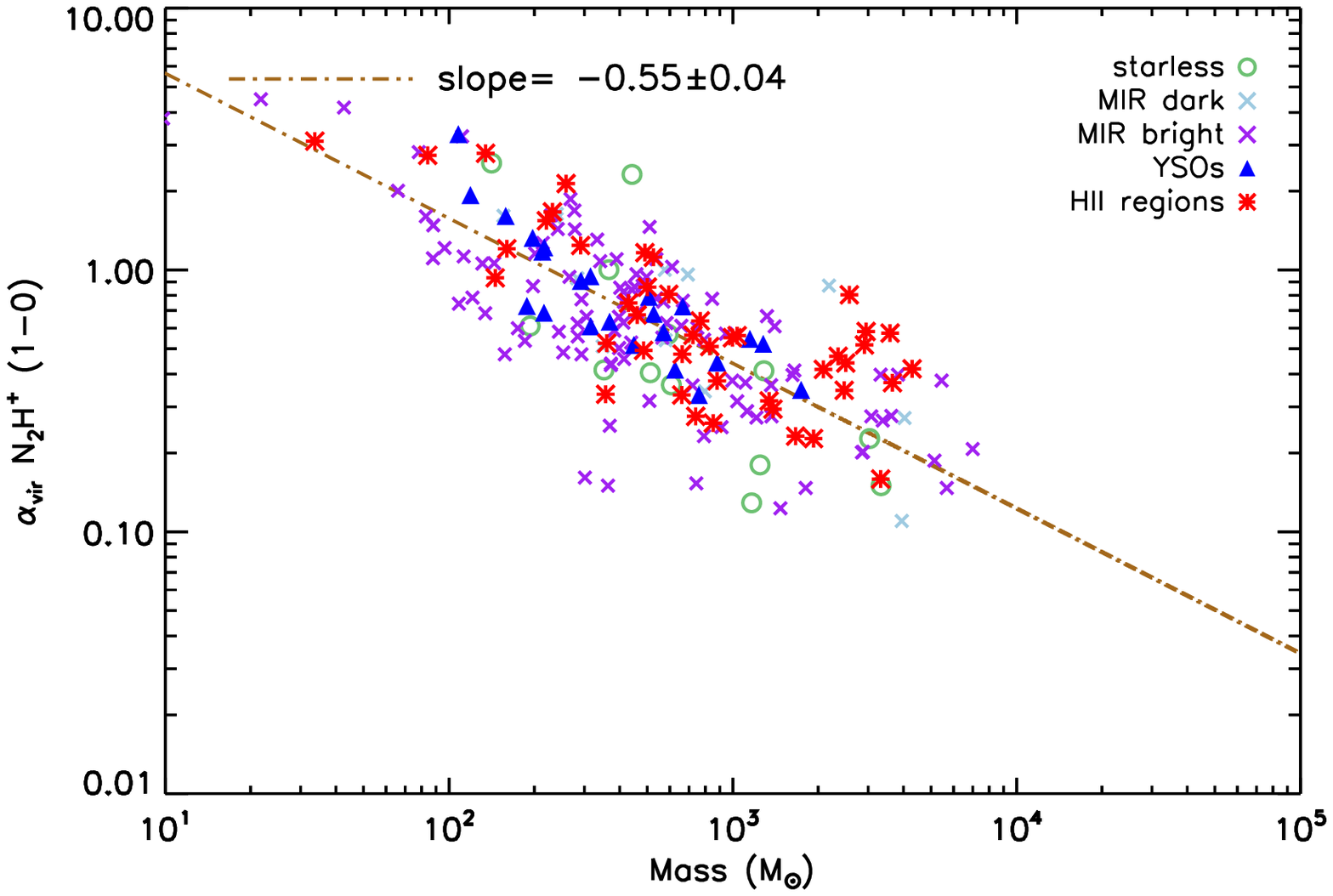} 
\caption{\textit{From the top:} same of Figures \ref{fig:mass_radius} and \ref{fig:mass_virial} but assuming the dust parameters estimated using \Hyp. The slopes of the diagrams are consistent with the findings discussed in the paper.}
\label{fig:mass_radius_Larson_Hyper}
\end{figure}

To further investigate this point, in Figure \ref{fig:mass_radius_Larson_Hyper} we show the \avir-mass diagram and the mass-radius diagram obtained from the results of the \Hyp\ photometry. The values of the slopes $\alpha$ and $\delta$ are consistent with the values derived in Section \ref{sec:Larson_third}. Also, the slope of the \avir-mass diagram is still determined from the slope of the mass-radius diagram, according to Equation \ref{eq:avir_mass_slope_no_larson}. Given a mass-radius slope of $\delta=2.28\pm0.11$, the expected \avir-mass slope would be $\alpha=-0.56\pm0.03$, in agreement with the result showed in Figure \ref{fig:mass_radius_histo_Hyper}, $\alpha=-0.55\pm0.04$.



\section{Clump properties}\label{app:clump_properties}

\begin{center}
\begin{table*}
\centering
\begin{tabular}{c|c|c|c|c|c|c|c|c|c|c}
\hline
\hline
  G304.673+00.256  & 18:15:46.0  & -16:39:08  &    0.23  &    19.9  &     229  &    6538  &  0.29  &   1.2  &   1.7  &          MIR dark \\
  G305.095+00.250  & 18:16:00.6  & -16:04:45  &    0.26  &    14.5  &     474  &     439  &  0.48  &   0.9  &   0.5  &          starless \\
  G305.137+00.068  & 18:27:09.6  & -12:42:37  &    0.41  &    17.0  &    4219  &    1738  &  1.68  &   1.4  &   0.2  &        MIR bright \\
  G305.196+00.033  & 18:26:00.4  & -11:52:21  &    0.45  &    27.9  &    2789  &  273915  &  0.92  &   1.4  &   0.4  &        HII region \\
  G305.201+00.227  & 13:05:31.2  & -62:29:59  &    0.46  &    18.2  &    1424  &    2854  &  0.45  &   1.9  &   1.3  &          MIR dark \\
  G305.562+00.014  & 13:05:38.9  & -62:14:40  &    0.34  &    31.1  &    1685  &   54175  &  0.98  &   1.3  &   0.4  &               YSO \\
  G305.822-00.114  & 13:06:34.3  & -62:33:49  &    0.28  &    18.7  &    1155  &    3216  &  1.00  &   1.1  &   0.3  &        MIR bright \\
  G307.560-00.587  & 13:10:13.3  & -62:32:33  &    0.39  &    32.1  &    2692  &  208376  &  1.19  &   1.9  &   0.6  &        HII region \\
  G308.688+00.529  & 13:10:42.3  & -62:43:16  &    0.27  &    12.9  &    1287  &    2580  &  1.19  &   0.9  &   0.2  &               YSO \\
  G308.754+00.549  & 13:11:14.1  & -62:45:05  &    0.35  &    23.2  &     795  &   19769  &  0.43  &   1.3  &   0.8  &          MIR dark \\
  G309.116+00.139  & 13:11:16.7  & -62:46:38  &    0.32  &    16.4  &     558  &    2777  &  0.37  &   0.8  &   0.4  &        MIR bright \\
  G309.235-00.458  & 13:11:09.1  & -62:33:25  &    0.19  &    14.2  &     709  &    1457  &  1.32  &   0.7  &   0.2  &        MIR bright \\
  G309.382-00.134  & 13:14:26.8  & -62:44:26  &    0.26  &    25.3  &     642  &    8596  &  0.66  &   1.4  &   0.8  &        MIR bright \\
  G309.422-00.622  & 13:16:33.2  & -62:49:42  &    0.18  &    17.8  &     421  &   11177  &  0.86  &   1.1  &   0.6  &               YSO \\
  G310.014+00.390  & 13:16:43.6  & -62:58:31  &    0.16  &    26.7  &     412  &  207612  &  1.05  &   1.0  &   0.5  &               YSO \\
  G310.373-00.303  & 13:16:48.7  & -62:50:36  &    0.23  &    15.8  &     781  &    1088  &  1.03  &   1.0  &   0.3  &        MIR bright \\
  G311.044+00.687  & 13:17:15.7  & -62:42:24  &    0.18  &    15.9  &     295  &    1054  &  0.58  &   0.9  &   0.5  &        MIR bright \\
  G311.511-00.455  & 13:32:31.2  & -63:05:17  &    0.15  &    23.2  &     183  &   27953  &  0.51  &   0.9  &   0.8  &               YSO \\
  G311.556+00.331  & 13:36:32.6  & -62:49:04  &    0.52  &    13.6  &    4027  &    1346  &  1.00  &   1.6  &   0.4  &        MIR bright \\
  G311.627+00.265  & 13:40:27.1  & -61:47:47  &    0.23  &    21.2  &     529  &    6741  &  0.69  &   1.2  &   0.7  &        MIR bright \\
  G312.070+00.081  & 13:40:58.0  & -61:45:43  &    0.08  &    24.8  &      52  &    4947  &  0.49  &   1.4  &   3.4  &        MIR bright \\
  G312.330-00.088  & 13:44:39.9  & -62:05:35  &    0.18  &    15.6  &     210  &    2129  &  0.43  &   1.3  &   1.8  &        MIR bright \\
  G312.596+00.045  & 13:46:45.1  & -62:38:58  &    0.29  &    29.9  &    1308  &  126752  &  1.03  &   1.1  &   0.3  &        HII region \\
  G314.219+00.272  & 13:47:24.4  & -62:18:07  &    0.34  &    39.5  &    1462  &   35723  &  0.86  &   1.4  &   0.5  &        HII region \\
  G314.257+00.413  & 13:48:38.6  & -62:46:08  &    0.25  &    16.5  &     387  &    1090  &  0.42  &   1.1  &   0.9  &          starless \\
  G314.993+00.095  & 13:51:38.0  & -61:39:08  &    0.14  &    19.1  &      80  &    1284  &  0.27  &   0.9  &   1.7  &        MIR bright \\
  G316.085-00.674  & 13:56:01.7  & -62:14:16  &    0.13  &    18.3  &     112  &   13664  &  0.42  &   1.2  &   2.0  &        MIR bright \\
  G316.140-00.504  & 13:59:22.2  & -61:06:30  &    0.47  &    20.9  &    5187  &   42134  &  1.54  &   1.0  &   0.1  &        HII region \\
  G316.586-00.811  & 14:05:45.9  & -62:04:50  &    0.15  &    21.6  &     180  &    4741  &  0.52  &   1.2  &   1.4  &               YSO \\
  G316.779-00.098  & 14:04:16.0  & -61:18:55  &    0.18  &    16.8  &     264  &    1186  &  0.53  &   1.4  &   1.5  &     ext. HII reg. \\
  G317.408+00.110  & 14:04:22.4  & -61:19:26  &    0.16  &    18.3  &     705  &    2972  &  1.74  &   1.1  &   0.3  &     ext. HII reg. \\
  G317.467-00.067  & 14:04:59.4  & -61:21:27  &    0.18  &    15.3  &     235  &     443  &  0.50  &   0.8  &   0.6  &        MIR bright \\
  G317.701+00.110  & 14:08:58.2  & -61:24:22  &    0.15  &    20.3  &     453  &   56352  &  1.30  &   1.2  &   0.6  &        MIR bright \\
  G317.868-00.152  & 14:09:09.7  & -61:24:21  &    0.14  &    20.8  &     426  &    1947  &  1.52  &   1.6  &   1.0  &        MIR bright \\
  G318.050+00.087  & 14:08:49.6  & -61:12:24  &    0.14  &    40.0  &     206  &  303692  &  0.67  &   0.9  &   0.7  &        HII region \\
  G320.162+00.910  & 14:11:27.5  & -61:29:23  &    0.17  &    11.4  &     445  &     144  &  0.98  &   0.7  &   0.2  &        MIR bright \\
  G320.247+00.403  & 14:13:14.9  & -61:16:52  &    0.21  &    18.0  &    1220  &   54661  &  1.84  &   0.9  &   0.2  &        MIR bright \\
  G320.285-00.309  & 14:25:13.1  & -60:31:41  &    0.45  &    25.0  &    3304  &   47480  &  1.07  &   1.7  &   0.5  &        HII region \\
  G320.382+00.178  & 14:25:05.3  & -60:22:52  &    0.87  &    16.5  &    6828  &   18366  &  0.60  &   1.3  &   0.3  &     ext. HII reg. \\
  G321.380-00.300  & 14:26:26.5  & -60:38:29  &    0.52  &    20.3  &    2908  &   33446  &  0.71  &   1.2  &   0.3  &        HII region \\
  G321.756+00.029  & 14:31:34.8  & -60:24:35  &    0.10  &    15.4  &     126  &     117  &  0.85  &   1.1  &   1.0  &        MIR bright \\
  G321.935-00.007  & 14:39:06.0  & -60:31:50  &    0.09  &    28.2  &     189  &    3678  &  1.45  &   1.3  &   0.9  &        MIR bright \\
  G322.520+00.637  & 14:42:11.7  & -60:41:02  &    0.16  &    19.6  &     308  &   21877  &  0.80  &   1.0  &   0.7  &        MIR bright \\
  G323.444+00.094  & 14:42:02.2  & -60:30:32  &    0.21  &    22.0  &     375  &    7890  &  0.57  &   0.9  &   0.5  &        HII region \\
  G323.458-00.081  & 14:46:23.4  & -60:35:47  &    0.23  &    31.5  &     929  &  164125  &  1.16  &   1.4  &   0.6  &        HII region \\
  G324.200+00.120  & 14:45:20.1  & -59:52:09  &    0.56  &     8.5  &   40990  &  370894  &  8.57  &   1.7  &   0.0  &        HII region \\
  G324.923-00.570  & 14:45:17.6  & -59:25:53  &    0.23  &    31.5  &     599  &  196146  &  0.77  &   1.2  &   0.7  &        HII region \\
  G326.340+00.505  & 14:49:07.9  & -59:24:44  &    0.30  &    15.9  &     727  &     501  &  0.55  &   1.5  &   1.0  &        MIR bright \\
  G326.427+00.913  & 14:50:59.2  & -59:50:09  &    0.17  &    16.6  &     346  &     582  &  0.83  &   0.8  &   0.3  &          starless \\
  G326.449-00.749  & 14:50:09.5  & -59:32:44  &    0.20  &    24.8  &     566  &   20554  &  0.97  &   1.4  &   0.8  &        HII region \\
  G326.472-00.377  & 14:51:11.9  & -59:16:59  &    0.19  &    27.5  &     724  &  247169  &  1.28  &   1.4  &   0.6  &        HII region \\
  G326.566+00.197  & 14:53:16.8  & -59:26:29  &    0.23  &    19.5  &     346  &   11734  &  0.44  &   1.4  &   1.4  &        MIR bright \\
  G326.653+00.618  & 14:53:43.0  & -59:08:50  &    0.23  &    13.5  &     794  &     482  &  0.98  &   1.6  &   0.9  &          starless \\
  G326.657+00.594  & 15:00:55.4  & -58:58:50  &    0.41  &    17.7  &    4635  &   41599  &  1.82  &   2.0  &   0.4  &     ext. HII reg. \\
  G326.661+00.519  & 15:04:56.2  & -57:25:28  &    0.14  &    23.7  &     334  &   11284  &  1.15  &   1.3  &   0.8  &               YSO \\
  G326.671+00.554  & 15:11:01.7  & -58:39:36  &    0.14  &    24.2  &     454  &    3557  &  1.61  &   1.3  &   0.6  &        MIR bright \\
  G326.722+00.613  & 15:07:21.1  & -57:49:21  &    0.18  &    25.0  &     643  &   49042  &  1.29  &   1.4  &   0.6  &        HII region \\
  G326.754+00.603  & 15:10:18.8  & -58:25:11  &    0.19  &    16.6  &     123  &     289  &  0.22  &   0.9  &   1.4  &          MIR dark \\
  G326.772-00.125  & 15:09:05.1  & -57:57:06  &    0.23  &    12.4  &    1270  &    5477  &  1.57  &   1.1  &   0.3  &        MIR bright \\
  G326.781-00.242  & 15:09:41.4  & -58:00:25  &    0.23  &    19.4  &     794  &    5268  &  1.03  &   1.1  &   0.4  &               YSO \\
\hline
\end{tabular}
\caption{Properties of the 213 clumps analyzed in this work. Col. 1: Clump name; Col. 2-3: Clump coordinates; Col. 4: Clump radius, defined by the \Cut\ fit at 250\mum; Col. 5-7: Clump temperature, mass and luminosity as obtained from the SED fitting; Col. 8: Clump surface density; Col. 9: Velocity dispersion obtained from the \n2h\ $(1-0)$ emission; Col. 10: virial parameter. Col. 11: Clump evolutionary phase determined as discussed in Section \ref{sec:clump_classification}.}
\label{tab:clump_parameters}
\end{table*}
\end{center}

\begin{center}
\begin{table*}
\centering
\begin{tabular}{c|c|c|c|c|c|c|c|c|c|c}
\hline
\hline
 Clump & RA & Dec & Radius & Temperature & Mass & Luminosity & $\Sigma$ & $\sigma$ & $\alpha_{vir}$ & evol. phase\\
  & ($\deg$) & ($\deg$) & (pc) & (K) & (M\sun) & (L\sun) & (g cm$^{-2}$) & km s$^{-1}$ &  & \\
\hline
  G326.795+00.382  & 15:11:54.4  & -58:09:51  &    0.21  &    19.0  &     622  &   29026  &  0.95  &   0.8  &   0.3  &        MIR bright \\
  G326.797+00.511  & 15:14:40.9  & -58:11:49  &    0.30  &    12.1  &     583  &     151  &  0.42  &   1.2  &   0.9  &        MIR bright \\
  G326.880-00.105  & 15:16:48.5  & -58:09:48  &    0.16  &    17.9  &     133  &   10907  &  0.33  &   0.7  &   0.6  &        MIR bright \\
  G326.919-00.305  & 15:17:23.0  & -57:50:47  &    0.18  &    12.3  &     763  &     462  &  1.56  &   1.3  &   0.5  &        MIR bright \\
  G326.975-00.030  & 15:18:26.5  & -57:21:57  &    0.20  &    13.1  &     905  &     588  &  1.46  &   1.3  &   0.5  &        MIR bright \\
  G326.987-00.031  & 15:19:43.0  & -57:18:04  &    0.18  &    17.5  &     569  &    4530  &  1.16  &   1.3  &   0.7  &        MIR bright \\
  G327.120+00.510  & 15:20:48.0  & -56:26:42  &    0.22  &    36.8  &     396  &  201059  &  0.53  &   1.2  &   1.0  &               YSO \\
  G327.167-00.356  & 15:28:31.5  & -56:23:11  &    0.25  &    10.5  &    1596  &     363  &  1.74  &   1.8  &   0.6  &        MIR bright \\
  G327.238-00.516  & 15:29:19.5  & -56:31:21  &    0.21  &    14.1  &     341  &     584  &  0.49  &   1.3  &   1.2  &        MIR bright \\
  G327.266-00.538  & 15:30:57.3  & -56:15:00  &    0.18  &    25.5  &     146  &    4319  &  0.29  &   0.7  &   0.7  &        MIR bright \\
  G327.272-00.574  & 15:32:51.8  & -55:56:05  &    0.14  &    22.0  &     234  &    1032  &  0.85  &   1.2  &   0.9  &        HII region \\
  G327.393+00.199  & 15:34:57.5  & -55:27:24  &    0.26  &    22.9  &    1090  &    6232  &  1.07  &   1.4  &   0.6  &               YSO \\
  G327.403+00.444  & 15:39:57.7  & -56:04:10  &    0.26  &    29.6  &    2148  &   92426  &  2.17  &   1.7  &   0.4  &        HII region \\
  G327.710-00.394  & 15:38:33.7  & -55:27:56  &    0.25  &    17.5  &    1015  &    8302  &  1.09  &   1.1  &   0.3  &        MIR bright \\
  G327.732-00.387  & 15:43:22.5  & -54:21:33  &    0.26  &    16.7  &    2039  &    8855  &  2.08  &   1.3  &   0.2  &        MIR bright \\
  G327.825-00.650  & 15:42:09.3  & -53:58:47  &    0.23  &    21.8  &     388  &   38923  &  0.50  &   1.3  &   1.1  &        MIR bright \\
  G327.947-00.113  & 15:49:18.7  & -55:16:51  &    0.15  &    18.0  &     194  &   49642  &  0.56  &   1.1  &   1.2  &               YSO \\
  G328.140-00.432  & 15:47:50.0  & -54:58:31  &    0.19  &    25.6  &     195  &    3483  &  0.35  &   0.7  &   0.6  &        MIR bright \\
  G328.256-00.413  & 15:45:53.2  & -54:27:50  &    0.12  &    14.6  &     115  &     239  &  0.53  &   0.8  &   0.9  &        MIR bright \\
  G328.899+00.350  & 15:43:36.1  & -53:57:47  &    0.24  &    17.2  &     637  &    8804  &  0.70  &   1.2  &   0.7  &        MIR bright \\
  G328.960+00.566  & 15:44:33.3  & -54:05:25  &    0.61  &    22.6  &    5584  &   81472  &  0.99  &   1.4  &   0.2  &        HII region \\
  G329.184-00.315  & 15:44:01.4  & -53:58:45  &    0.18  &    40.0  &     131  &  186015  &  0.27  &   1.3  &   2.9  &        MIR bright \\
  G329.422-00.164  & 15:44:35.3  & -54:04:40  &    0.34  &    15.8  &    2509  &   19422  &  1.42  &   1.0  &   0.2  &        HII region \\
  G329.467+00.516  & 15:44:42.9  & -54:05:42  &    0.24  &    18.4  &     510  &    8577  &  0.58  &   1.2  &   0.7  &        MIR bright \\
  G329.468+00.503  & 15:45:02.8  & -54:09:06  &    0.16  &    28.2  &     238  &   61959  &  0.63  &   1.1  &   1.0  &        MIR bright \\
  G329.524+00.084  & 15:44:57.2  & -54:07:08  &    0.25  &    23.8  &     888  &   70965  &  0.94  &   1.5  &   0.7  &        MIR bright \\
  G330.283+00.492  & 15:44:59.1  & -54:02:18  &    0.28  &    23.0  &     782  &   22014  &  0.66  &   1.2  &   0.6  &        HII region \\
  G330.673-00.375  & 15:45:12.0  & -54:01:49  &    0.29  &    16.0  &     910  &    1873  &  0.73  &   1.2  &   0.5  &        MIR bright \\
  G330.677-00.403  & 15:48:23.6  & -54:35:28  &    0.22  &    19.5  &     725  &    4882  &  0.99  &   1.5  &   0.8  &     ext. HII reg. \\
  G330.820-00.509  & 15:48:55.3  & -54:40:39  &    0.27  &    12.2  &     428  &    1558  &  0.40  &   0.7  &   0.4  &        MIR bright \\
  G330.876-00.384  & 15:46:20.8  & -54:10:42  &    0.21  &    21.2  &    1303  &  124968  &  1.98  &   1.3  &   0.3  &        HII region \\
  G330.927-00.407  & 15:45:48.5  & -54:04:31  &    0.22  &    19.6  &     812  &    7995  &  1.16  &   0.9  &   0.3  &        HII region \\
  G330.958-00.273  & 15:48:53.2  & -54:30:26  &    0.27  &    20.0  &    1029  &   50090  &  0.91  &   1.2  &   0.4  &        MIR bright \\
  G331.132-00.245  & 15:49:56.4  & -54:38:26  &    0.25  &    31.3  &    2328  &  201288  &  2.56  &   2.0  &   0.5  &        HII region \\
  G331.133-00.525  & 15:49:03.5  & -54:23:38  &    0.26  &    23.5  &     578  &   37152  &  0.56  &   1.3  &   0.9  &     ext. HII reg. \\
  G331.230-00.226  & 15:49:07.8  & -54:23:04  &    0.41  &    13.2  &     907  &    2479  &  0.36  &   1.1  &   0.6  &          MIR dark \\
  G331.273-00.375  & 15:49:06.9  & -54:21:53  &    0.41  &    11.2  &    2561  &    1568  &  1.00  &   1.0  &   0.2  &          starless \\
  G331.340+00.019  & 15:47:32.7  & -53:52:38  &    0.15  &    18.4  &     120  &   13810  &  0.35  &   0.7  &   0.8  &        MIR bright \\
  G331.342-00.346  & 15:51:29.3  & -54:31:27  &    0.30  &    22.9  &     986  &   24380  &  0.75  &   0.9  &   0.3  &               YSO \\
  G331.434-00.284  & 15:52:34.4  & -54:36:19  &    0.23  &    19.4  &     296  &    7997  &  0.37  &   1.0  &   0.9  &          MIR dark \\
  G331.505-00.343  & 15:52:49.7  & -54:36:19  &    0.15  &    18.0  &     176  &    8049  &  0.54  &   0.7  &   0.4  &        MIR bright \\
  G331.512-00.103  & 15:53:00.9  & -54:37:34  &    0.38  &    27.2  &    2938  &  261659  &  1.33  &   1.6  &   0.4  &               YSO \\
  G331.531-00.101  & 15:50:18.7  & -53:57:03  &    0.35  &    17.1  &    3595  &    4855  &  1.90  &   1.8  &   0.4  &        MIR bright \\
  G331.570-00.229  & 15:49:19.6  & -53:45:12  &    0.19  &    12.6  &     453  &     737  &  0.86  &   0.9  &   0.4  &        MIR bright \\
  G331.625+00.527  & 15:54:33.0  & -54:12:35  &    0.19  &    18.1  &     459  &    2836  &  0.82  &   1.2  &   0.7  &        MIR bright \\
  G331.638+00.501  & 15:54:38.0  & -54:11:23  &    0.31  &     9.0  &    3098  &     120  &  2.14  &   0.8  &   0.1  &          starless \\
  G331.693-00.216  & 15:56:15.8  & -54:19:58  &    0.28  &    16.1  &     463  &     920  &  0.40  &   1.2  &   0.9  &        MIR bright \\
  G331.708+00.583  & 15:53:09.6  & -53:40:25  &    0.50  &    27.8  &    2856  &    6588  &  0.76  &   1.9  &   0.7  &        MIR bright \\
  G331.723-00.203  & 15:54:34.6  & -53:50:41  &    0.15  &    17.1  &     282  &    7124  &  0.81  &   1.4  &   1.2  &        MIR bright \\
  G331.857-00.125  & 15:56:57.7  & -53:57:46  &    0.21  &    14.3  &     850  &    7133  &  1.24  &   1.1  &   0.3  &          MIR dark \\
  G331.884+00.061  & 15:52:42.6  & -53:09:47  &    0.29  &    15.3  &    1605  &    9743  &  1.23  &   1.6  &   0.5  &        MIR bright \\
  G332.094-00.421  & 15:57:28.5  & -53:52:24  &    0.18  &    27.2  &     874  &   39754  &  1.77  &   1.2  &   0.3  &               YSO \\
  G332.240-00.043  & 15:54:06.5  & -53:11:38  &    0.17  &    19.2  &     435  &    3139  &  1.06  &   1.3  &   0.8  &        MIR bright \\
  G332.278-00.546  & 15:57:28.3  & -52:52:38  &    0.16  &    13.7  &     485  &     653  &  1.26  &   1.0  &   0.4  &        MIR bright \\
  G332.294-00.094  & 15:56:51.3  & -52:40:19  &    0.17  &    22.4  &     925  &   16856  &  2.08  &   1.0  &   0.2  &        HII region \\
  G332.469-00.523  & 16:01:10.0  & -53:16:00  &    0.20  &    19.2  &     890  &   89373  &  1.41  &   1.3  &   0.4  &               YSO \\
  G332.543-00.124  & 16:01:47.1  & -53:11:41  &    0.22  &    18.8  &     712  &    2525  &  0.96  &   0.7  &   0.2  &        HII region \\
  G332.558-00.592  & 16:03:32.4  & -53:09:26  &    0.20  &    14.3  &     416  &    1381  &  0.70  &   1.4  &   1.1  &        MIR bright \\
  G332.604-00.168  & 16:02:20.2  & -52:55:18  &    0.15  &    19.6  &     169  &    1535  &  0.52  &   1.3  &   1.6  &        MIR bright \\
  G332.681-00.008  & 15:59:36.7  & -52:22:53  &    0.26  &     9.7  &    1243  &     117  &  1.20  &   0.7  &   0.1  &          starless \\
  G332.695-00.613  & 15:59:40.7  & -52:23:27  &    0.17  &    34.7  &     345  &   22181  &  0.84  &   1.3  &   0.9  &        MIR bright \\
  G332.826-00.549  & 16:01:45.2  & -52:40:13  &    0.32  &    21.2  &    5561  &  424112  &  3.60  &   2.1  &   0.3  &        HII region \\

\hline
\end{tabular}
\caption{Table \ref{tab:clump_parameters} cointinues}
\end{table*}
\end{center}

\begin{center}
\begin{table*}
\centering
\begin{tabular}{c|c|c|c|c|c|c|c|c|c|c}
\hline
\hline
 Clump & RA & Dec & Radius & Temperature & Mass & Luminosity & $\Sigma$ & $\sigma$ & $\alpha_{vir}$ & evol. phase \\
  & ($\deg$) & ($\deg$) & (pc) & (K) & (M\sun) & (L\sun) & (g cm$^{-2}$) & km s$^{-1}$ &  & \\
\hline
  G332.959+00.775  & 16:00:08.2  & -51:37:04  &    0.17  &    23.1  &     183  &   38152  &  0.41  &   1.3  &   1.9  &        HII region \\
  G333.029-00.062  & 16:03:43.6  & -51:51:45  &    0.14  &    27.0  &     172  &  122708  &  0.60  &   1.5  &   2.2  &        HII region \\
  G333.052+00.030  & 16:09:22.6  & -52:14:48  &    0.23  &    23.7  &     272  &  115420  &  0.33  &   0.8  &   0.7  &     ext. HII reg. \\
  G333.130-00.563  & 16:09:31.3  & -52:15:52  &    0.21  &    21.0  &    2145  &    7452  &  3.32  &   1.6  &   0.3  &        MIR bright \\
  G333.182-00.396  & 16:10:40.6  & -52:14:37  &    0.44  &    15.5  &    1799  &    9652  &  0.62  &   1.0  &   0.3  &          starless \\
  G333.185-00.092  & 16:10:23.1  & -52:06:59  &    0.20  &    21.1  &     413  &   59749  &  0.72  &   1.2  &   0.8  &        MIR bright \\
  G333.202-00.045  & 16:10:44.7  & -52:05:50  &    0.20  &    14.0  &     477  &    9106  &  0.80  &   1.0  &   0.5  &        MIR bright \\
  G333.203+00.295  & 16:10:17.9  & -51:58:41  &    0.16  &    15.5  &     164  &     762  &  0.43  &   0.9  &   0.9  &        MIR bright \\
  G333.234-00.062  & 16:10:59.8  & -51:50:23  &    0.23  &    40.0  &     593  &   15015  &  0.79  &   1.6  &   1.1  &        MIR bright \\
  G333.340-00.128  & 16:09:15.2  & -51:32:36  &    0.34  &    19.3  &    1024  &    5899  &  0.58  &   0.9  &   0.3  &        HII region \\
  G333.449-00.183  & 16:12:15.2  & -52:02:28  &    0.26  &    15.5  &     128  &    2745  &  0.13  &   1.9  &   8.0  &          starless \\
  G333.466-00.165  & 16:11:21.9  & -51:45:30  &    0.15  &    27.6  &     870  &   17027  &  2.43  &   1.4  &   0.4  &     ext. HII reg. \\
  G333.480-00.225  & 16:12:14.6  & -51:50:17  &    0.20  &    15.8  &     417  &    6797  &  0.69  &   1.2  &   0.8  &          starless \\
  G333.528-00.493  & 16:10:49.1  & -51:30:10  &    0.26  &    10.1  &    1180  &     110  &  1.17  &   0.8  &   0.2  &        MIR bright \\
  G333.561-00.023  & 16:12:26.4  & -51:46:13  &    0.17  &    16.0  &     294  &     523  &  0.66  &   1.4  &   1.3  &          MIR dark \\
  G333.670-00.352  & 16:12:35.8  & -51:39:42  &    0.15  &    15.0  &     131  &    1559  &  0.39  &   1.4  &   2.7  &          starless \\
  G333.755-00.231  & 16:13:11.8  & -51:39:19  &    0.22  &    14.0  &     461  &     319  &  0.64  &   1.2  &   0.8  &        MIR bright \\
  G333.759+00.363  & 16:12:10.6  & -51:28:32  &    0.16  &    21.2  &     136  &    6690  &  0.36  &   1.4  &   2.6  &        MIR bright \\
  G333.774-00.258  & 16:12:15.0  & -51:27:35  &    0.19  &    15.0  &     374  &     530  &  0.69  &   1.4  &   1.1  &        MIR bright \\
  G334.026-00.048  & 16:12:59.9  & -51:31:40  &    0.29  &    13.7  &    1388  &   37586  &  1.10  &   1.3  &   0.4  &        HII region \\
  G334.344+00.049  & 16:09:57.3  & -50:56:20  &    0.82  &    22.2  &    4023  &   58523  &  0.39  &   1.5  &   0.5  &     ext. HII reg. \\
  G334.656-00.286  & 16:10:07.9  & -50:56:54  &    0.17  &    24.0  &     128  &   65924  &  0.31  &   0.7  &   0.8  &        MIR bright \\
  G334.746+00.505  & 16:13:30.9  & -51:26:07  &    0.18  &    17.3  &     311  &     659  &  0.62  &   1.2  &   1.0  &        MIR bright \\
  G335.221-00.345  & 16:10:06.3  & -50:50:24  &    0.15  &    19.3  &     227  &    2210  &  0.64  &   1.1  &   1.0  &        MIR bright \\
  G335.284-00.134  & 16:10:01.6  & -50:49:29  &    0.15  &    16.4  &     319  &    9853  &  0.92  &   1.2  &   0.8  &        MIR bright \\
  G335.349+00.413  & 16:13:36.2  & -51:24:19  &    0.25  &    13.2  &     338  &    1445  &  0.37  &   0.8  &   0.6  &        MIR bright \\
  G335.427-00.239  & 16:13:51.8  & -51:15:21  &    0.16  &    22.1  &     302  &    3876  &  0.81  &   1.5  &   1.3  &        MIR bright \\
  G335.591+00.184  & 16:13:11.3  & -51:05:52  &    0.18  &    10.3  &     467  &     624  &  0.93  &   1.0  &   0.4  &          starless \\
  G335.688-00.813  & 16:16:16.7  & -51:18:22  &    0.31  &    17.0  &    1587  &    5194  &  1.11  &   1.0  &   0.2  &        MIR bright \\
  G335.790+00.174  & 16:15:17.2  & -50:55:58  &    0.16  &    27.9  &     951  &  194865  &  2.38  &   1.5  &   0.4  &        MIR bright \\
  G337.134+00.007  & 16:17:41.7  & -51:16:02  &    0.59  &    14.8  &    5823  &    1619  &  1.13  &   1.5  &   0.2  &        MIR bright \\
  G337.174-00.059  & 16:15:45.4  & -50:55:52  &    0.60  &    11.8  &   11423  &   53948  &  2.14  &   1.1  &   0.1  &     ext. HII reg. \\
  G337.705-00.054  & 16:16:42.9  & -50:50:14  &    0.63  &    29.2  &   17125  &  201807  &  2.91  &   2.3  &   0.2  &     ext. HII reg. \\
  G337.845-00.376  & 16:18:26.6  & -51:07:08  &    0.14  &    40.0  &      96  &  434742  &  0.35  &   0.9  &   1.4  &        HII region \\
  G337.933-00.506  & 16:17:01.5  & -50:46:47  &    0.40  &    17.2  &    3415  &   26465  &  1.45  &   1.0  &   0.1  &          MIR dark \\
  G337.973-00.519  & 16:19:09.6  & -51:06:17  &    0.26  &    12.5  &     220  &    3181  &  0.22  &   0.9  &   1.2  &          MIR dark \\
  G337.995+00.077  & 16:17:29.5  & -50:46:10  &    0.48  &    23.6  &    2705  &  222684  &  0.77  &   1.4  &   0.4  &        HII region \\
  G338.066-00.070  & 16:17:08.5  & -50:36:08  &    0.21  &    14.3  &     599  &    3048  &  0.89  &   1.2  &   0.6  &     ext. HII reg. \\
  G338.281+00.541  & 16:19:52.0  & -51:01:29  &    0.23  &    18.6  &     847  &    8074  &  1.02  &   1.3  &   0.6  &               YSO \\
  G338.325+00.154  & 16:20:12.5  & -50:53:09  &    0.12  &    22.5  &     114  &   79901  &  0.50  &   1.3  &   2.0  &        HII region \\
  G338.423-00.410  & 16:14:59.8  & -49:50:39  &    0.14  &    16.4  &     115  &    6746  &  0.39  &   0.9  &   1.2  &        MIR bright \\
  G338.461-00.244  & 16:21:22.7  & -50:52:54  &    0.21  &    25.1  &     511  &   12294  &  0.75  &   0.5  &   0.1  &        MIR bright \\
  G338.867-00.479  & 16:18:56.8  & -50:23:50  &    0.15  &     9.9  &     337  &      48  &  0.97  &   1.2  &   0.8  &        MIR bright \\
  G338.917+00.382  & 16:18:39.4  & -50:18:55  &    0.01  &    23.2  &       1  &   13928  &  0.96  &   1.3  &  12.8  &        HII region \\
  G338.927+00.632  & 16:20:47.8  & -50:38:42  &    0.27  &    19.3  &    2765  &   11858  &  2.60  &   1.7  &   0.3  &        MIR bright \\
  G338.935-00.062  & 16:21:35.8  & -50:40:50  &    0.15  &    24.7  &     176  &    8420  &  0.49  &   1.4  &   2.0  &               YSO \\
  G339.105+00.148  & 16:21:06.5  & -50:31:43  &    0.27  &    23.8  &     709  &  163227  &  0.63  &   1.4  &   0.8  &        HII region \\
  G339.284+00.134  & 16:19:46.0  & -50:18:32  &    0.22  &    20.1  &     515  &    1311  &  0.69  &   1.5  &   1.2  &        MIR bright \\
  G339.398-00.415  & 16:19:38.7  & -50:15:50  &    0.23  &    14.6  &     328  &     577  &  0.41  &   1.0  &   0.8  &               YSO \\
  G339.476+00.185  & 16:18:09.7  & -50:01:17  &    0.53  &    26.0  &    4147  &   80996  &  0.97  &   1.8  &   0.5  &     ext. HII reg. \\
  G339.622-00.122  & 16:21:18.2  & -50:30:15  &    0.17  &    24.1  &     407  &   13979  &  0.98  &   1.1  &   0.6  &               YSO \\
  G339.834+00.633  & 16:19:51.2  & -50:15:10  &    0.23  &    11.2  &     660  &    1880  &  0.87  &   0.4  &   0.1  &        MIR bright \\
  G339.924-00.084  & 16:21:42.5  & -50:28:06  &    0.31  &    18.6  &    1859  &   58883  &  1.26  &   1.3  &   0.3  &               YSO \\
  G340.055-00.244  & 16:19:28.9  & -50:04:41  &    0.14  &    23.8  &     875  &   36319  &  2.89  &   1.6  &   0.5  &        HII region \\
  G340.273-00.212  & 16:20:36.9  & -50:13:35  &    0.24  &    33.4  &    1182  &    7577  &  1.37  &   1.2  &   0.4  &     ext. HII reg. \\
  G340.307-00.377  & 16:20:07.7  & -50:04:46  &    0.23  &    12.6  &    1010  &     400  &  1.22  &   1.1  &   0.3  &          MIR dark \\
  G340.311-00.436  & 16:21:20.6  & -50:11:18  &    0.50  &    10.5  &    4239  &    2616  &  1.11  &   1.0  &   0.1  &        MIR bright \\
  G340.401-00.378  & 16:21:20.2  & -50:09:47  &    0.51  &    11.7  &    4758  &    1680  &  1.21  &   1.3  &   0.2  &        MIR bright \\
  G340.431-00.372  & 16:21:40.1  & -50:11:45  &    0.17  &    17.0  &     258  &    1284  &  0.56  &   1.1  &   1.0  &               YSO \\
  G340.785-00.097  & 16:23:04.1  & -50:20:58  &    0.47  &    33.4  &    2442  &  161404  &  0.73  &   1.3  &   0.4  &        MIR bright \\
  G340.878-00.374  & 16:22:22.7  & -50:11:51  &    0.19  &    21.4  &     974  &  125591  &  1.83  &   1.6  &   0.6  &        HII region \\
  G341.215-00.236  & 16:21:08.6  & -49:59:44  &    0.23  &    21.2  &     655  &   92934  &  0.83  &   1.1  &   0.5  &        HII region \\

\hline
\end{tabular}
\caption{Table \ref{tab:clump_parameters} cointinues}
\end{table*}
\end{center}

\begin{center}
\begin{table*}
\centering
\begin{tabular}{c|c|c|c|c|c|c|c|c|c|c}
\hline
\hline
 Clump & RA & Dec & Radius & Temperature & Mass & Luminosity & $\Sigma$ & $\sigma$ & $\alpha_{vir}$ & evol. phase \\
  & ($\deg$) & ($\deg$) & (pc) & (K) & (M\sun) & (L\sun) & (g cm$^{-2}$) & km s$^{-1}$ &  & \\
\hline
  G341.218-00.213  & 16:23:02.8  & -50:08:55  &    0.16  &    32.2  &     365  &   18395  &  0.96  &   1.4  &   1.0  &               YSO \\
  G341.282-00.295  & 16:22:53.9  & -50:00:21  &    0.26  &    15.2  &     848  &    5444  &  0.85  &   1.4  &   0.7  &        MIR bright \\
  G342.369+00.140  & 16:20:19.1  & -49:34:51  &    0.05  &    19.4  &      27  &   22213  &  0.68  &   1.3  &   3.5  &        MIR bright \\
  G342.415+00.412  & 16:23:06.0  & -50:00:36  &    0.56  &    14.2  &    3398  &   10881  &  0.73  &   0.6  &   0.1  &        MIR bright \\
  G342.484+00.183  & 16:23:17.2  & -49:40:59  &    0.63  &    22.2  &    6698  &   19631  &  1.12  &   1.1  &   0.1  &        MIR bright \\
  G342.706+00.125  & 16:21:37.1  & -49:23:28  &    0.04  &    28.8  &      30  &   52852  &  1.60  &   1.6  &   3.4  &        HII region \\
  G342.822+00.382  & 16:24:14.2  & -49:23:25  &    0.42  &    13.5  &    1594  &    2519  &  0.60  &   1.2  &   0.4  &        MIR bright \\
  G342.959-00.318  & 16:27:02.6  & -49:24:00  &    0.17  &    20.6  &     213  &   44215  &  0.51  &   1.2  &   1.2  &               YSO \\
  G343.134-00.484  & 16:23:58.3  & -48:46:58  &    0.22  &    13.6  &     321  &     128  &  0.45  &   1.4  &   1.5  &        MIR bright \\
  G343.501+00.025  & 16:27:26.2  & -49:12:34  &    0.03  &    14.8  &      14  &   13364  &  1.41  &   1.1  &   2.6  &        MIR bright \\
  G343.503-00.015  & 16:29:41.5  & -49:01:58  &    0.18  &    23.2  &     719  &  461495  &  1.47  &   1.3  &   0.5  &        HII region \\
  G343.520-00.519  & 16:29:01.3  & -48:50:27  &    0.16  &    21.3  &     348  &    3066  &  0.86  &   1.2  &   0.9  &               YSO \\
  G343.689-00.018  & 16:26:55.0  & -48:24:58  &    0.16  &    20.0  &     275  &    1224  &  0.71  &   1.0  &   0.6  &        MIR bright \\
  G343.720-00.223  & 16:30:05.7  & -48:48:42  &    0.16  &    21.3  &     295  &    3602  &  0.75  &   0.9  &   0.5  &        MIR bright \\
  G343.737-00.113  & 16:30:57.9  & -48:43:45  &    0.14  &    14.6  &     343  &     436  &  1.15  &   1.2  &   0.7  &        MIR bright \\
  G343.756-00.163  & 16:28:55.0  & -48:24:01  &    0.15  &    23.8  &     680  &   14171  &  2.12  &   1.2  &   0.4  &        MIR bright \\
  G343.938+00.097  & 16:33:43.6  & -49:00:47  &    0.11  &    14.9  &      51  &     444  &  0.27  &   1.0  &   2.6  &        MIR bright \\
  G344.101-00.661  & 16:29:47.1  & -48:15:49  &    0.14  &    21.5  &     281  &    6584  &  0.97  &   1.6  &   1.6  &        MIR bright \\
  G344.221-00.594  & 16:35:06.2  & -48:46:14  &    0.16  &    34.1  &     336  &   37800  &  0.92  &   1.3  &   1.0  &        HII region \\
  G344.246-00.670  & 16:33:29.5  & -48:03:43  &    0.05  &    10.6  &      24  &      91  &  0.58  &   1.3  &   4.0  &          starless \\
  G344.726-00.541  & 16:34:13.2  & -48:06:15  &    0.14  &    10.2  &     292  &      50  &  0.95  &   0.9  &   0.5  &          starless \\
  G345.132-00.175  & 16:34:11.1  & -47:33:24  &    0.09  &    18.4  &      98  &     747  &  0.80  &   1.1  &   1.3  &        MIR bright \\
  G345.144-00.217  & 16:34:38.7  & -47:36:28  &    0.11  &    16.8  &     107  &     896  &  0.62  &   1.0  &   1.1  &        MIR bright \\
  G345.718+00.818  & 16:33:40.1  & -47:23:28  &    0.09  &    18.5  &     207  &    1069  &  1.75  &   1.1  &   0.7  &               YSO \\
  G346.078-00.056  & 16:36:17.2  & -47:40:46  &    0.60  &    21.8  &    4806  &   83689  &  0.89  &   1.5  &   0.3  &        HII region \\
  G347.967-00.434  & 16:35:58.7  & -47:23:36  &    0.41  &    18.9  &    4779  &  118251  &  1.86  &   1.5  &   0.2  &          MIR dark \\
  G348.171+00.465  & 16:36:18.9  & -47:23:17  &    0.07  &    16.7  &     168  &     390  &  2.03  &   1.1  &   0.6  &        MIR bright \\
  G348.181+00.482  & 16:36:25.5  & -47:24:26  &    0.08  &    22.1  &     337  &    1448  &  3.41  &   1.0  &   0.3  &        MIR bright \\
  G349.092+00.106  & 16:36:15.4  & -47:19:02  &    0.47  &    40.0  &    1277  &  142104  &  0.38  &   1.7  &   1.2  &        MIR bright \\

\hline
\end{tabular}
\caption{Table \ref{tab:clump_parameters} cointinues}
\end{table*}
\end{center}

\section{\hco\ (1-0) spectra}\label{app:hco_spectra}
\hco $(1-0)$ spectra of the 21 clumps with double-peaked blue-asymmetries. The red lines in the plot are the double-Gaussian fits, and the blue crosses are the positions of the 2 identified peaks and in correspondence of the dip between the peaks.

\begin{figure*}
\centering
\includegraphics[width=8cm]{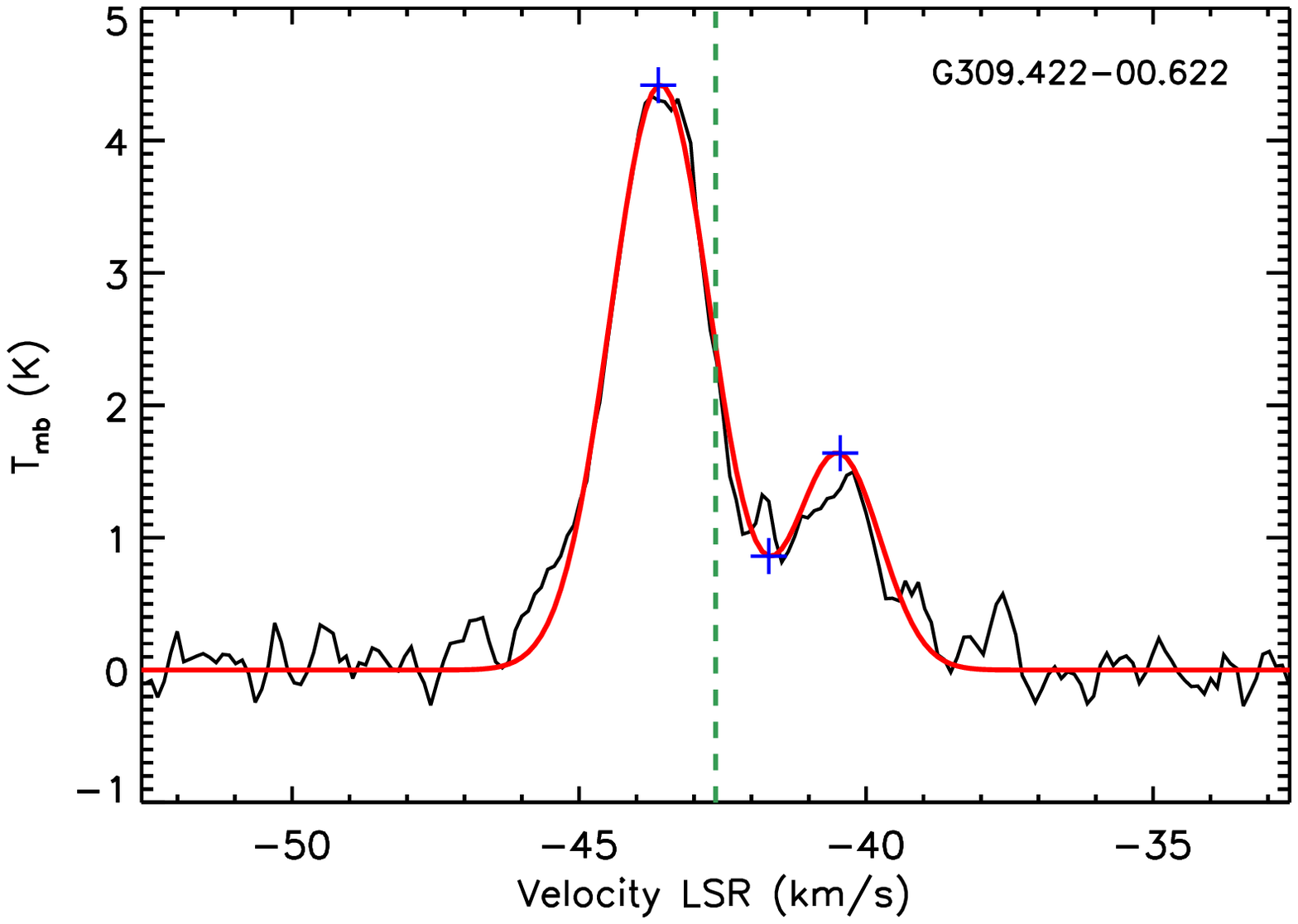} 
\includegraphics[width=8cm]{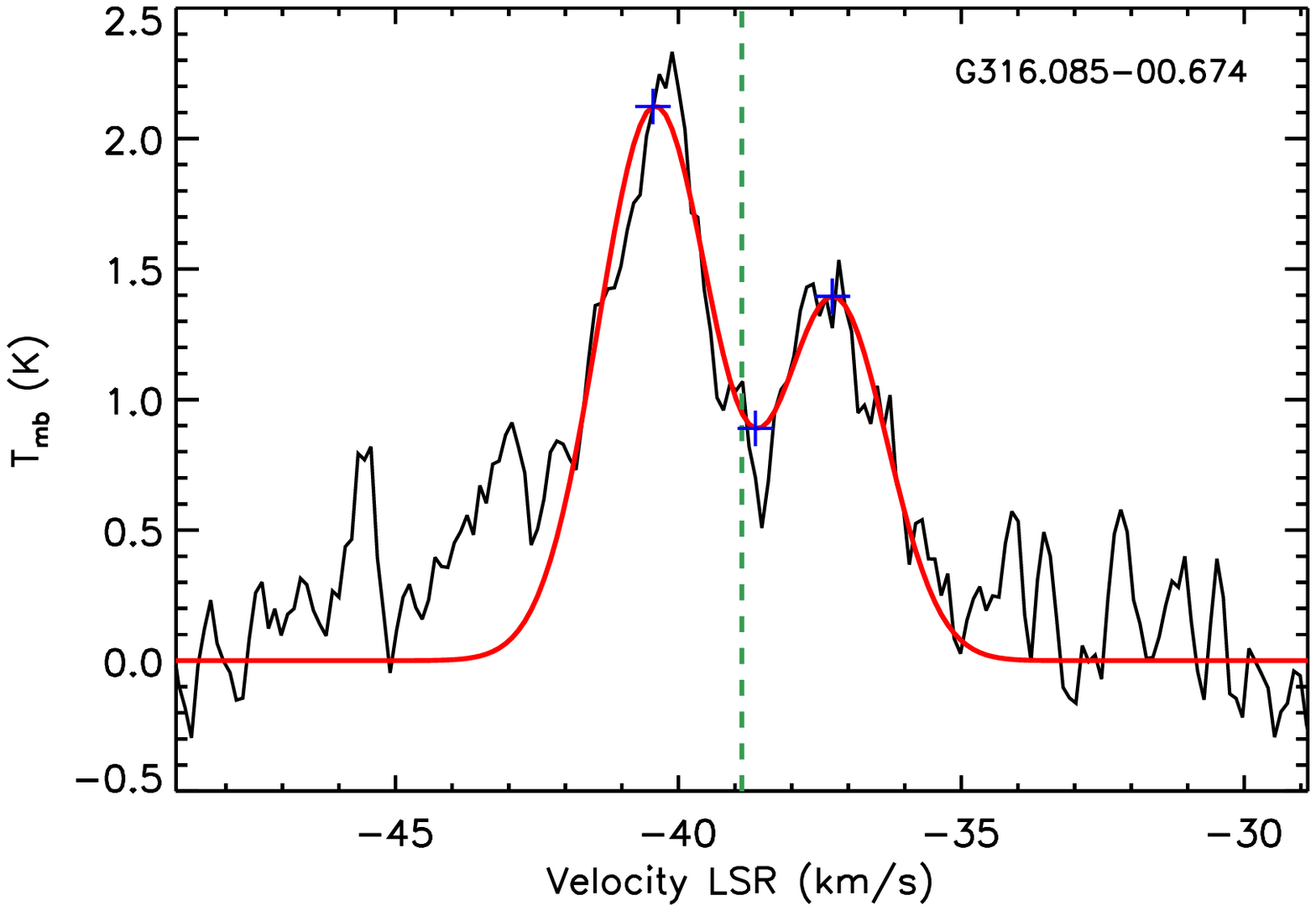} 
\includegraphics[width=8cm]{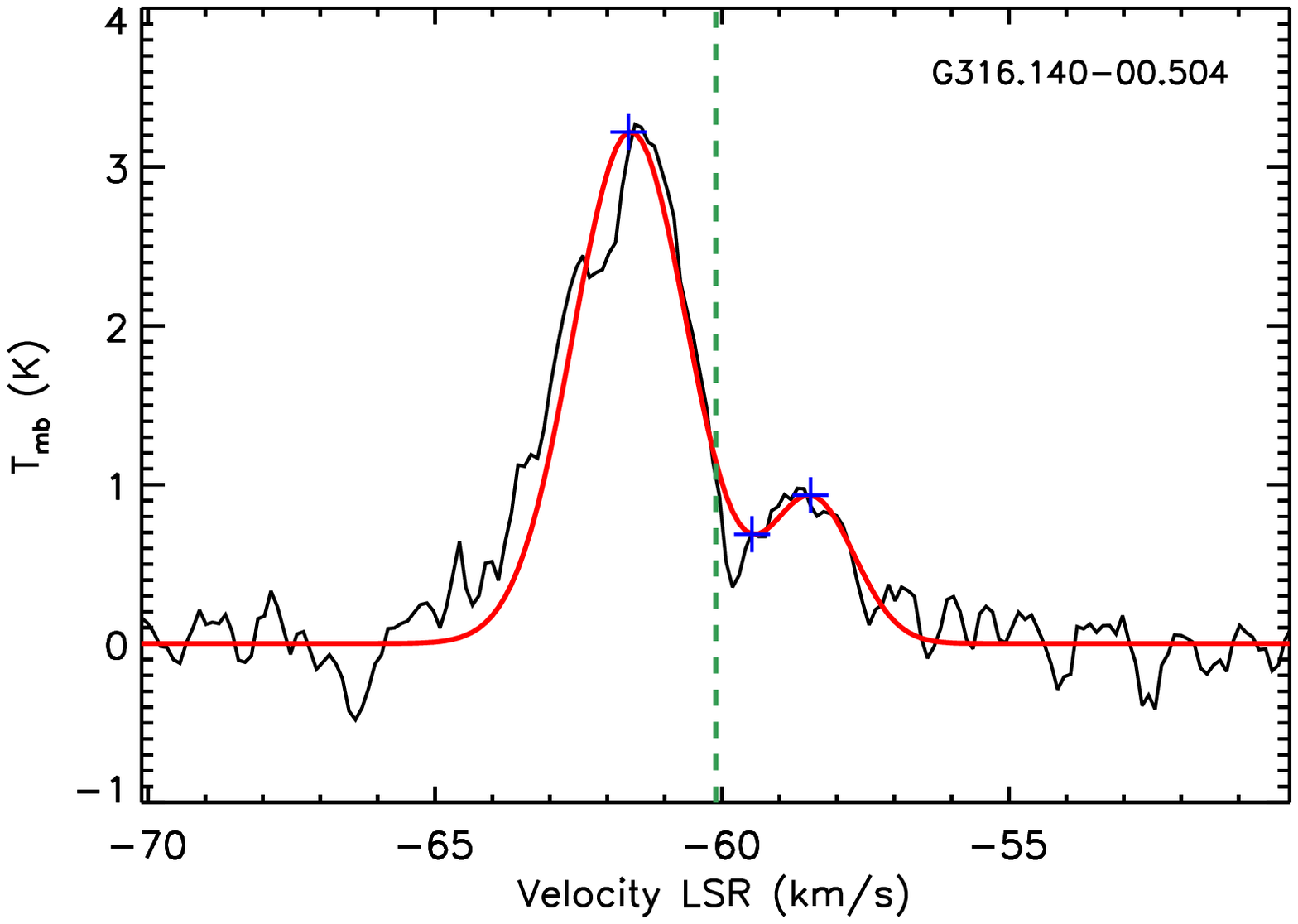} 
\includegraphics[width=8cm]{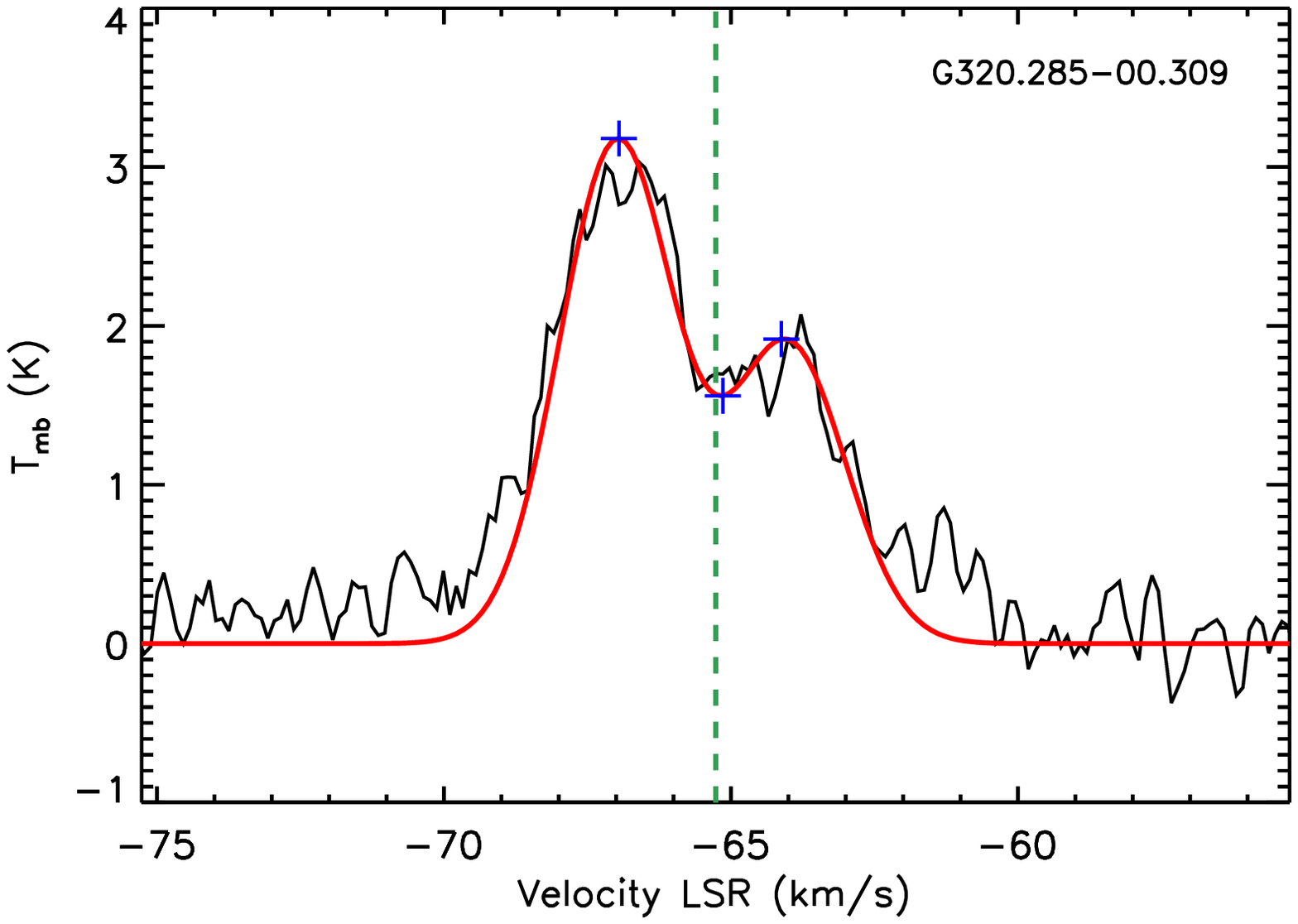}
\includegraphics[width=8cm]{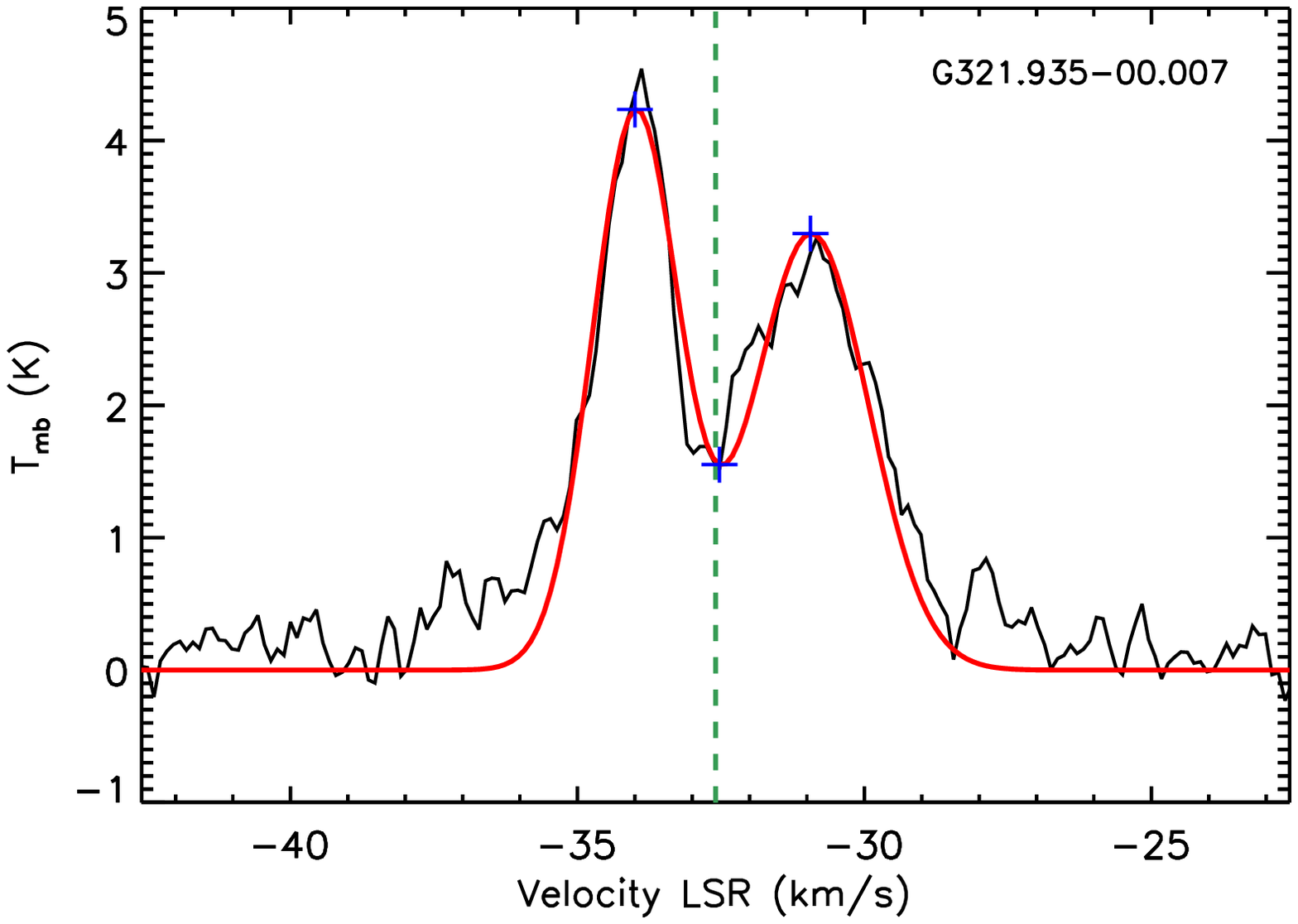}
\includegraphics[width=8cm]{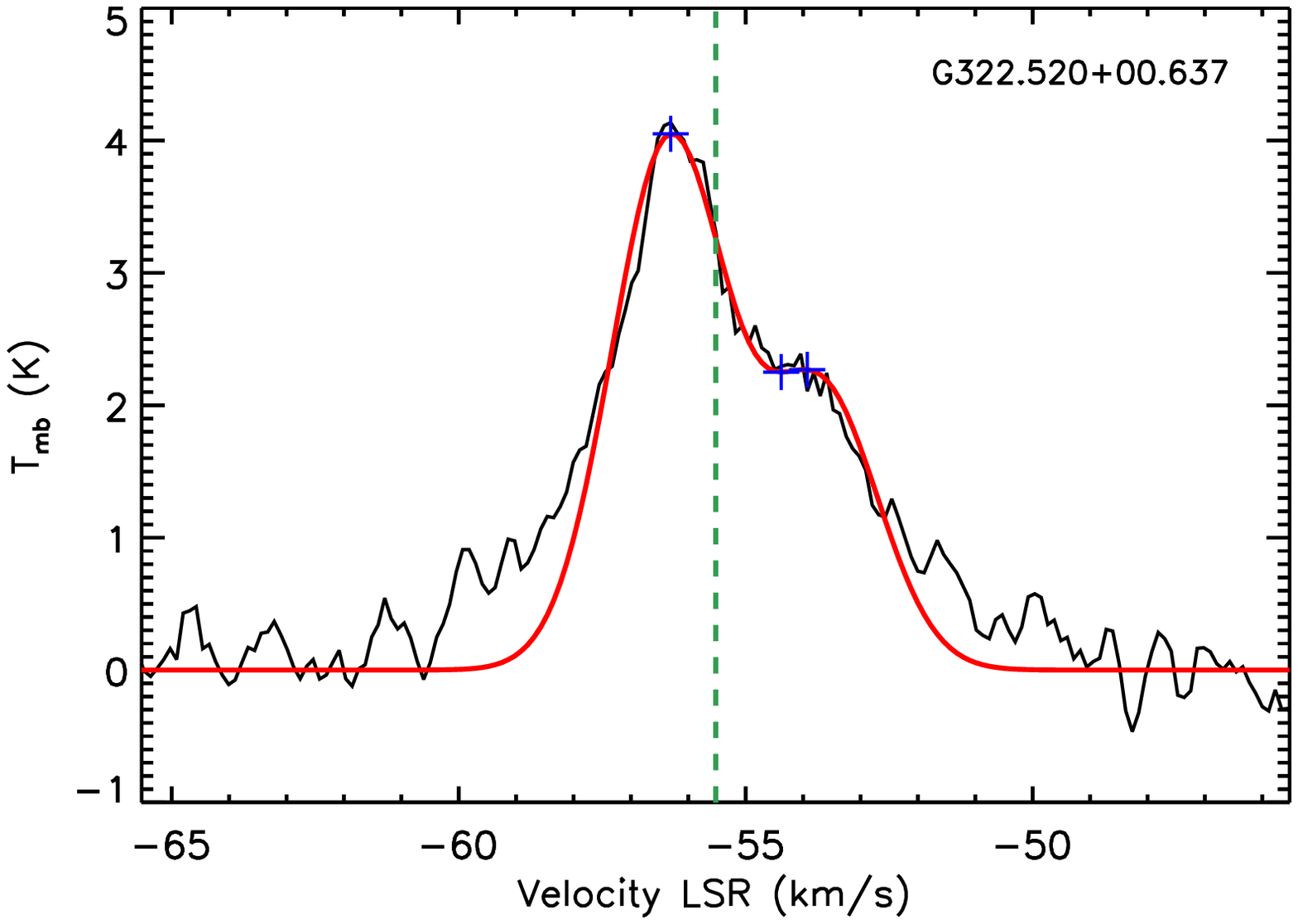}
\includegraphics[width=8cm]{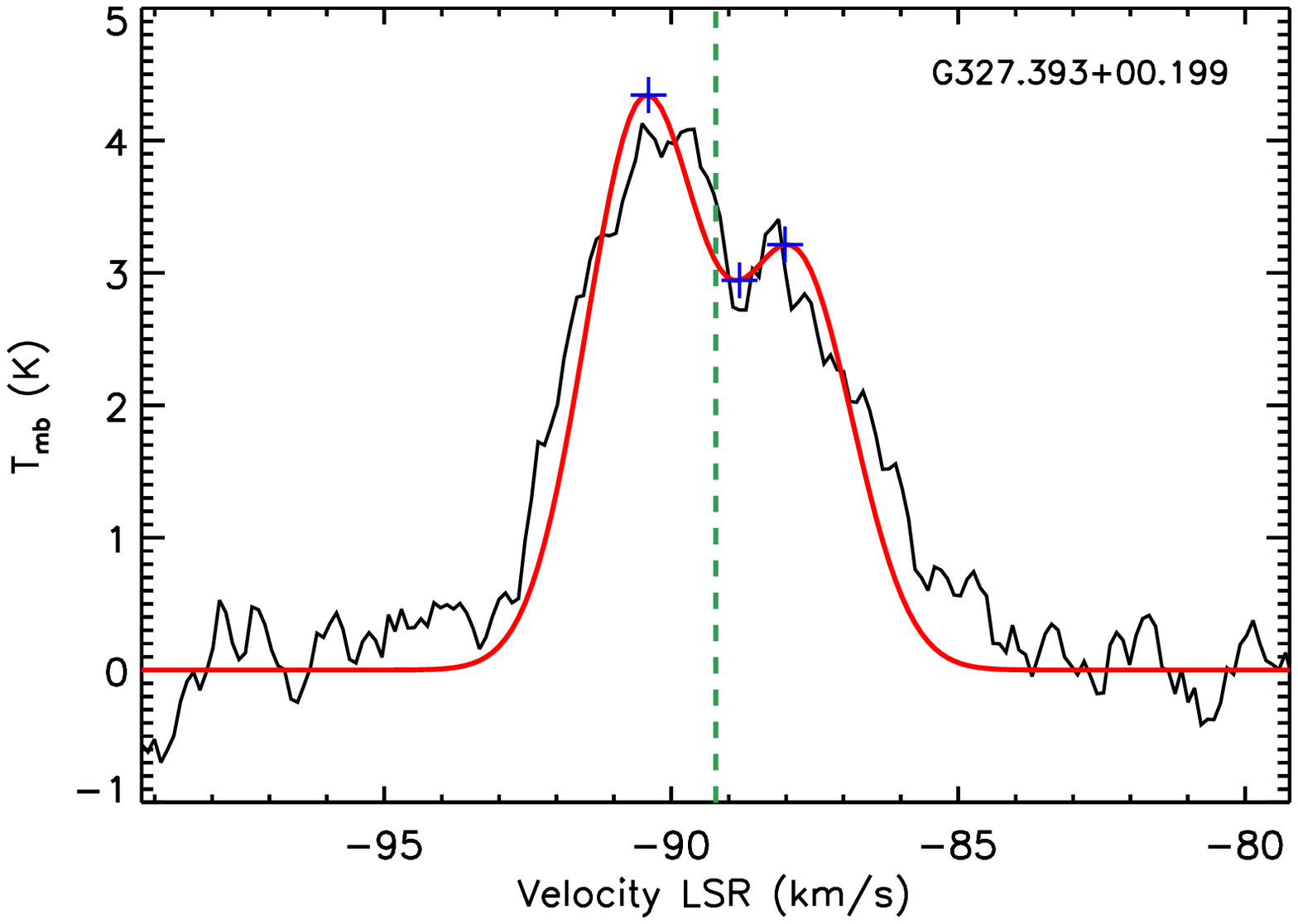}
\includegraphics[width=8cm]{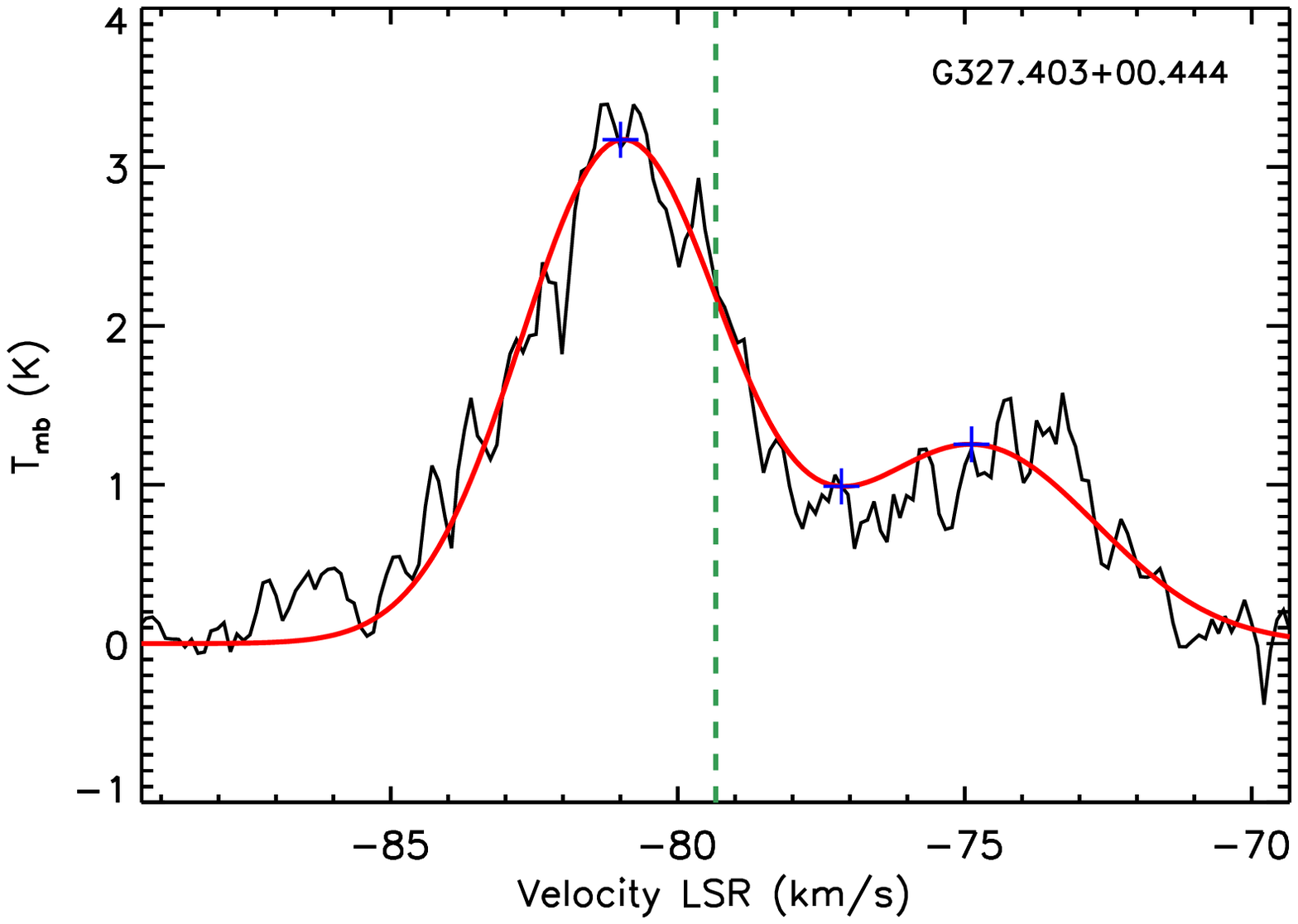} 
\label{fig:hco_spectra}
\end{figure*}

\begin{figure*}
\centering
\includegraphics[width=8cm]{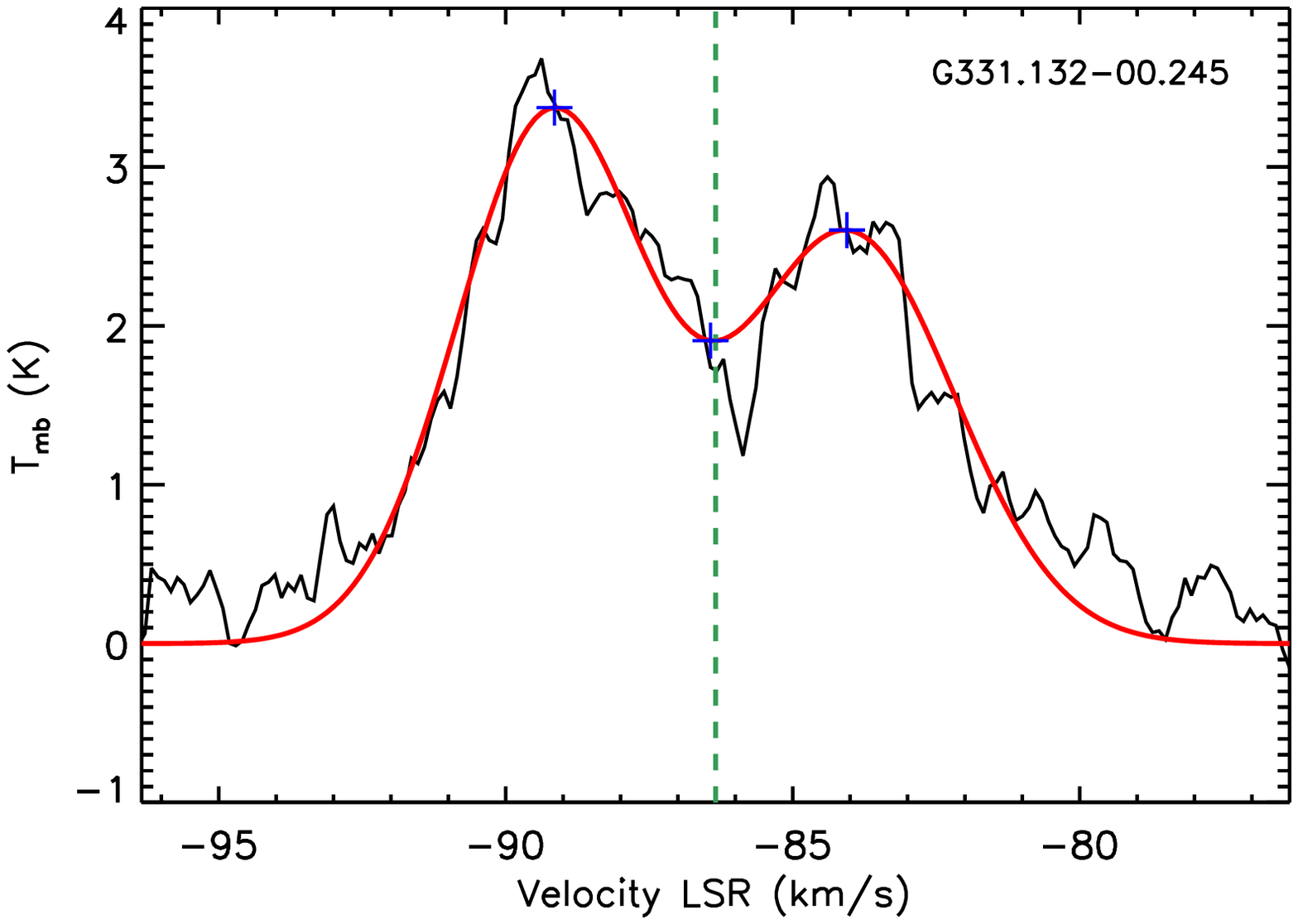}
\includegraphics[width=8cm]{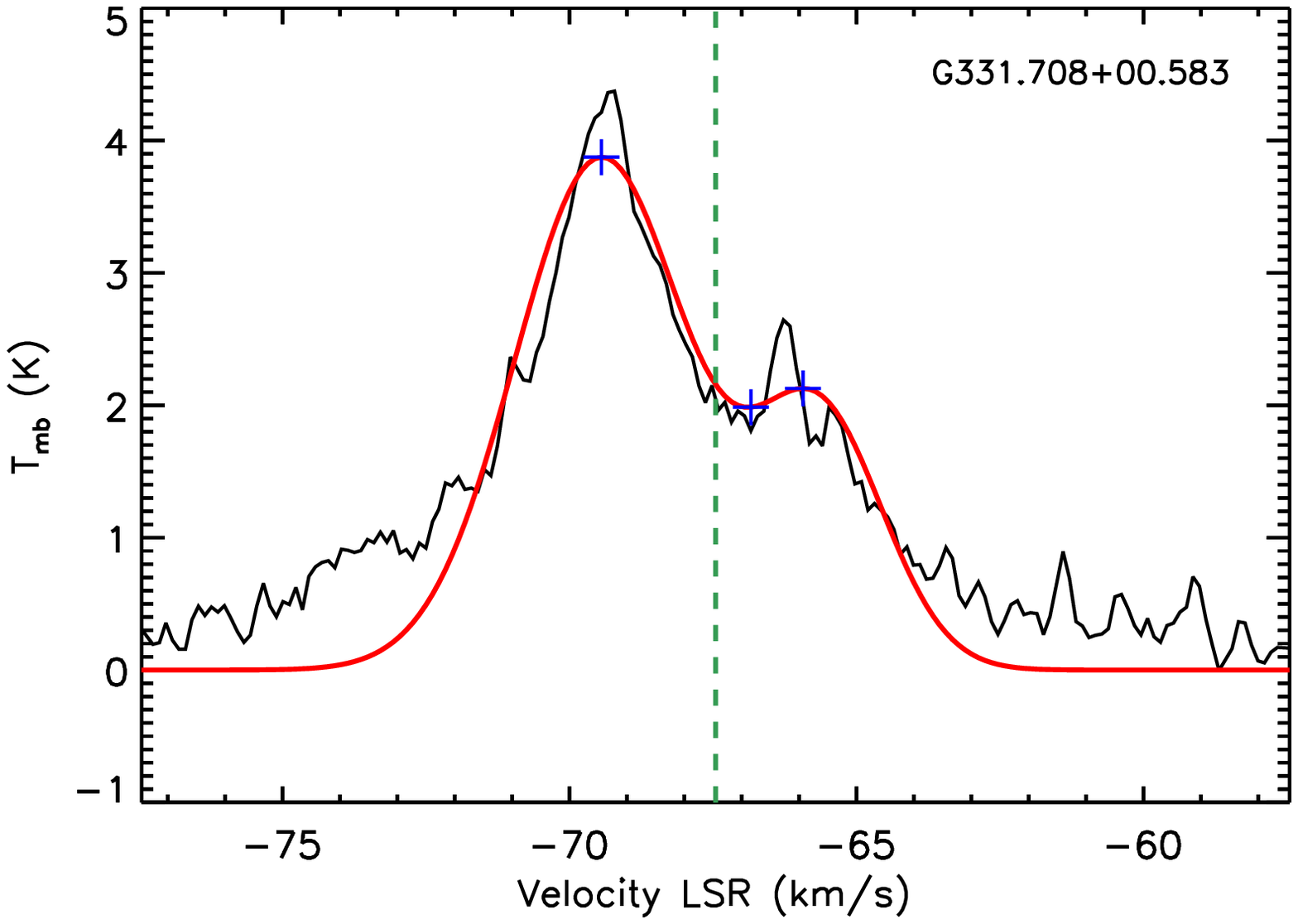}
\includegraphics[width=8cm]{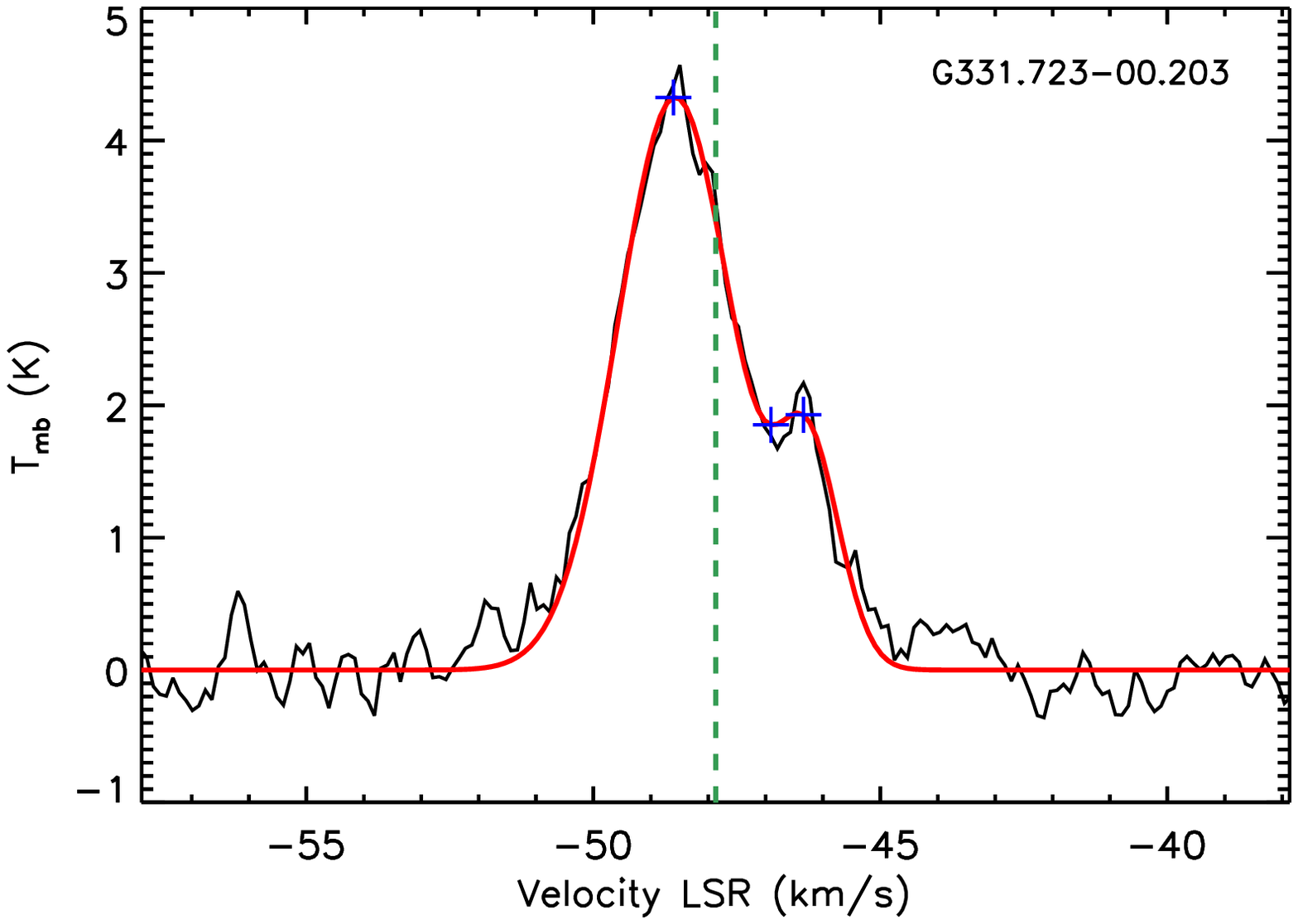} 
\includegraphics[width=8cm]{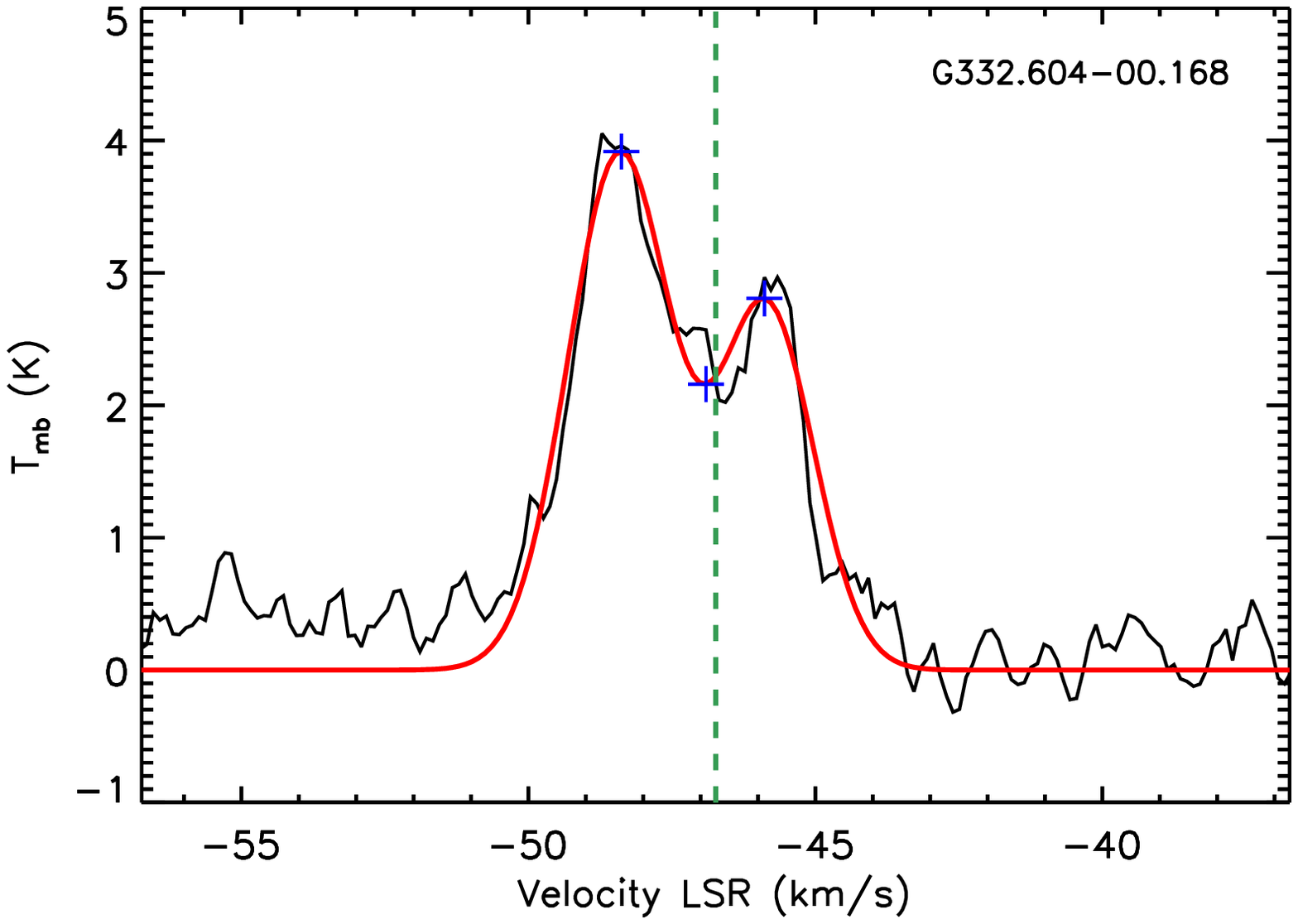} 
\includegraphics[width=8cm]{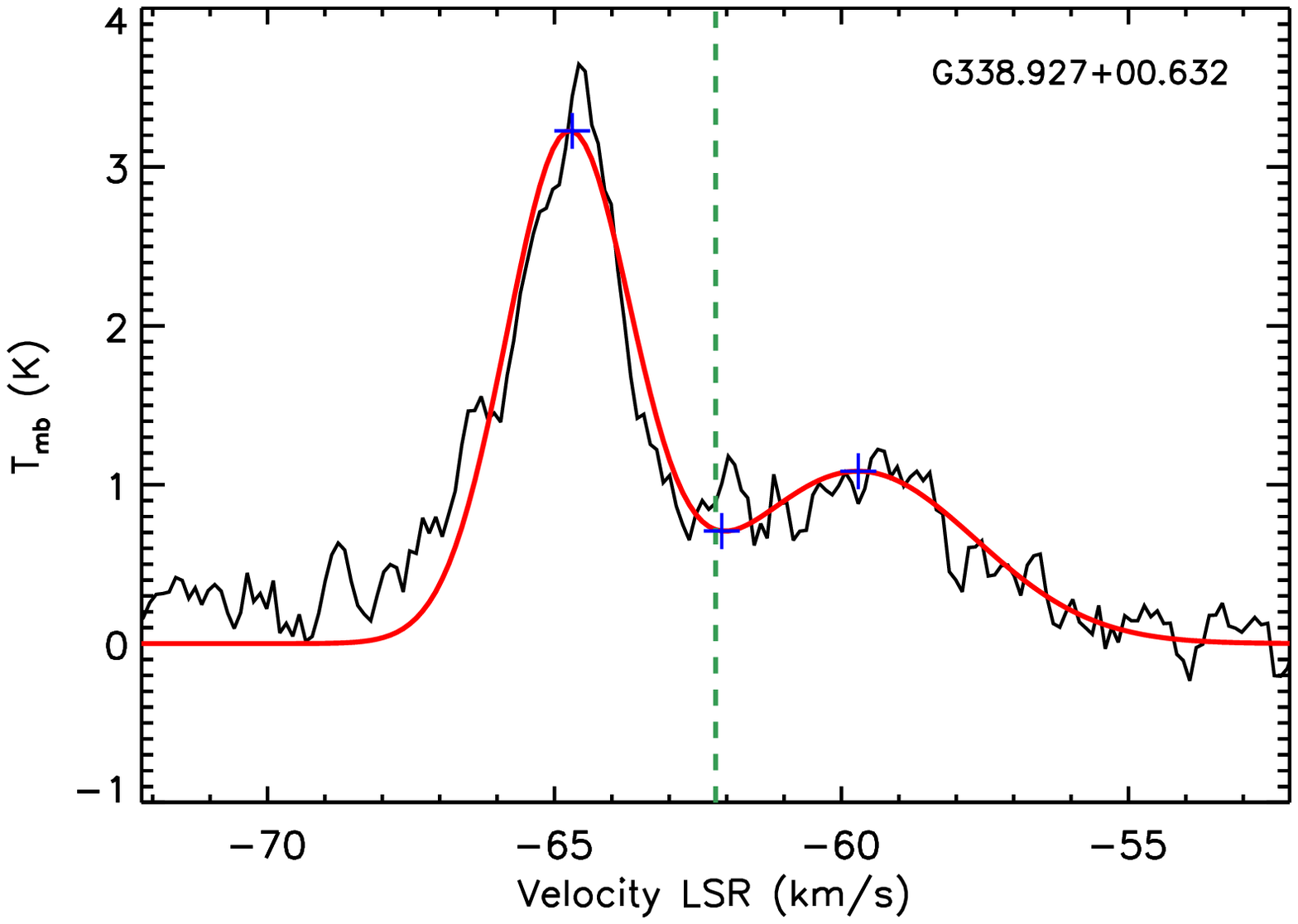}
\includegraphics[width=8cm]{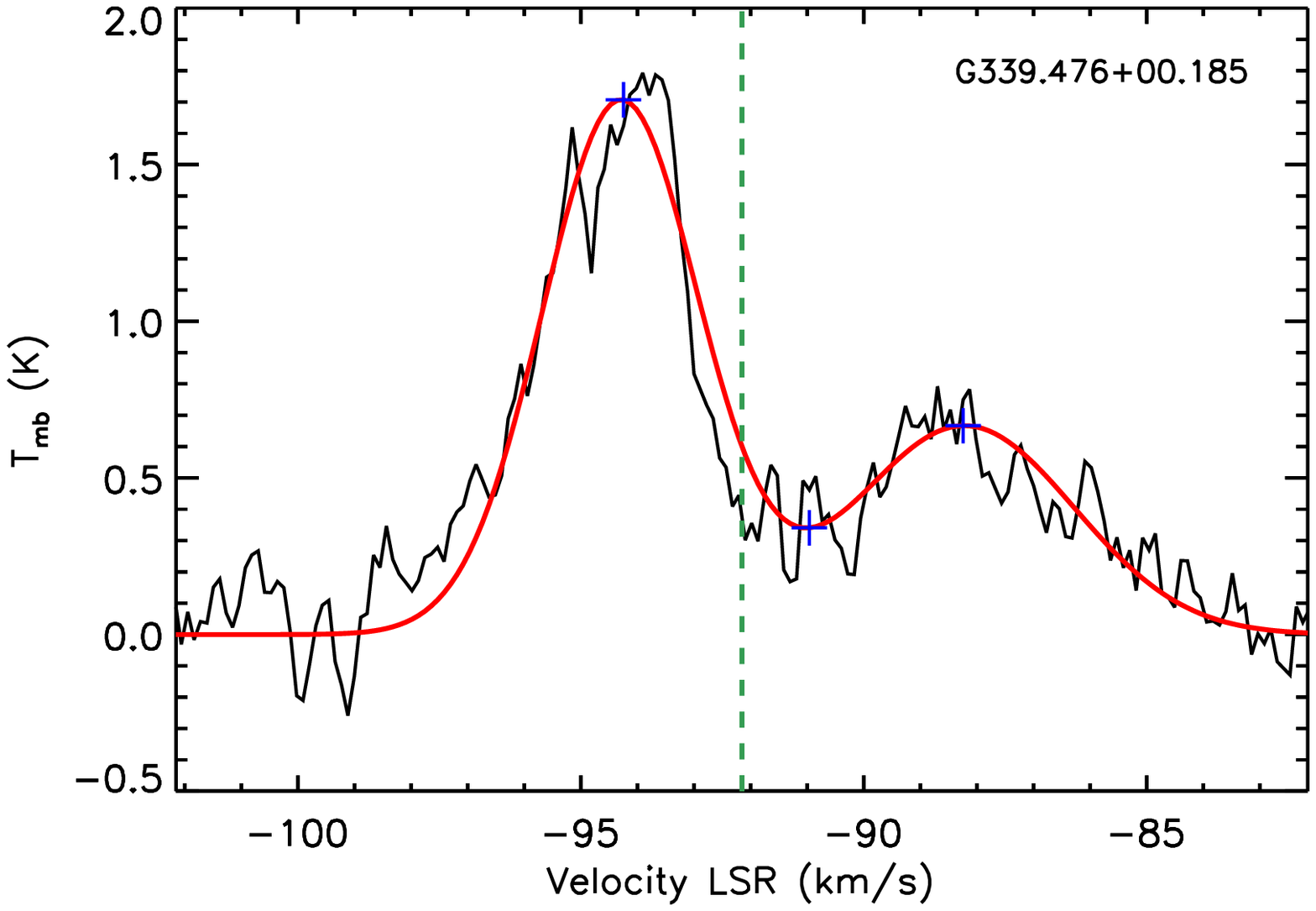}
\includegraphics[width=8cm]{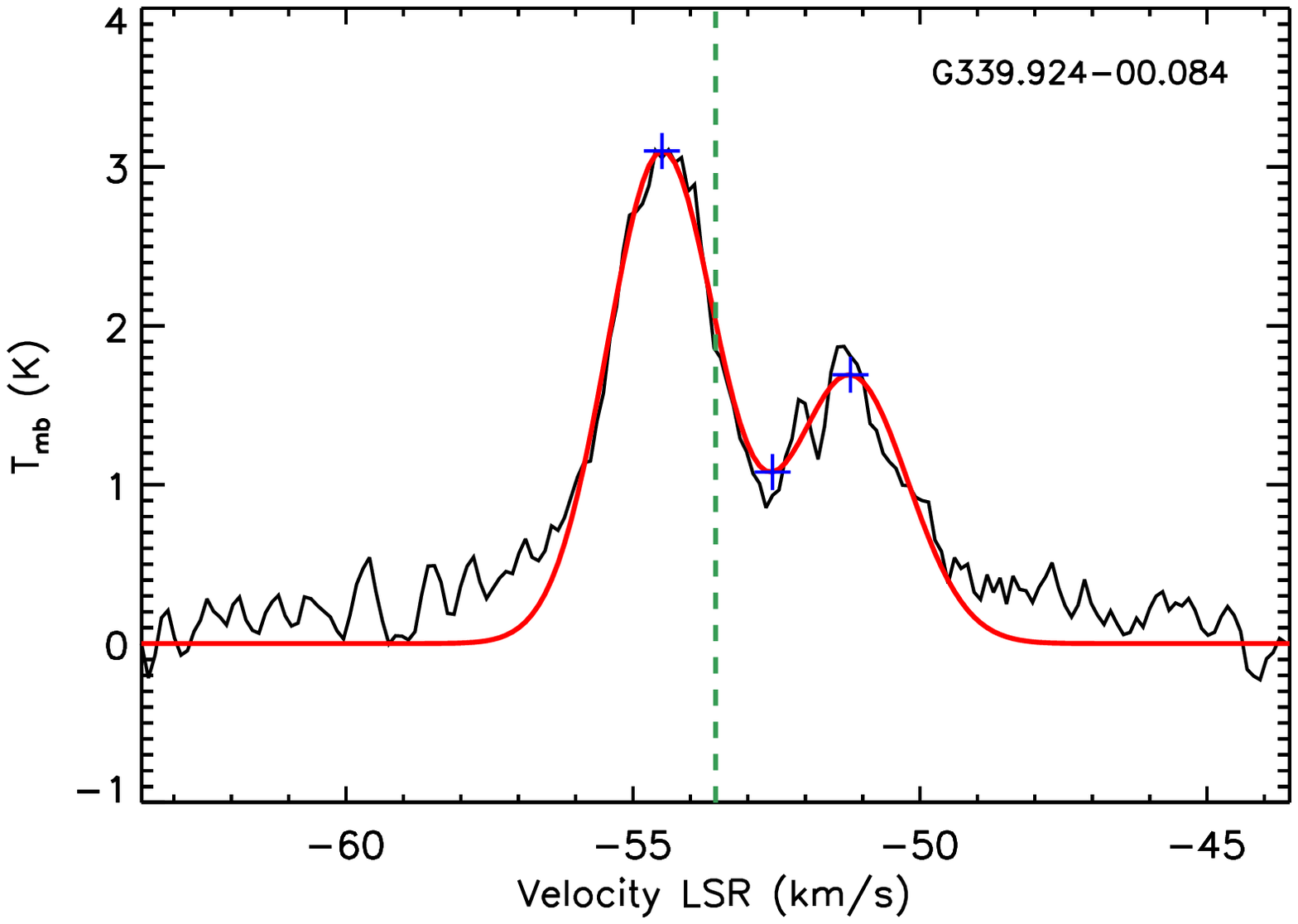}
\includegraphics[width=8cm]{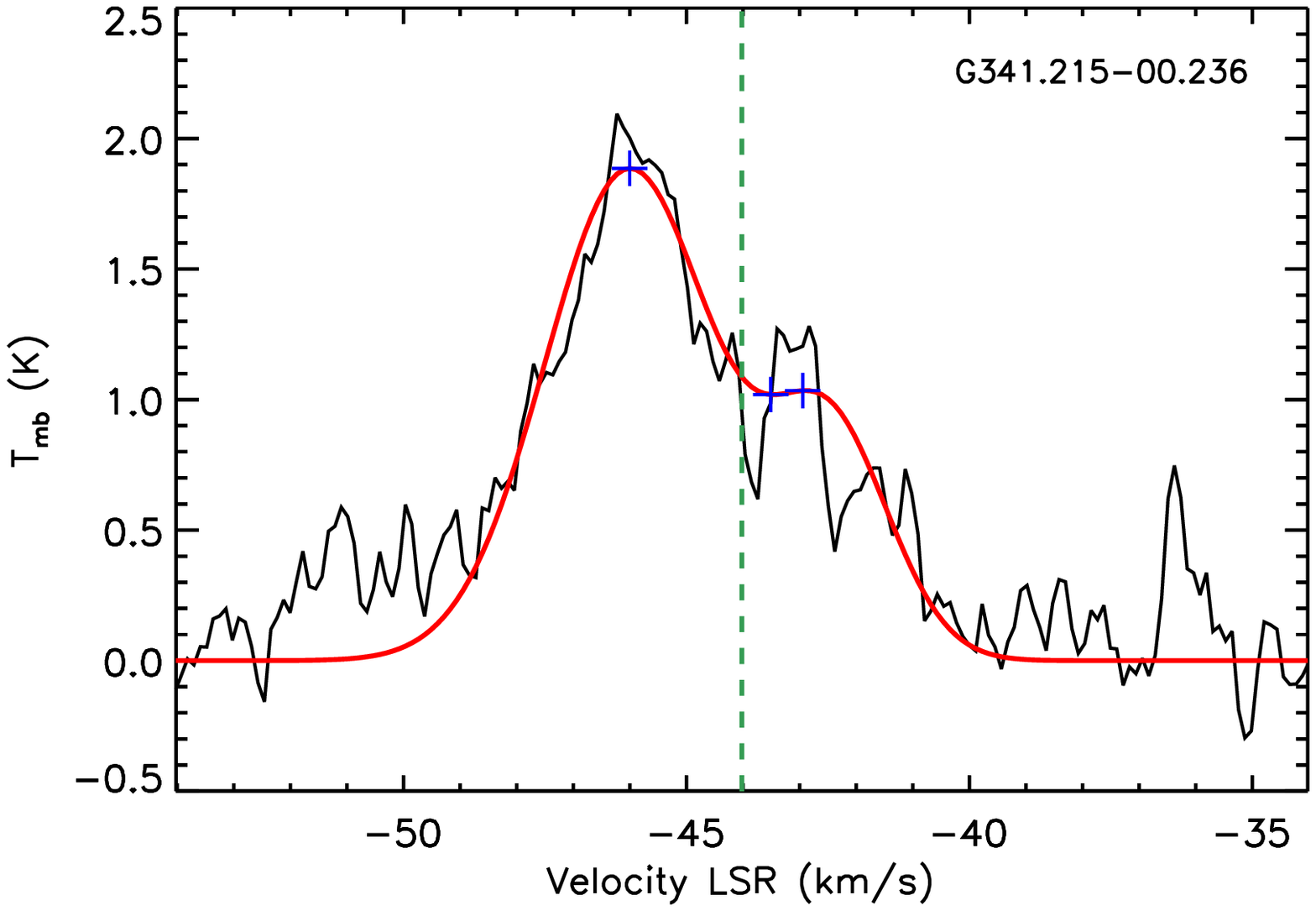} 
\end{figure*}

\begin{figure*}
\centering
\includegraphics[width=8cm]{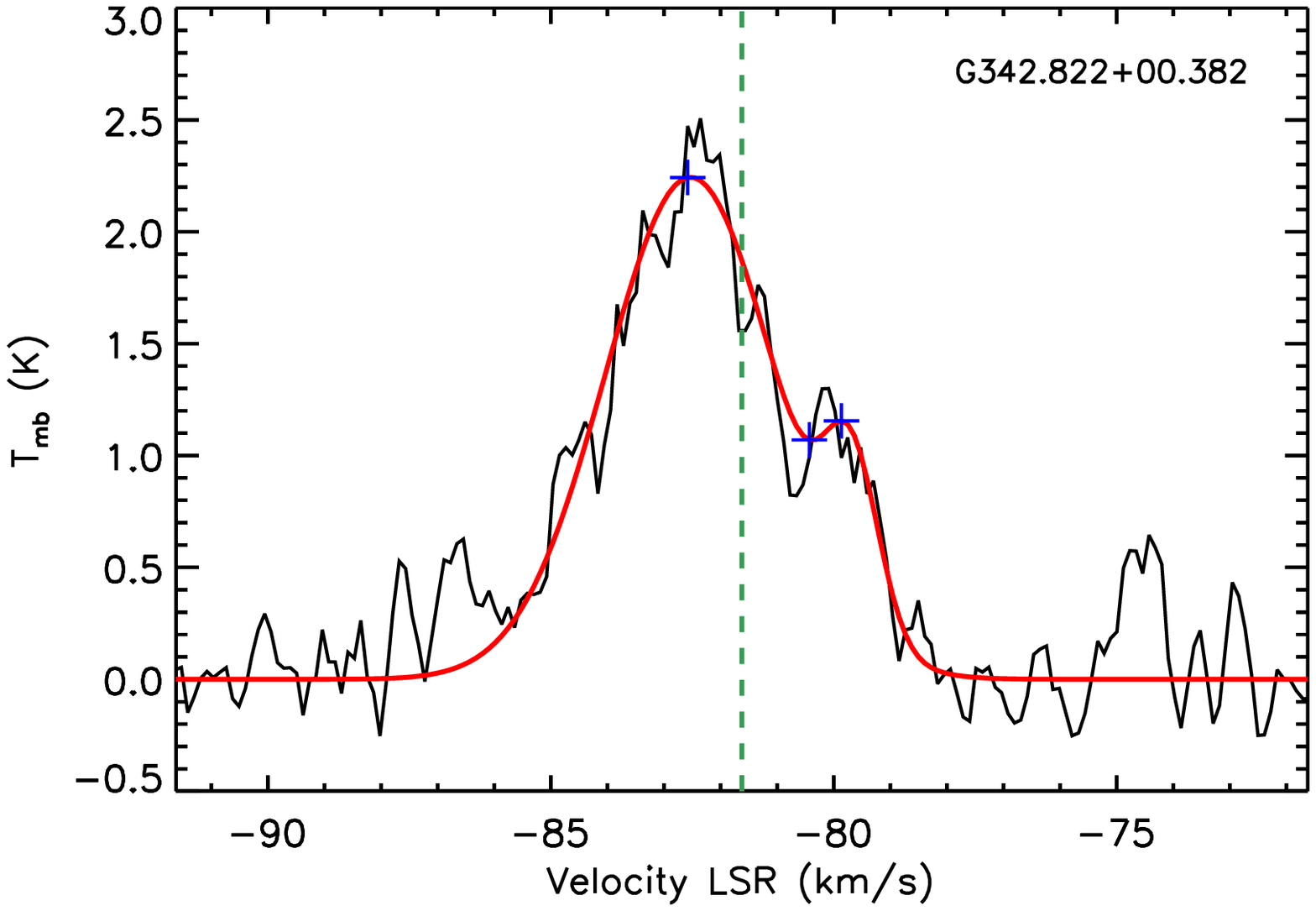}
\includegraphics[width=8cm]{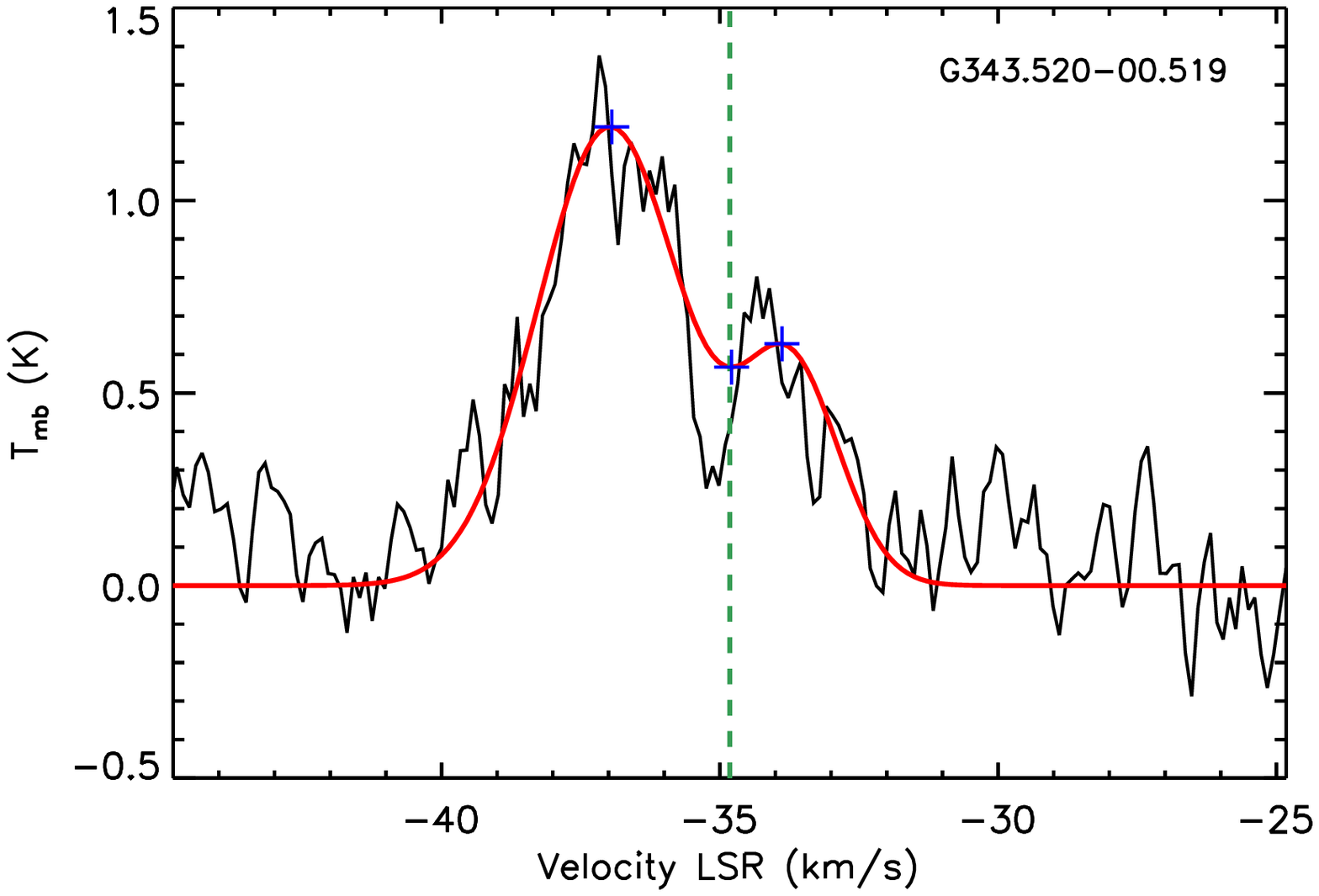} 
\includegraphics[width=8cm]{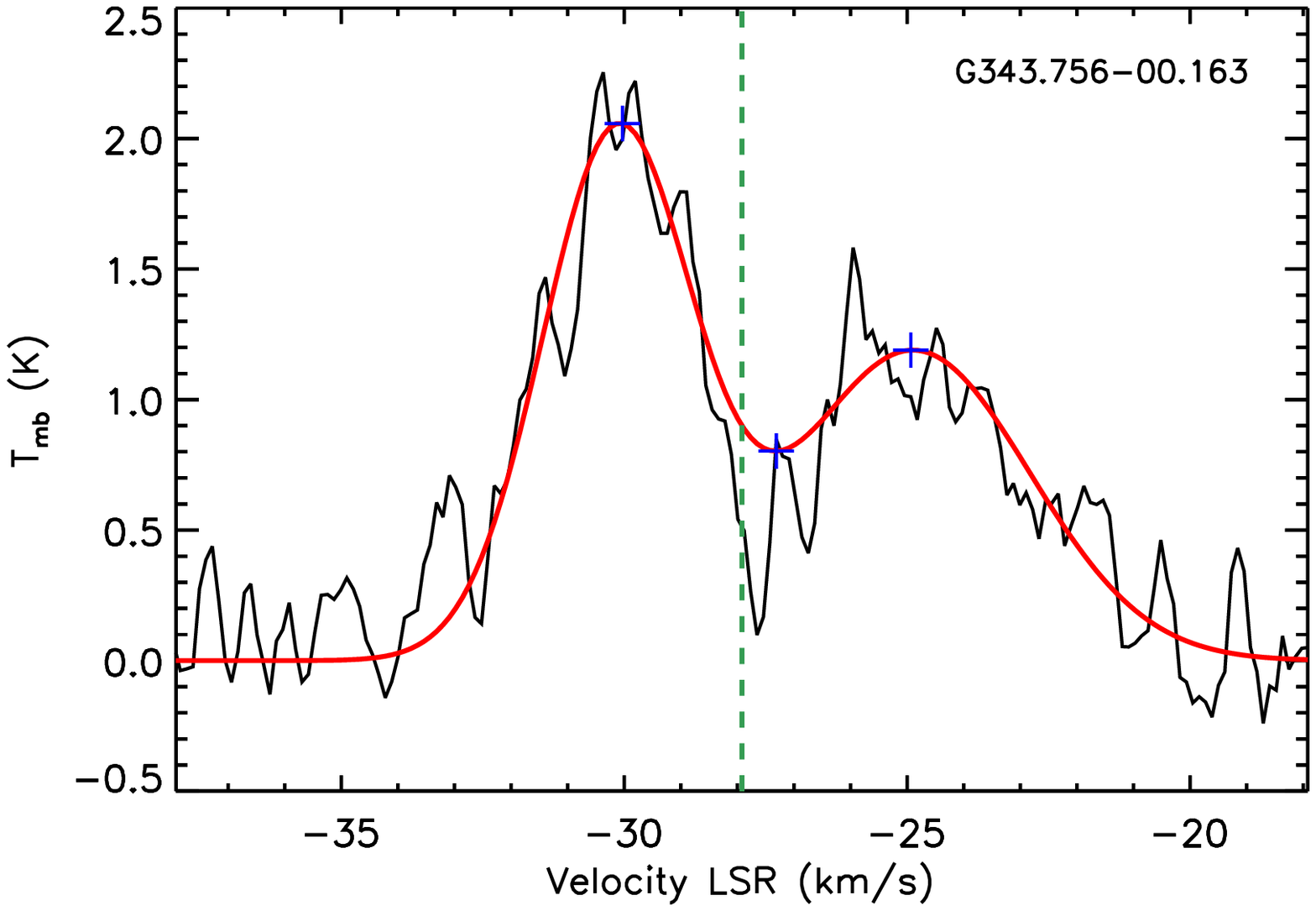} 
\includegraphics[width=8cm]{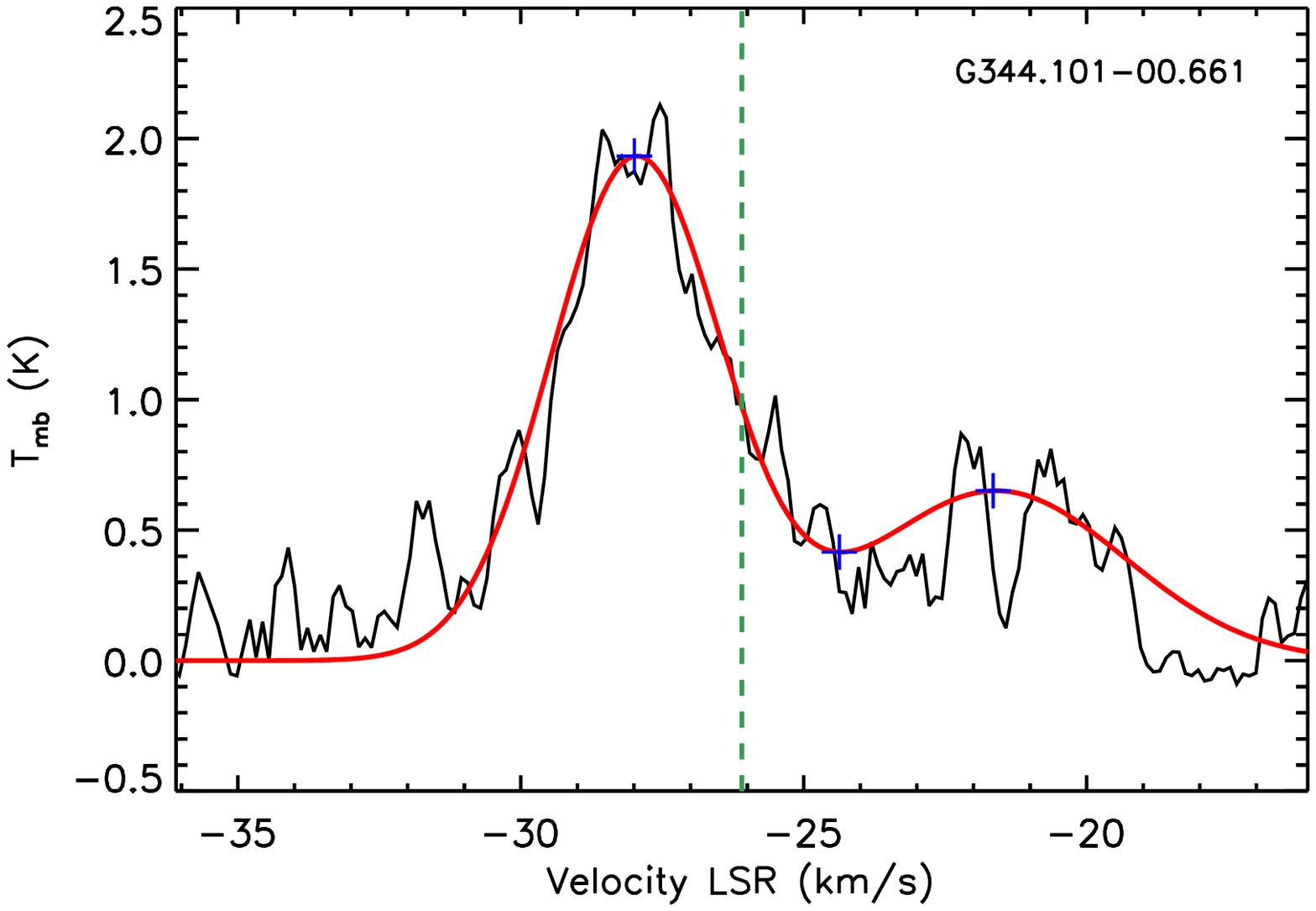}
\includegraphics[width=8cm]{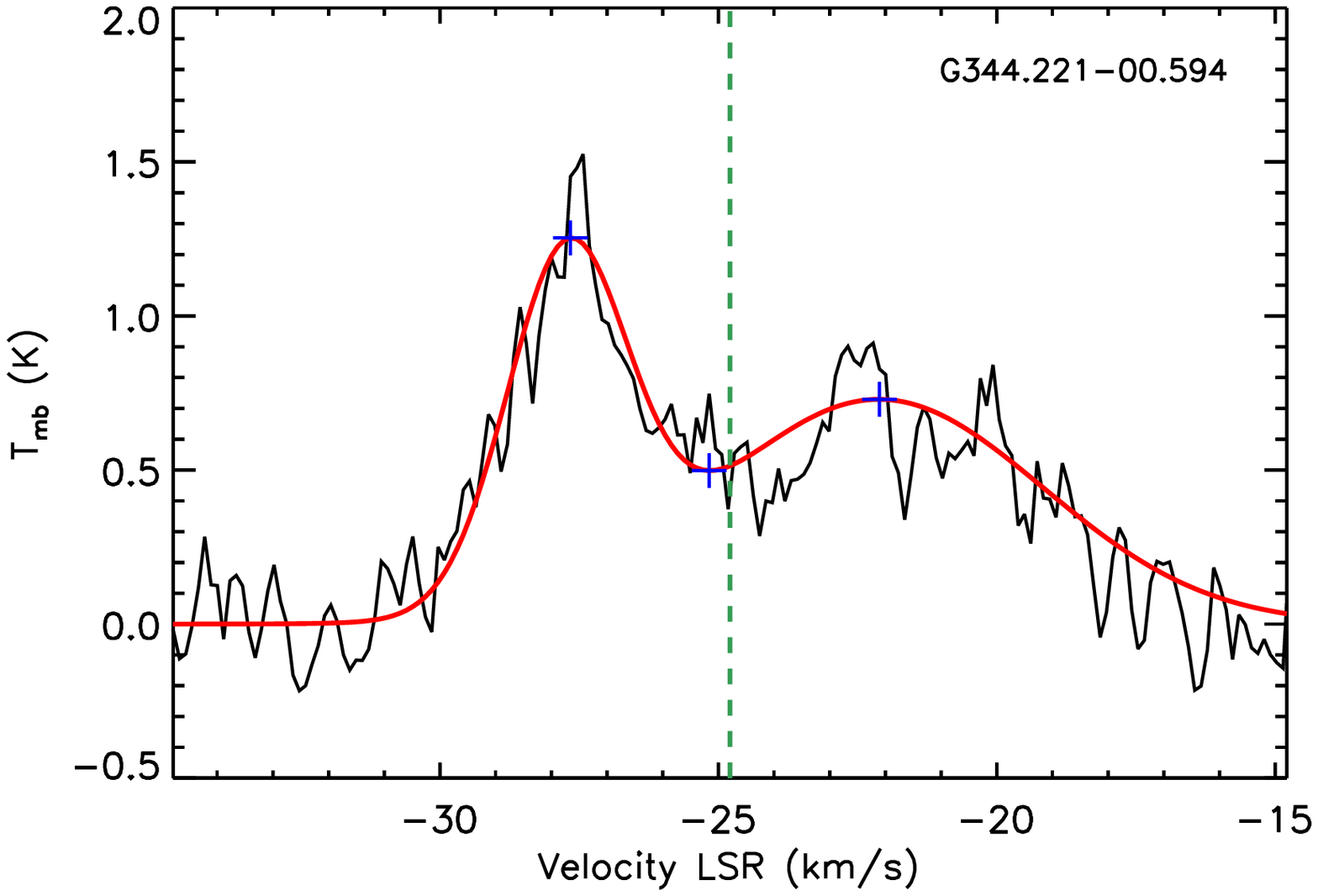}
\end{figure*}

\end{document}